\documentclass[a4paper,11pt]{article}
\pdfoutput=1 

\usepackage{jheppub} 

\usepackage[T1]{fontenc}
\usepackage{multirow,bbold,slashed,wasysym,tikzfeynman,subcaption,comment}

\allowdisplaybreaks
 
\definecolor{nicered}{rgb}{0.7,0.1,0.1} 
\definecolor{nicegreen}{rgb}{0.1,0.5,0.1}
\definecolor{niceblue}{rgb}{0.0,0.1,0.7}
\hypersetup{colorlinks,citecolor=niceblue,linkcolor=niceblue,urlcolor=niceblue}

\usepackage[normalem]{ulem}

\def \bm#1{\mbox{\boldmath$#1$\unboldmath}}
\def \beq{\begin{equation}}
\def \eeq{\end{equation}}
\def \bea{\begin{eqnarray}}
\def \eea{\end{eqnarray}}

\def \GeV{${\rm GeV}$}
\def \TeV{${\rm TeV}$}
\def \PeV{${\rm PeV}$}

\def \ZeV{${\rm ZeV}$}
\def \NLOPS{NLO$+$PS}

\title{DIS-secting the universe with high-energy neutrinos}

\author[a]{Rhorry Gauld,}
\author[a,b]{Lorenzo Guerra,}
\author[a]{Ulrich Haisch}
\author[a,c]{and Alessio Vorona}

\affiliation[a]{Max Planck Institute for Physics, \\ Boltzmannstr.~8, 85748 Garching, Germany}
\affiliation[b]{Technische Universität München, \\ James-Franck-Strasse 1, 85748 Garching, Germany}
\affiliation[c]{Nikhef, Theory Group, Science Park 105, \\ 1098 XG, Amsterdam, The Netherlands}

\emailAdd{rgauld@mpp.mpg.de}
\emailAdd{guerra@mpp.mpg.de}
\emailAdd{haisch@mpp.mpg.de}
\emailAdd{a.vorona@uva.nl}

\preprint{MPP-2026-205} 

\abstract{We revisit the impact of electroweak (EW) radiative corrections on the kinematic distributions of charged leptons in neutrino-induced charged-current (CC) deep inelastic scattering~(DIS). These corrections are particularly relevant for observables sensitive to collinear photon radiation, such as the energy spectrum of high-energy muons detected in neutrino telescopes. To study these effects, we implement full next-to-leading~order~(NLO)~EW corrections to CC~DIS in a flexible Monte Carlo framework, supplemented with higher-order logarithmically enhanced final-state QED radiation and coherent scattering contributions. We compare our results to exclusive NLO~QCD plus parton shower~(PS) predictions obtained with an existing {\tt POWHEG-BOX} event generator, examining different PS models and settings. Based on this comparison, we provide recommendations for the use of the {\tt POWHEG-BOX} event generator in experimental analyses. We~further assess the impact of radiative corrections on the correlation between the observed muon energy and the incoming neutrino energy, and on the mean inelasticity averaged over a realistic neutrino~flux. These corrections can exceed $10\%$, and above about $10^4\,{\rm GeV}$ they receive sizeable contributions from photon-initiated and coherent $W$-boson production, which are not included in the event generator.}

\begin{document} 
\maketitle
\flushbottom

\section{Introduction}
\label{sec:introduction}

Existing neutrino telescopes probe interactions over energies from a few~\GeV~to a few hundred~\PeV~\cite{IceCube:2011ucd,Formaggio:2012cpf,ANTARES:2013iuz,IceCube:2020acn,IceCube:2020wum,Ackermann:2022rqc,Baikal-GVD:2022fis}, with atmospheric and astrophysical fluxes providing broad coverage for precision measurements and indirect searches for physics beyond the Standard Model~(BSM)~\cite{Anchordoqui:2018qom,Arguelles:2022tki,Feng:2022inv}. ANTARES~\cite{ANTARES:2011hfw} and IceCube~\cite{IceCube:2008qbc} pioneered techniques to reconstruct neutrino energies, arrival directions, and flavours, now complemented by KM3NeT~\cite{KM3Net:2016zxf} and Baikal-GVD~\cite{Baikal-GVD:2018isr}, strengthening the global observational network. Future experiments employing diverse detection strategies aim for unprecedented sensitivity to high-energy phenomena across particle physics, astrophysics, and cosmology~\cite{GRAND:2018iaj,Aguilar:2019jay,Romero-Wolf:2020pzh,IceCube-Gen2:2020qha,PUEO:2020bnn,POEMMA:2020ykm,IceCube-Gen2:2021rkf}, targeting neutrinos at ultra-high energies approaching the~\ZeV~scale~\cite{Ackermann:2022rqc}.

At lower energies, up to the~\TeV~scale, accelerator-based experiments provide complementary insights. Neutrino-matter scattering measurements, such as charm production at NuTeV~\cite{NuTeV:2007uwm}, have probed hadron structure, while forward-physics facilities including SND@LHC~\cite{SNDLHC:2022ihg,SNDLHC:2023pun} and FASER$\nu$~\cite{FASER:2019dxq,FASER:2020gpr,FASER:2023zcr} have reported their first neutrino results. Upcoming initiatives, such as SHiP under the HI-ECN3 project~\cite{SHiP:2015vad,SHiP:2025ows} and the Forward Physics Facility (FPF) at the high-luminosity LHC~\cite{Anchordoqui:2021ghd,Feng:2022inv}, will further explore neutrino interactions with high precision. By combining telescope and accelerator-based observations, the field achieves a unique synergy: telescopes probe extreme astrophysical energies, while accelerator-based experiments provide high-statistics, controlled measurements, together spanning an unprecedented neutrino energy range.

Accurate simulations of neutrino-hadron interactions, particularly deep inelastic scattering~(DIS), are essential to interpret these measurements. The structure-function approach provides the standard framework for high-energy DIS, expressing cross sections in terms of structure functions that encode the momentum distributions of quarks and gluons in the nucleon. These functions capture the strong-interaction dynamics probed by the exchanged electroweak~(EW) gauge boson and can be computed as convolutions of parton distribution functions~(PDFs) with perturbatively calculated coefficient functions in QCD~\cite{SanchezGuillen:1990iq,vanNeerven:1991nn,Zijlstra:1991qc,Zijlstra:1992qd,Zijlstra:1992kj,vanNeerven:1999ca,Moch:1999eb,Moch:2004xu,Vermaseren:2005qc,Blumlein:2022gpp,Basdew-Sharma:2022vya}. QCD predictions for neutrino-nucleon DIS have steadily improved from leading order~(LO)~\cite{Dicus:2001kb,Gandhi:1998ri,Gluck:1998js,Cooper-Sarkar:2007zsa,Connolly:2011vc} to next-to-leading order~(NLO)~\cite{Cooper-Sarkar:2011jtt} and recently to next-to-next-to-leading order~(NNLO)~\cite{Bertone:2018dse,Candido:2023utz,Xie:2023suk}. Effects of heavy-quark masses, nuclear modifications of~PDFs, small Bjorken-$x_B$ resummation~\cite{Bertone:2018dse,Candido:2023utz,Xie:2023suk}, QED or EW corrections~\cite{DeRujula:1979grv,Sigl:1997cq,Seckel:1997kk,Diener:2003ss,Diener:2005me,Alikhanov:2015kla,Ballett:2018uuc,Gauld:2019pgt,Zhou:2019vxt,Zhou:2019frk,Garcia:2020jwr,Xie:2023qbn,Plestid:2024bva,Weigel:2024gzh}, target mass corrections~\cite{Kretzer:2003iu}, and effects beyond collinear factorisation~\cite{Jalilian-Marian:2003ghc,Fiore:2005wf,Block:2013nia,Albacete:2015zra,Goncalves:2015fua,Arguelles:2015wba,Le:2024wvy} have also been studied.

The structure-function approach is appealing for its inclusive treatment of QCD radiation, but it is limited to DIS cross sections differential only in quantities such as the leptonic momentum transfer $Q^2$ and/or the Bjorken fraction $x_B$. As a result, it does not provide an exclusive description of the radiation produced in the scattering process, which is crucial for experiments reconstructing the energy and direction of the incoming neutrino. To~obtain a fully exclusive description of DIS, public Monte Carlo~(MC) event generators incorporating~NLO~QCD accuracy have been developed using {\tt POWHEG-BOX}~\cite{Banfi:2023mhz,Buonocore:2024pdv,FerrarioRavasio:2024kem} or~{\tt Sherpa}~\cite{Sherpa:2024mfk}, and applied, for instance, to neutrino scattering at the LHC~\cite{vanBeekveld:2024ziz}. Fiducial~DIS~cross sections can now generally be predicted with QCD uncertainties of a few percent, depending on the neutrino energy, which is a prerequisite for fully exploiting data from next-generation neutrino experiments.

Kinematic distributions of identified charged leptons in charged-current~(CC) DIS, however, can be significantly distorted --- exceeding $10\%$ --- by hard, collinear photon emissions~\cite{Plestid:2024bva}. Experiments such as IceCube and FASER$\nu$ identify outgoing charged leptons to reconstruct neutrino interactions, but the momentum of a bare lepton is not collinear safe, as photon emission alters the measured momentum. Such emissions lead to logarithmically enhanced QED corrections, which must be taken into account. This is particularly important for through-going track events at large-volume neutrino detectors, where a high-energy muon produced outside or near the detector traverses the volume and the observed track provides a proxy for the muon's energy and its direction, which are then used to infer the incoming neutrino's energy and arrival direction. For starting-track events, the interaction vertex lies inside the detector and the inelasticity is reconstructed from the energies of the hadronic cascade and the muon track. In both cases, a collinear photon radiated off~the muon does not contribute to the reconstructed muon energy, so that the relevant observables are those of the bare muon. The~dominant corrections to these observables originate from final-state radiation~(FSR) of photons from the outgoing charged lepton in the CC DIS process. Beyond this leading effect, higher-order QED contributions, exact NLO~EW corrections~\cite{Diener:2003ss,Diener:2005me}, as well as resonantly enhanced channels~\cite{Seckel:1997kk,Alikhanov:2015kla,Ballett:2018uuc,Gauld:2019pgt,Zhou:2019vxt,Zhou:2019frk} can generate additional effects ranging from the percent level to more than $10\%$ in specific regions of phase~space.

The primary goal of this article is to provide precise predictions for the kinematic distributions of charged leptons produced in high-energy CC~DIS. To this end, we perform a fully differential calculation that includes both NLO~QCD and~EW (QCD$+$EW) corrections. The calculation is implemented in a dedicated fixed-order MC~program and supplemented with higher-order, logarithmically enhanced QED effects through lepton fragmentation functions~(FFs)~\cite{Lipatov:1974qm,Altarelli:1977zs,Arbuzov:2002pp,Arbuzov:2002cn,Arbuzov:2002rp,Bigi:2023cbv}. In addition, effects arising from coherent and diffractive scattering off~nuclei and nucleons are taken into account adopting the formalism discussed in~Refs.~\cite{Ballett:2018uuc,Gauld:2019pgt,Garcia:2020jwr}. Details of the theoretical framework and calculations are presented in~Section~\ref{sec:calculation}. The~DIS kinematics and their collinear-factorisation implementation, including QED radiation, dressed leptons, and the setup used in the NLO QCD$+$EW calculation, are summarised in~Section~\ref{sec:setup}. A comprehensive study of fixed-order neutral-current~(NC) and CC~DIS rates for~(mostly) inclusive observables over a wide energy range is presented in~Section~\ref{sec:fixedorder}. In~Section~\ref{sec:comparison}, we compare our fixed-order NLO~QCD$+$EW results with the publicly available NLO~QCD plus parton-shower~(\NLOPS) {\tt POWHEG-BOX} event generator~\cite{Banfi:2023mhz,FerrarioRavasio:2024kem}, interfaced with~{\tt Pythia~8.3}~\cite{Bierlich:2022pfr}, where QED~effects are modelled by the parton shower~(PS). By exploring different PS models and settings and benchmarking them against our new calculation, we provide recommendations for the optimal use of the existing {\tt POWHEG-BOX} event generator in experimental analyses. In~Section~\ref{sec:realistic}, we present fixed-order predictions for the inelasticity distribution over a wide range of neutrino energies, which encode the correlation between the observed muon energy and the incoming neutrino energy. In~Section~\ref{sec:astroflux}, we compare them with the corresponding {\tt POWHEG-BOX} predictions and implement an astrophysical neutrino flux in the event generator. The radiative corrections can exceed $10\%$, and above about $10^4\,{\rm GeV}$ they receive sizeable contributions from photon-induced and coherent $W$-boson production, which are not included in the event generator. Our conclusions and outlook, including the mean inelasticity averaged over the atmospheric neutrino flux, are presented in~Section~\ref{sec:conclusions} with additional technical details relegated to the appendices.

\section{Calculation of QED and EW corrections}
\label{sec:calculation}

Throughout this study, we compare various theoretical predictions for high-energy neutrino-matter scattering processes relevant to large-volume neutrino detectors. The dominant contribution to the scattering rate arises from DIS, while subleading effects originate from neutrino-nucleon, neutrino-nucleus, and neutrino-electron interactions --- see, for example,~Ref.~\cite{Garcia:2020jwr} for a discussion of the different types of contributions. In the following, we present the elements of our new calculation of the QED and EW corrections to the CC~DIS process $\nu_\ell + A \to \ell^- + X$, where $A$ denotes the nucleus in the initial state and $X$ the hadronic final state. The~corresponding calculation for an incident antineutrino is obtained from the neutrino case via CP~symmetry at the level of the partonic amplitudes, while the NC DIS process is treated in an analogous manner.

\subsection[NLO~QCD\texorpdfstring{$+$}{+}EW effects]{NLO QCD$\bm{+}$EW effects} 
\label{sec:DIScalc}

The fixed-order radiative corrections to the DIS processes are computed in a fully differential manner using the dipole subtraction method for both massless and massive particles~\cite{Catani:1996vz,Catani:2002hc}. The calculation is carried out within the framework of collinear factorisation according~to
\beq \label{eq:dsigma}
d \sigma =\sum_i \int \! dx \, f_i (x,\mu_F^2) \, d \hat{\sigma} \left( p_1, x \hspace{0.25mm} p; \mu_F, \mu_R \right)\,.
\eeq
Here, the sum runs over all parton types $i$ in the nucleon, $f_i(x,\mu_F^2)$ denotes the PDF of parton $i$ carrying momentum fraction $x$, and $d\hat{\sigma}\left(p_1, x \hspace{0.25mm} p; \mu_F, \mu_R\right)$ is the partonic differential~DIS cross section with $p_1$ ($p$) the four-momentum of the incoming neutrino (nucleon), such that $x \hspace{0.25mm} p$ is the momentum of the incoming parton. The~factorisation and renormalisation scales are denoted by $\mu_F$ and $\mu_R$, respectively. 

To achieve NLO~QCD$+$EW accuracy, the partonic differential DIS cross section is expanded as
\beq \label{eq:sigma_expanded}
d \hat{\sigma} = d \hat{\sigma}_{\rm LO} + \frac{\alpha_s}{4\pi} \, d \hat{\sigma}_{{\rm NLO}, {\rm QCD}} + \frac{\alpha}{4\pi} \, d \hat{\sigma}_{{\rm NLO}, {\rm EW}} \, ,
\eeq
where the arguments of the partonic differential cross sections have been suppressed for brevity, and $\alpha_s$ and $\alpha$ denote the QCD and QED coupling constants, respectively. The~LO contribution corresponds to the tree-level $2 \to 2$ process. 

The NLO terms in~Eq.~(\ref{eq:sigma_expanded}) arise from the sum of virtual and real corrections, denoted by $V$ and $R$, together with their corresponding subtraction counterterms, $V_s$ and $R_s$. Making no distinction between the NLO QCD and EW contributions, we write
\beq \label{eq:sigma_NLO}
\int \! d \hat{\sigma}_{\rm NLO} = \int_2 \left( d \hat{\sigma}_{V} + d \hat{\sigma}_{V_s} \right) + \int_3 \left( d \hat{\sigma}_{R} + d \hat{\sigma}_{R_s} \right) \,.
\eeq
Here, $\int_i$ denotes integration over the corresponding $i$-particle phase space, emphasising that the virtual and real contributions (and their subtraction terms) are integrated over independent two- and three-particle phase spaces. For initial-state partons, $d\hat{\sigma}_{V_s}$ also contains the collinear counterterms associated with the renormalisation of the PDFs, which involve an additional convolution in the momentum fraction of the incoming parton.

\begin{figure}[t!]
\centering
\begin{subfigure}{0.32\textwidth}
\centering
\begin{tikzpicture}[line width=1.25 pt, scale=0.8]
	\draw[fermion] (-3.5,2.5) -- (-2,2);
	\draw[fermion] (-2,2) -- (-1,2);
	\draw[vector] (-2,2) -- (-1.5,0.25);
	\draw[vector] (-1,2) -- (-0.4,0);
	\draw[fermion] (-1,2) -- (0.25,2.25);
	\draw[fermion] (-1.5,0.25) -- (-0.4,0);
	\draw[fermion] (-0.4,0) -- (0.5,-0.5);
	\draw[fermion] (-3.5,-0.25) -- (-1.5,0.25);
	\node at (0.7,2.35) {$\mu^-$};
	\node at (-3.85,2.5) {$\nu_\mu$};
	\node at (-2.25,1) {$W$};	
	\node at (-0.2,0.9) {$\gamma$};
	\node at (-3.8,-0.35) {$q$};
	\node at (0.85,-0.6) {$q^{\hspace{0.25mm} \prime}$};	
\end{tikzpicture}
\end{subfigure}
\hfill
\begin{subfigure}{0.32\textwidth}
\centering
\begin{tikzpicture}[line width=1.25 pt, scale=0.8]
	\draw[fermion] (-3.5,2.5) -- (-2,2);
	\draw[fermion] (-2,2) -- (-0.7,2.3);
	\draw[fermion] (-0.7,2.3) -- (0.4,2.7);
	\draw[vector] (-2,2) -- (-1.5,0.25);
	\draw[vector] (-1.65,1) -- (-0.7,2.3);
	\draw[fermion] (-1.5,0.25) -- (0.5,-0.25);
	\draw[fermion] (-3.5,-0.25) -- (-1.5,0.25);
	\node at (0.9,2.8) {$\mu^-$};
	\node at (-3.85,2.5) {$\nu_\mu$};
	\node at (-2.25,1) {$W$};	
	\node at (-0.75,1.35) {$Z$};
	\node at (-3.8,-0.35) {$q$};
	\node at (0.8,-0.35) {$q^{\hspace{0.25mm} \prime}$};	
\end{tikzpicture}
\end{subfigure}
\hfill
\begin{subfigure}{0.32\textwidth}
\centering
\begin{tikzpicture}[line width=1.25 pt, scale=0.8]
	\draw[fermion] (-3.5,2.5) -- (-2,2);
	\draw[fermion] (-2,2) -- (0,2.5);
	\draw[vector] (-2,2) -- (-1.5,0.25);
	\draw[dashed] (-1.8,1.55) .. controls (-0.8,1.2) .. (-1.55,0.5);
	\draw[fermion] (-1.5,0.25) -- (0.5,-0.25);
	\draw[fermion] (-3.5,-0.25) -- (-1.5,0.25);
	\node at (0.55,2.6) {$\mu^-$};
	\node at (-3.85,2.5) {$\nu_\mu$};
	\node at (-2.25,1) {$W$};	
	\node at (-0.7,1.25) {$h$};
	\node at (-3.8,-0.35) {$q$};
	\node at (0.8,-0.35) {$q^{\hspace{0.25mm} \prime}$};	
\end{tikzpicture}
\end{subfigure}
\vspace{2mm} 
\caption{Representative Feynman diagrams depicting virtual EW corrections to the CC~DIS process $\nu_\mu + q \to \mu^- + q^{\hspace{0.25mm} \prime}$, involving the exchange of photons ($\gamma$), $W$, $Z$, or Higgs~($h$) bosons. Counterterm contributions, accounted for in the renormalisation, are~omitted.}
\label{fig:virtual}
\end{figure}
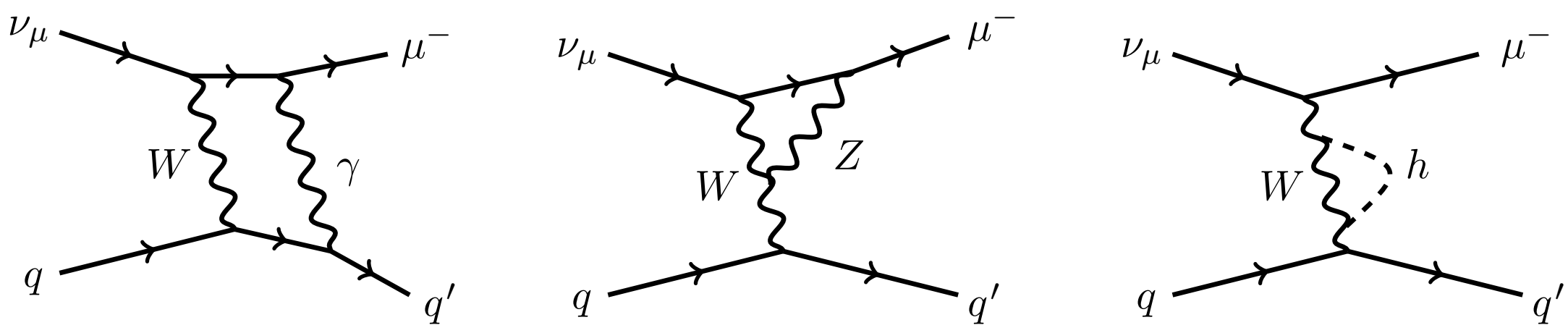

\paragraph{LO contributions:} The relevant squared amplitudes for the DIS subprocesses are computed analytically with the aid of the {\tt FeynArts}~\cite{Hahn:2000kx}, {\tt FeynCalc}~\cite{Mertig:1990an,Shtabovenko:2016sxi,Shtabovenko:2020gxv,Shtabovenko:2023idz}, and {\tt FormCalc}~\cite{Hahn:1998yk,Hahn:2016ebn} packages, employing a complex treatment of couplings and propagators. The full dependence on the quark and charged-lepton masses is retained in this calculation, allowing the results to be applied in an arbitrary heavy-flavour scheme. In~practice, we adopt a variable flavour number scheme~(VFNS)~\cite{Aivazis:1993pi} with up to five active quark flavours. Quark-mass~effects become most relevant for the~CC~DIS subprocess $\nu_{\ell} + b \to \ell^- + t$, which becomes kinematically accessible for $E_{\nu_\ell} \gtrsim m_t^2/(2 \hspace{0.25mm} m_N) \simeq 16 \, {\rm TeV}$ but is numerically relevant only at very high neutrino energies, $E_{\nu_{\ell}} \gtrsim 100 \, {\rm TeV}$. A more detailed discussion of these effects can be found in~Appendix~\ref{app:Top}.

\paragraph{Virtual corrections:} The renormalised virtual NLO QCD corrections are relatively straightforward to compute and have been obtained analytically using {\tt FeynArts}, {\tt FeynCalc}, and {\tt FormCalc}. In contrast, the calculation of the virtual EW corrections, including the required renormalisation counterterms, is more involved and has been discussed in detail in~Ref.~\cite{Diener:2003ss}. Representative Feynman diagrams for these corrections in the CC~DIS process $\nu_\mu + q \to \mu^- + q^{\hspace{0.25mm} \prime}$ are shown in~Figure~\ref{fig:virtual}. 

In this work, we derive analytic expressions for the virtual corrections in terms of one-loop scalar integrals, which we evaluate using {\tt OneLOop}~\cite{vanHameren:2010cp}. We have also employed equivalent expressions obtained directly with {\tt OpenLoops}~\cite{Buccioni:2019sur} and/or {\tt Recola~2}~\cite{Denner:2017wsf}. The calculation is performed consistently within the complex-mass scheme~\cite{Denner:2005fg,Denner:2019vbn,Denner:2006ic}, and the electromagnetic coupling is taken in the $G_F$ scheme~\cite{Denner:2000bj}. The same coupling is used for the emission of real photons and in photon-induced channels. Adopting $\alpha(0)$ for these on-shell photons instead would reduce the corresponding corrections by about $3.7\%$ of their~size. In~this setup, the EW input-scheme parameters are the Fermi constant~$G_F$, extracted from muon decay, together with the masses and widths of the EW gauge bosons, $m_W$, $\Gamma_W$, $m_Z$, and $\Gamma_Z$. These define, via tree-level relations, the remaining EW parameters such as the electromagnetic coupling and the weak mixing angle. Below, we comment on the numerical impact of adopting alternative EW input schemes.

\begin{figure}[t!]
\centering
\begin{subfigure}{0.32\textwidth}
\centering
\begin{tikzpicture}[line width=1.25 pt, scale=0.8]
	\draw[fermion] (-3.5,2.5) -- (-2,2);
	\draw[fermion] (-2,2) -- (-1,2);
	\draw[vector] (-2,2) -- (-1.5,0.25);
	\draw[vector] (-1,2) -- (0.25,1.5);
	\draw[fermion] (-1,2) -- (0.25,2.25);
	\draw[fermion] (-1.5,0.25) -- (0.5,-0.25);
	\draw[fermion] (-3.5,-0.25) -- (-1.5,0.25);
	\node at (0.7,2.35) {$\mu^-$};
	\node at (-3.85,2.5) {$\nu_\mu$};
	\node at (-2.25,1) {$W$};	
	\node at (0.6,1.4) {$\gamma$};
	\node at (-3.8,-0.35) {$q$};
	\node at (0.8,-0.35) {$q^{\hspace{0.25mm} \prime}$};	
\end{tikzpicture}
\end{subfigure}
\hfill
\begin{subfigure}{0.32\textwidth}
\centering
\begin{tikzpicture}[line width=1.25 pt, scale=0.8]
	\draw[fermion] (-3.5,2.5) -- (-2,2);
	\draw[fermion] (-2,2) -- (0,2.5);
	\draw[vector] (-2,2) -- (-1.5,0.25);
	\draw[vector] (-1.65,1) -- (0,1.1);
	\draw[fermion] (-1.5,0.25) -- (0.5,-0.25);
	\draw[fermion] (-3.5,-0.25) -- (-1.5,0.25);
	\node at (0.55,2.6) {$\mu^-$};
	\node at (-3.85,2.5) {$\nu_\mu$};
	\node at (-2.25,1) {$W$};	
	\node at (0.3,1.2) {$\gamma$};
	\node at (-3.8,-0.35) {$q$};
	\node at (0.8,-0.35) {$q^{\hspace{0.25mm} \prime}$};	
\end{tikzpicture}
\end{subfigure}
\hfill
\begin{subfigure}{0.32\textwidth}
\centering
\begin{tikzpicture}[line width=1.25 pt, scale=0.8]
	\draw[fermion] (-3.5,2.5) -- (-2,2);
	\draw[fermion] (-2,2) -- (0,2.5);
	\draw[vector] (-2,2) -- (-1.5,0.25);
	\draw[vector] (-0.5,0.2) -- (0.25,1.1);
	\draw[fermion] (-1.5,0.25) -- (-0.5,0.2);
	\draw[fermion] (-0.5,0.2) -- (0.5,-0.25);
	\draw[fermion] (-3.5,-0.25) -- (-1.5,0.25);
	\node at (0.55,2.6) {$\mu^-$};
	\node at (-3.85,2.5) {$\nu_\mu$};
	\node at (-2.25,1) {$W$};	
	\node at (0.4,1.35) {$\gamma$};
	\node at (-3.8,-0.35) {$q$};
	\node at (0.8,-0.35) {$q^{\hspace{0.25mm} \prime}$};	
\end{tikzpicture}
\end{subfigure}
\vspace{2mm} 
\caption{Examples of photonic FSR bremsstrahlung diagrams for the CC~DIS process $\nu_\mu + q \to \mu^- + q^{\hspace{0.25mm} \prime}$. Initial-state radiation (ISR) corrections and counterterm contributions, which account for the absorption of initial-state mass singularities into the PDFs, are not~shown.}
\label{fig:real}
\end{figure}
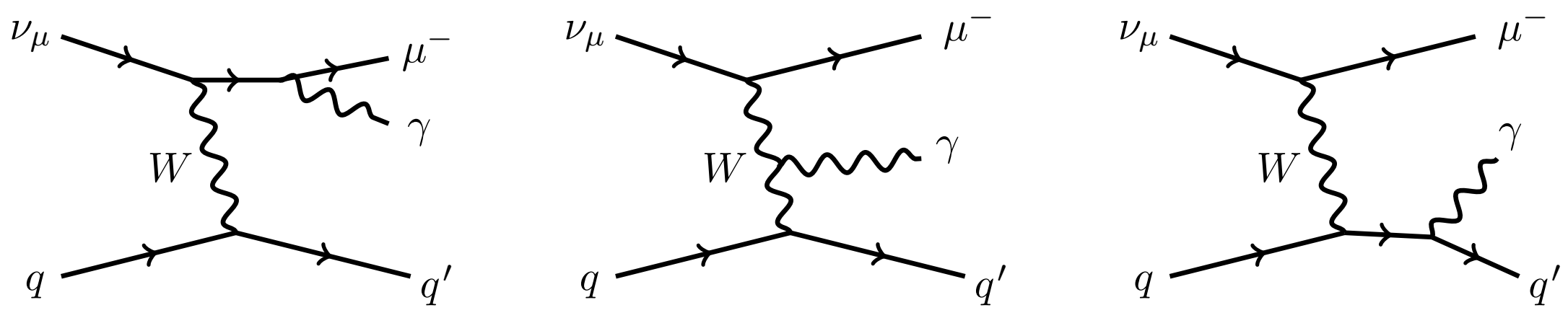

\paragraph{Real corrections:} The real-emission corrections to the relevant LO DIS subprocesses involve the emission of a gluon (photon) for the NLO QCD (QED) contributions, together with the corresponding crossed gluon- (photon-)initiated channels. Examples of photonic bremsstrahlung diagrams for the CC~DIS process $\nu_\mu + q \to \mu^- + q^{\hspace{0.25mm} \prime}$ are shown in~Figure~\ref{fig:real}. The corresponding squared amplitudes are computed using {\tt FeynArts}, {\tt FeynCalc}, and {\tt FormCalc}, and are numerically validated against results from {\tt OpenLoops} and {\tt Recola~2} at random phase-space points. 

Notice that for the NC DIS process $\nu_\ell + q \to \nu_\ell + q$, the real-emission squared amplitudes in QCD and QED are analytically related via
\beq \label{eq:NCDISmagic}
\overline{\sum} \, |\mathcal{M}_{\rm QED}|^2 = \frac{\alpha}{\alpha_s} \frac{Q_q^2}{C_F} \; \overline{\sum} \, |\mathcal{M}_{\rm QCD}|^2 \,,
\eeq
where $Q_q$ is the electric charge of the quark participating in the scattering, given in units of the electromagnetic coupling $e$, and $C_F = 4/3$ denotes the quadratic Casimir of the fundamental representation of QCD. The~symbol~$\overline{\sum}$ indicates that colour and spin indices are averaged over the initial states and summed over the final states. The relevant averaging factors for colour and spin are applied according to the type of initial-state parton: gluon~$\{8,2\}$, photon~$\{1,2\}$, or quark~$\{3,2\}$. With these averaging factors, Eq.~(\ref{eq:NCDISmagic}) applies to (anti)quark-initiated channels, while the relation between photon- and gluon-initiated channels receives an additional factor $N_c^2-1=8$ from the different colour averaging. In~consequence, Eq.~(\ref{eq:NCDISmagic}) can be used to cross-check the real-radiation corrections obtained in QED against those in~QCD.

For the~CC~DIS process, the squared amplitudes for QCD and QED emissions are instead structurally different and thus not related as in~Eq.~(\ref{eq:NCDISmagic}), for example due to photon FSR from the charged-lepton leg, as illustrated by the first Feynman diagram in~Figure~\ref{fig:real}. The~impact of these large logarithmically enhanced QED corrections will be studied in detail for differential distributions in later sections.

\paragraph{Subtraction terms:} For QCD corrections to DIS processes, examples of the construction of the subtraction terms $d \hat{\sigma}_{V_s}$ and $d \hat{\sigma}_{R_s}$ appearing in~Eq.~(\ref{eq:sigma_NLO}) are given in the appendices of~Refs.~\cite{Catani:1996vz,Catani:2002hc}. We implemented these formulae with minor adaptations for the neutrino-induced DIS processes considered here. The same formulae can also be adapted to the QED case, allowing the subtraction method to be applied to the EW corrections. While this procedure is discussed in~Ref.~\cite{Schonherr:2017qcj}, we recalculated the QCD-to-QED conversion factors for the implementation of $d \hat{\sigma}_{\rm NLO, EW}$ in~Eq.~\eqref{eq:sigma_expanded}, ensuring a consistent treatment of charge correlators and the normalisation between real and virtual subtraction terms. Alternatively, the subtraction terms presented in~Refs.~\cite{Dittmaier:1999mb,Dittmaier:2008md} could also be used.

We performed several cross-checks of the implementation of the subtraction terms entering~Eq.~(\ref{eq:sigma_NLO}). In particular, the singular limits of the real and real-subtraction terms were verified in kinematic regions affected by soft and collinear emissions. Additionally, the infrared~(IR) poles of the renormalised virtual corrections and their associated scale dependence were compared with those obtained using a generic implementation of the Catani--Seymour $I$-operator~\cite{Catani:1996vz} for both QCD and QED emissions. This implementation was further validated numerically against the {\tt OpenLoops} implementation of the $I$-operator for subprocesses involving massless leptons.

\paragraph{Validation of the implementation:} Beyond the technical consistency checks described above, several benchmarks have been used to validate the predictions at the cross-section level. In particular, the implementation of QCD corrections can be numerically validated at the level of the hadronic cross section by comparison with predictions obtained using the inclusive QCD structure-function approach~\cite{Dicus:2001kb,Gandhi:1998ri,Gluck:1998js,Cooper-Sarkar:2007zsa,Connolly:2011vc,Cooper-Sarkar:2011jtt,Bertone:2018dse,Candido:2023utz,Xie:2023suk}. In~practice, we have compared the results of these two approaches separately for both the massless and massive quark contributions to the cross section for all~CC and NC~DIS subprocesses. These comparisons were performed using an in-house implementation of the necessary QCD coefficient functions~\cite{SanchezGuillen:1990iq,vanNeerven:1991nn,Zijlstra:1991qc,Zijlstra:1992qd,Zijlstra:1992kj,vanNeerven:1999ca,Moch:1999eb,Moch:2004xu,Vermaseren:2005qc,Blumlein:2022gpp,Basdew-Sharma:2022vya} in $x$-space, and were additionally cross-checked against results obtained with the setup employed in~Ref.~\cite{Bertone:2018dse}, which uses {\tt APFEL}~\cite{Bertone:2013vaa} to compute structure-function predictions combined with two nested one-dimensional integrations over the DIS variables.

Several consistency checks were performed for our EW corrections. In NC DIS, the main differences relative to QCD arise from additional EW contributions evaluated with~{\tt Recola~2}. Setting these to zero and applying the QCD-to-QED rescaling factor of Eq.~(\ref{eq:NCDISmagic}) reproduces the expected QCD result. For~CC~DIS, new real-emission structures involving the charged lepton prevent a direct cross-check, so validation was performed in the $m_\ell \to 0$ limit, comparing the observed behaviour with the expected leading collinear corrections. Tests were carried out both differentially, for example, for the charged-lepton energy distribution, and for the total cross section. Differential tests probe factorised corrections associated with collinear photon radiation off~the charged lepton, described by the splitting $\ell^- \to \ell^- \gamma$, which vanish for observables inclusive in the lepton kinematics --- see the first Feynman diagram in~Figure~\ref{fig:real}. In contrast, processes such as $\nu_\ell + \gamma \to \ell^- + q \bar q^{\hspace{0.25mm} \prime}$ introduce a logarithmic dependence on the lepton mass through initial-state $\gamma \to \ell^+ \ell^-$ splittings, which persists in the total cross section. As an additional cross-check, we have performed an independent calculation of these corrections in the limit $m_{\ell}=0$, finding agreement with the $m_{\ell}\to 0$ limit of our results. This calculation required an adaptation of the Catani--Seymour subtraction formalism to the case of ``many identified particles'', allowing the calculation to be performed while keeping the momentum of the bare lepton identified. Further details of these cross-checks are provided in Appendices~\ref{app:glashow} and~\ref{app:FSR}. These tests further indicate that, for physical muon and tau masses, the code remains stable and efficient up to the highest considered energies, $E_{\nu_\ell} \simeq 1 \, {\rm EeV}$. For neutrino energies $E_{\nu_\ell} \simeq 100 \, {\rm TeV}$, it was also verified that the code remains stable even for the physical electron mass.

\subsection{Lepton fragmentation}
\label{sec:leptonfragmentation}

As discussed above, the full NLO~EW correction is cross-checked against a collinear approximation of the process $\ell^- \to \ell^- \gamma$ relative to the Born-level process. In~practice, the leading-logarithmic (LL) correction corresponding to this photonic FSR can be implemented~as
\beq \label{eq:sigma_LL}
d \hat{\sigma}_{{\rm LL}, {\rm QED}} = d \hat{\sigma}_{\rm LO} \, \int_0^1 \! dz \, \frac{\alpha}{2\pi} \, P^{(0)}_{\ell \ell}(z) \, \ln \left( \frac{\mu_h^2}{m_{\ell}^2} \right) \, \Theta_{\rm selection} \left ( z \hspace{0.25mm} p_{\ell} \right ) \,,
\eeq
where 
\beq \label{eq:P0}
P^{(0)}_{\ell \ell}(z) = \left[ \frac{1 + z^2}{1 - z} \right]_+ \,,
\eeq
is the LO lepton-to-lepton splitting function, describing the splitting $\ell \to \ell \gamma$. The symbol ``$+$'' denotes the standard plus-prescription, which regulates the soft singularity at $z \to 1$ by defining the integral in a distributional sense. This function obeys
\beq
\int_0^1 \! dz \, P^{(0)}_{\ell \ell}(z) = 0 \,,
\eeq
ensuring that logarithmic corrections vanish for observables inclusive in the lepton kinematics, in agreement with the Kinoshita--Lee--Nauenberg theorem~\cite{Kinoshita:1962ur,Lee:1964is}. In~Eq.~(\ref{eq:sigma_LL}), the function $\Theta_{\rm selection} \left (z \hspace{0.25mm} p_\ell \right )$ imposes kinematic cuts on the charged lepton after photon emission,~i.e. with the reduced momentum fraction~$z \hspace{0.25mm} p_\ell$, and the LO differential cross section~$d \hat{\sigma}_{\rm LO}$ is evaluated using massless lepton kinematics to capture the leading dependence on the lepton mass. The scale $\mu_h$ represents a hard scale of the process, which we identify with the partonic centre-of-mass~(COM) energy, i.e.~$\mu_h^2 = \hat{s}_{12}$ with $\hat{s}_{12}$ defined in~Eq.~(\ref{eq:sij}). While this choice is somewhat arbitrary --- since the massive fixed-order~NLO~EW correction, which includes the LL~contribution of~Eq.~\eqref{eq:sigma_LL}, does not depend on $\mu_h$ --- it is physically motivated: for observables sensitive to the charged-lepton energy, logarithmically enhanced corrections are significant due to the large mass hierarchy, and their impact increases with the partonic COM energy, which in turn grows with $E_{\nu_\ell}$.

Given the importance of the logarithmically enhanced corrections in~Eq.~\eqref{eq:sigma_LL}, which have been previously considered for CC~DIS in~Ref.~\cite{Plestid:2024bva}, it is important to account for higher-order effects as well as subleading terms in the lepton-mass logarithm $\ln \left (\mu_h/m_\ell \right )$. This~can be achieved using a lepton fragmentation approach, as applied for instance to muon decay~\cite{Arbuzov:2002pp,Arbuzov:2002cn,Arbuzov:2002rp} and to inclusive semi-leptonic $B$ decays~\cite{Bigi:2023cbv}. Alternatively, soft-photon corrections can be resummed via exponentiation, as in the structure-function method~\cite{Diener:2005me}. In this work, we implement the lepton~FF~up to $\mathcal{O}(\alpha^2)$, including both leading and subleading lepton-mass logarithms~\cite{Arbuzov:2002cn}. These corrections are applied factorised to both the LO and NLO QCD partonic CC~DIS cross sections:
\beq \label{eq:sigma_frag}
d \hat{\sigma}_{{\rm frag}, {\rm QED}} = d \hat{\sigma} \, \int_0^1 \! dz \, D_{\ell}\left( z, \mu_h^2 \right) \, \Theta_{\rm selection} \left( z \hspace{0.25mm} p_{\ell} \right) \,.
\eeq
Here, $D_{\ell}\left( z, \mu_h^2 \right)$ denotes the lepton~FF~given~in~Eq.~(10) of~Ref.~\cite{Arbuzov:2002cn}. Its~first-order correction is equivalent to that of the heavy-quark~FF~\cite{Mele:1990cw}, up to an overall QCD-to-QED conversion factor. We will also show numerical results obtained with the soft exponentiated part of Eq.~(4.2) of~Ref.~\cite{Diener:2005me}, which are labelled as $D_{\ell}^{\rm soft}$.
While the dominant effects arise from the lepton-mass logarithms $\ln \left (\mu_h/m_\ell \right )$, subleading corrections can also be relevant in specific regions of phase space, as will be discussed in more detail later.

\subsection{Coherent, diffractive, and resonant scattering contributions}
\label{sec:coherent}

\begin{figure}[t!]
\centering
\begin{subfigure}{0.32\textwidth}
\centering
\begin{tikzpicture}[line width=1.25 pt, scale=0.8]
	\draw[fermion] (-3.5,2.5) -- (-2.5,2.25);
	\draw[vector] (-2.5,2.25) -- (-0.5,2.5);
	\draw[fermion] (-0.5,2.5) -- (0.6,2.7);
	\draw[fermion] (0.5,1.9) -- (-0.5,2.5);
	\draw[vector] (-1,1.5) -- (-1.5,0.25);
	\draw[fermion] (-2.5,2.25) -- (-1,1.5);
	\draw[fermion] (-1,1.5) -- (0.25,1.3);
	\draw[fermion] (-1.5,0.25) -- (0.5,-0.25);
	\draw[solid] (-1.5,0.4) -- (0.5,-0.1);
	\draw[solid] (-1.5,0.1) -- (0.5,-0.4);
	\draw[fermion] (-3.5,-0.25) -- (-1.5,0.25);
	\draw[solid] (-3.5,-0.1) -- (-1.5,0.4);	
	\draw[solid] (-3.5,-0.4) -- (-1.5,0.1);	
	\node at (0.75,1.2) {$\mu^-$};
	\node at (-3.85,2.5) {$\nu_\mu$};
	\node at (0.9,1.8) {$\ell^+$};
	\node at (1,2.8) {$ \nu_\ell$};
	\node at (-1.7,1) {$\gamma$};	
	\node at (-2.2,1.5) {$\mu^-$};	
	\node at (-1.5,2.85) {$W$};	
	\node at (-3.9,-0.3) {$A$};
	\node at (0.9,-0.3) {$A^\prime$};	
	\node[draw, circle, fill=gray!40, minimum size=0.5cm] (blob) at (-1.5,0.25) {};
\end{tikzpicture}
\end{subfigure}
\hfill
\begin{subfigure}{0.32\textwidth}
\centering
\begin{tikzpicture}[line width=1.25 pt, scale=0.8]
	\draw[fermion] (-3.5,2.5) -- (-2.5,2.25);
	\draw[fermion] (-2.5,2.25) -- (0.25,2.5);
	\draw[vector] (-1,1.5) -- (-1.5,0.25);
	\draw[vector] (-1,1.5) -- (-2.5,2.25);
	\draw[vector] (-1,1.5) -- (0,1.5);
	\draw[fermion] (0,1.5) -- (0.5,1.85);
	\draw[fermion] (0.5,1.15) -- (0,1.5);
	\draw[fermion] (-1.5,0.25) -- (0.5,-0.25);
	\draw[solid] (-1.5,0.4) -- (0.5,-0.1);
	\draw[solid] (-1.5,0.1) -- (0.5,-0.4);
	\draw[fermion] (-3.5,-0.25) -- (-1.5,0.25);
	\draw[solid] (-3.5,-0.1) -- (-1.5,0.4);	
	\draw[solid] (-3.5,-0.4) -- (-1.5,0.1);	
	\node at (0.75,2.6) {$\mu^-$};
	\node at (-3.85,2.5) {$\nu_\mu$};
	\node at (0.9,1) {$\ell^+$};
	\node at (0.9,2) {$\nu_\ell$};
	\node at (-1.7,1) {$\gamma$};	
	\node at (-2.2,1.5) {$W$};	
	\node at (-0.5,1) {$W$};	
	\node at (-3.9,-0.3) {$A$};
	\node at (0.9,-0.3) {$A^\prime$};	
	\node[draw, circle, fill=gray!40, minimum size=0.5cm] (blob) at (-1.5,0.25) {};
\end{tikzpicture}
\end{subfigure}
\hfill
\begin{subfigure}{0.32\textwidth}
\centering
\begin{tikzpicture}[line width=1.25 pt, scale=0.8]
 \draw[fermion] (-3.5,2.5) -- (-2,2.35);
	\draw[fermion] (-2,2.35) -- (0.3,2.7);
	\draw[vector] (-2,2.35) -- (-1.75,1.6);
	\draw[vector] (-1.5,0.25) -- (-1.75,1);
	\draw[fermion] (-1.75,1) -- (-1.75,1.6);
	\draw[fermion] (-0.25,0.9) -- (-1.75,1);
	\draw[fermion] (-1.75,1.5) -- (-0.25,1.7);
	\draw[fermion] (-1.5,0) -- (0.5,-0.25);
	\draw[solid] (-1.5,0.15) -- (0.5,-0.1);
	\draw[solid] (-1.5,-0.15) -- (0.5,-0.4);
	\draw[fermion] (-3.5,-0.25) -- (-1.5,0);
	\draw[solid] (-3.5,-0.1) -- (-1.5,0.15);	
	\draw[solid] (-3.5,-0.4) -- (-1.5,-0.15);	
	\node at (0.8,2.8) {$\mu^-$};
	\node at (-3.85,2.5) {$\nu_\mu$};
	\node at (0.25,0.9) {$\ell^+$};
	\node at (-2.2,1.3) {$\ell^+$};
	\node at (0.25,1.7) {$\nu_\ell$};
	\node at (-2.1,0.5) {$\gamma$};	
	\node at (-2.4,1.9) {$W$};
	\node at (-3.9,-0.3) {$A$};
	\node at (0.9,-0.3) {$A^\prime$};	
	\node[draw, circle, fill=gray!40, minimum size=0.5cm] (blob) at (-1.5,0.0) {};
\end{tikzpicture}
\end{subfigure}
\vspace{4mm} 
\caption{Example Feynman diagrams for coherent and diffractive scattering in the~CC~process $\nu_\mu + A \to \mu^- + \ell^+ + \nu_\ell + A^\prime$, where $A$ and $A^\prime$ denote a nucleus or nucleon. The first two diagrams correspond to $W$-boson production, while the third shows neutrino trident production via photon exchange.}
\label{fig:coherent}
\end{figure}
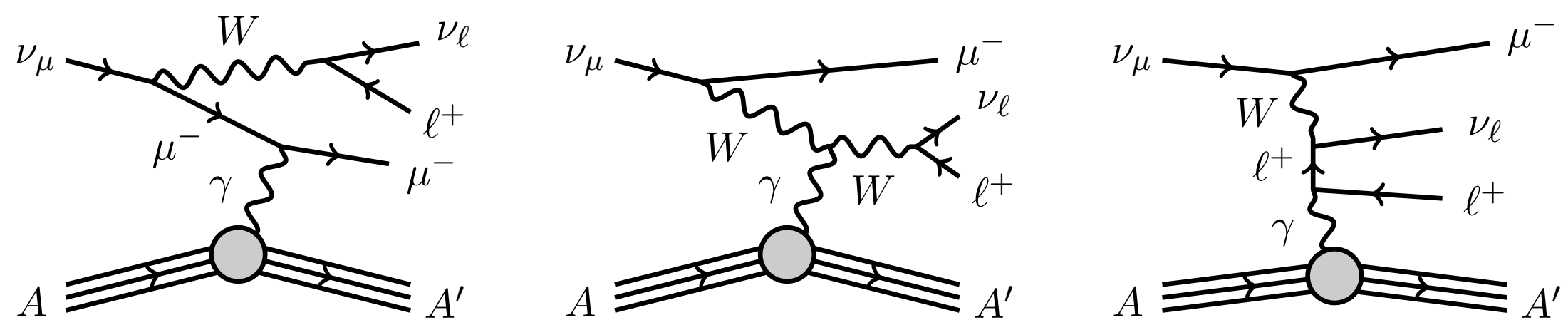

The DIS processes discussed above provide the dominant contribution to the neutrino-matter scattering rate. Nevertheless, additional interactions with hadronic matter can also be phenomenologically relevant. In particular, coherent effects proportional to the square of the nuclear charge, $Z^2$, arise when the momentum transfer is sufficiently small for the neutrino to probe the electromagnetic field of the nucleus as a whole~\cite{Ballett:2018uuc,Zhou:2019vxt}. At low neutrino energies, such interactions have been studied extensively in the context of coherent and diffractive scattering off~nuclei and individual nucleons~\cite{Ballett:2018uuc}. At higher energies, however, these interactions can also become numerically relevant in CC processes involving an intermediate $W$ boson that becomes resonantly enhanced~\cite{Seckel:1997kk}. Representative coherent and diffractive scattering processes are shown in~Figure~\ref{fig:coherent}. To describe these channels exclusively, we employ the fully differential implementation of coherent and diffractive neutrino scattering developed in~Refs.~\cite{Gauld:2019pgt,Garcia:2020jwr}, based on the formalism of~Ref.~\cite{Ballett:2018uuc}. Before discussing the implementation details of these scattering processes, we first outline the general physical features they share with photon-induced DIS subprocesses, as this will help clarify several of the results presented in the following sections.

{\bf Common behaviour of photon-induced processes:} As discussed in~Section~\ref{sec:DIScalc}, the EW corrections to the DIS process introduce several structures that are absent in the case of QCD corrections. One such contribution arises in the photon-induced channel through the exchange of a $W$ boson in the $s$-channel, which can be understood as the crossed version of the first Feynman diagram shown in~Figure~\ref{fig:real}, with both the photon and the incoming quark crossed. Such contributions are proportional to the photon~PDF and are therefore naively suppressed due to the smallness of the photon~PDF, which is itself suppressed by a factor of ${\cal O}(\alpha)$ relative to the quark~PDFs. However, corrections in this channel can receive both an initial-state collinear enhancement associated with the charged-lepton mass and an additional resonant enhancement from intermediate on-shell $W$-boson production in the final-state quark system. The~combination of these effects can lead to changes of several percent in the predicted cross section and can also modify the behaviour of the inelasticity distribution of the final-state charged leptons. Similar physical effects are also relevant for coherent and diffractive scattering predictions. It is therefore useful to discuss these effects first in the context of the DIS process before turning to the specific details of our implementation of coherent and diffractive scattering~processes.

The total cross-section computation in the photon-induced channel receives contributions from numerous subprocesses. Since we are primarily interested in processes involving a charged lepton in the final state, we restrict the discussion to subprocesses of the form $\nu_{\ell} + \gamma \to \ell^- + f \bar f^{\hspace{0.25mm} \prime}$. A key point is that, if the COM energy of the neutrino-photon system is sufficiently large, the squared amplitude for this subprocess receives a resonant enhancement when the invariant mass of the final-state fermion pair is close to the $W$-boson mass, i.e. $m_{f \bar f^\prime} \simeq m_W$, corresponding to on-shell $W$-boson production. Numerically, the resonant contribution to the cross section dominates whenever it is kinematically accessible, i.e. when the squared COM energy of the neutrino-photon system, $s_{\nu_\ell \gamma}$, satisfies $s_{\nu_\ell \gamma} \geq (m_{\ell} + m_W)^2$. In this regime, the total rate can be well approximated by treating the $W$ boson as an on-shell particle. If desired, $W$-boson decays can be straightforwardly incorporated using the narrow-width approximation~(NWA).

The approximation for the total DIS cross section in this channel takes the form
\beq \label{eq:sigmavgDIS}
\sigma_{\nu_\ell \gamma,{\rm DIS}}^{N} = \int_{x_0}^1 dx \, f_{\gamma}^{N} \left (x, \mu_F^2 \right ) \hat \sigma_{\nu_\ell \gamma \to \ell^- W^+} \left ( x, E_{\nu_\ell} \right ) \,,
\eeq
where $x_0 = (m_{\ell} + m_W)^2/(2 \hspace{0.25mm} m_N E_{\nu_\ell})$ and $f_{\gamma}^{N} \left (x, \mu_F^2 \right )$ denotes the photon~PDF of a target nucleon $N$, which in the {\tt LUXqed} approach contains both elastic and inelastic contributions. For sufficiently high neutrino energies, the partonic cross section can be approximated as~\cite{Seckel:1997kk,Zhou:2019vxt,Zhou:2019frk}
\beq \label{eq:sigmavNpartonic}
\hat{\sigma}_{\nu_\ell \gamma \to \ell^- W^+} \left ( x, E_{\nu_\ell} \right ) \simeq 2 \sqrt{2} \hspace{0.25mm} \alpha \hspace{0.25mm} G_F \left [ 1 + \frac{m_W^2}{4 \hspace{0.125mm} m_N E_{\nu_\ell} x} \ln \left ( \frac{ 2 \hspace{0.25mm} m_N E_{\nu_\ell} x}{m_\ell^2} \right ) \right ] \,.
\eeq
Notice that the $t$-channel lepton propagator in the left-hand diagram of~Figure~\ref{fig:coherent} becomes singular as $t \to 0$ if the lepton mass $m_\ell$ is neglected. In principle, the calculation can be performed either retaining a finite lepton mass or in the massless approximation. We have carried out fully differential calculations for the off-shell process in both setups, finding agreement in the limit $m_\ell \to 0$, and a detailed comparison and validation of the two approaches is provided in~Appendix~\ref{app:glashow}. Here, we instead comment on two main features of the total cross-section prediction for the partonic process in~Eq.~\eqref{eq:sigmavNpartonic}: first, the cross section depends logarithmically on the charged-lepton mass, implying different results for electron, muon, and tau final states, with the correction being largest (smallest) for electrons~(taus); second, this logarithmic enhancement is power-suppressed by the partonic COM energy and therefore becomes increasingly damped at high energies.

The coherent and diffractive scattering processes discussed in the following involve kinematics that differ significantly from those of the DIS processes. At the same time, the electromagnetic form factors of nucleons and nuclei exhibit behaviour distinct from that of the inelastic photon~PDF. Nevertheless, the qualitative features of the partonic cross section $\nu_{\ell} + \gamma \to \ell^- + W$ are common to all of these scattering processes. We now provide details of the implementation of coherent and diffractive processes used in this work.

{\bf Coherent scattering on the nucleus:} At low momentum transfer, $Q \lesssim 0.1 \, {\rm GeV}$, neutrinos can scatter coherently off~the electromagnetic field of an entire nucleus. In this regime, the nucleus behaves as a single, coherent source, and the scattering amplitudes from individual protons add constructively, leading to a characteristic $Z^2$ enhancement of the cross section. At very high energies, $E_{\nu_\ell} \gtrsim m_W^2 R_A/4 \simeq 25\,{\rm TeV}$, where $m_W \simeq 80.4 \,{\rm GeV}$ is the $W$-boson mass and $R_A \simeq 3 \,{\rm fm} \simeq 1/(0.07 \, {\rm GeV})$ denotes a typical nuclear radius, this coherent interaction can produce an on-shell $W$ boson. The corresponding Feynman diagrams are shown in the left and middle panels of~Figure~\ref{fig:coherent}.

The full off-shell calculation for both coherent and diffractive scattering processes has been presented in~Ref.~\cite{Zhou:2019vxt}, with earlier studies~\cite{Lovseth:1971vv,Belusevic:1987cw,Gaidaenko:2000hg} based on the equivalent photon approximation~\cite{Fermi:1924tc,vonWeizsacker:1934nji,Williams:1934ad}. Calculations applicable at low energies within the SM have been performed in~Refs.~\cite{Altmannshofer:2014pba,Magill:2016hgc,Ge:2017poy,Ballett:2018uuc,Altmannshofer:2019zhy}. We also note that, in accelerator-based neutrino experiments, neutrino trident production provides a powerful probe of BSM physics~\cite{Altmannshofer:2014pba,Magill:2016hgc,Ge:2017poy,Altmannshofer:2019zhy}. At high energies, instead, the on-shell $W$-boson approximation has been shown to capture the dominant contribution to the cross section~\cite{Zhou:2019vxt,Zhou:2019frk}.

In this work, we adopt the on-shell $W$-boson approximation, using the scattering formalism of~Ref.~\cite{Ballett:2018uuc}, and include the $W$-boson decays in the NWA. Non-resonant contributions, such as the trident diagram shown in the right panel of~Figure~\ref{fig:coherent}, are therefore not included. Our implementation is fully differential and allows arbitrary selection cuts on the final-state particles. This is essential for the present analysis, since it enables both cuts on the final-state charged leptons and the computation of inelasticity distributions. Within the NWA, charged leptons produced in the hard scattering can moreover be separated from those arising from the $W$-boson decay. Final-state muons therefore receive two types of contributions: primary muons, which are produced only in muon-neutrino-initiated scattering, and decay muons from $W^+ \to \mu^+ \nu_\mu$, which arise in the scattering of all three neutrino flavours. In~Section~\ref{sec:exclusive}, we refer to the primary contribution as ``coherent primary'' and to the sum of the primary and decay contributions as ``coherent full''. For the latter, we assume equal fluxes for the three neutrino flavours, $\nu_e:\nu_\mu:\nu_\tau = 1:1:1$. Events with two muons are treated inclusively, i.e.~each muon contributes separately to the observables considered. If only muon neutrinos were included, the difference between the ``coherent full'' and ``coherent primary'' predictions would be smaller by a factor of about~$3$.

For ice or water targets, as encountered in neutrino telescopes, the maximum ratio of coherent $W$-boson production to CC DIS reaches approximately $3\%$ for $\nu_\mu$, and is slightly higher (lower) for $\nu_e$ ($\nu_\tau$). While this is a relatively small effect, these contributions are relevant for analyses aiming at percent-level precision. Further details are given in~Section~\ref{sec:exclusive}.

{\bf Elastic and diffractive scattering:} For momentum transfers $Q \simeq 1 \, {\rm GeV}$, neutrinos can scatter off~the electromagnetic field of bound nucleons within the nucleus. In this case, the formalism of~Ref.~\cite{Ballett:2018uuc} can also be readily applied to $W$-boson production by introducing an appropriate hadronic tensor expressed in terms of the electric and magnetic Sachs form factors. However, when the photon~PDF is formulated within the LUXqed framework, it naturally incorporates both inelastic and elastic contributions. The elastic component is directly determined from the proton's electric and magnetic Sachs form factors, which are themselves fitted to low-momentum-transfer $e^- p$ scattering data~\cite{A1:2013fsc}. Consequently, this elastic contribution enters the neutrino-nucleon cross-section calculations through the DIS framework discussed previously. In principle, one should also account for contributions from bound neutrons. However, numerically this effect is negligible and has therefore not been included in the results shown in~Section~\ref{sec:exclusive}.

{\bf Scattering on atomic electrons:} In addition to the scattering processes off~nuclei and nucleons discussed earlier, incoming neutrinos can also interact with atomic electrons. This contribution is usually negligible, except for high-energy electron antineutrinos, where scattering on atomic electrons is resonantly enhanced at $E_{\bar \nu_e} \simeq m_W^2/(2 \hspace{0.25mm} m_e) \simeq 6.3 \, {\rm PeV}$ due to the production of an on-shell $W$-boson, known as the Glashow resonance~\cite{Glashow:1960zz}. An~event consistent with this signature was observed by IceCube~\cite{IceCube:2021rpz}. A precision calculation incorporating the complete NLO~QCD$+$EW corrections and higher-order photonic~ISR effects was presented in~Ref.~\cite{Gauld:2019pgt}, while possible atomic broadening effects were investigated in~Ref.~\cite{Loewy:2014zva}. The detailed description of signatures associated with this process is beyond the scope of the present work. Nevertheless, the development and public release of dedicated simulation tools could prove valuable for future detector upgrades, such as IceCube-Gen2~\cite{IceCube-Gen2:2020qha}, where substantially larger event samples are expected.

\section{Prologue to MC~simulations}
\label{sec:setup}

Before presenting any numerical results, we summarise the DIS kinematics and their parton-level implementation in collinear factorisation, including real QED emissions that require a dedicated treatment of lepton-based observables. A discussion of IR-safe dressed lepton definitions is also included, which is essential for ensuring collinear safety in the reference scales used for PDF evaluations. We also summarise the PDF sets, as well as the factorisation and renormalisation scales and input parameter choices used in the NLO~QCD$+$EW calculation.

\subsection{DIS kinematics}
\label{sec:kinematics}

To begin, we discuss the kinematics of the DIS process. The~CC~case is taken as an example, as it involves additional complications due to QED emissions compared to the~NC~case. The hadron-level kinematics for the scattering process can be summarised as
\beq \label{eq:2to2}
\nu_\ell (p_1) + N(p) \to \ell^- (p_3) + X(p_X) \,,
\eeq
with $\nu_\ell$ ($\ell^-$) denoting the incoming (outgoing) lepton, and the labels $N$ and $X$ referring to the struck nucleon and the outgoing hadronic system, excluding the charged lepton $\ell^-$. 

It is standard to introduce the momentum-transfer variable in terms of the lepton kinematics, defined as
\beq \label{eq:Q2q}
Q^2 = -q^2 \,, \qquad q = p_{1} - p_{3} \,,
\eeq
where, in the absence of photon radiation off~the charged lepton, $Q^2$ denotes the virtuality of the exchanged EW boson. The usual DIS kinematic variables, including the Bjorken scaling variable $x_B$ and the inelasticity variable $y$, are then defined~as
\beq \label{eq:kin1}
s = \left(p_{1} + p \right)^2\,, \qquad x_B = \frac{Q^2}{2 \hspace{0.25mm} q \cdot p}\,, \qquad y = \frac{q \cdot p}{p_{1} \cdot p}\,.
\eeq
Expressed in the nucleon rest frame, the variables in~Eqs.~(\ref{eq:kin1}) take the form
\beq \label{eq:kin2}
s = 2 m_N E_{\nu_\ell} + m_N^2\,, \qquad x_B = \frac{Q^2}{2 \hspace{0.25mm} m_N E_{\nu_\ell} \hspace{0.125mm} y} \,, \qquad y = 1 - \frac{E_{\ell^-}}{E_{\nu_\ell}} \,,
\eeq
where $E_{\nu_\ell}$ and $E_{\ell^-}$ denote the energies of the incoming and outgoing lepton, respectively. To avoid unnecessary clutter, we will hereafter often drop the flavour label and charge assignment in $E_{\nu_\ell}$ and $E_{\ell^-}$, and simply use $E_{\nu}$ and $E_{\ell}$ to denote the corresponding energies, which should not lead to any confusion.

The differential DIS calculation is performed within collinear factorisation, as discussed in~Section~\ref{sec:DIScalc}. 
These computations are carried out at parton level in terms of a set of partonic momenta $\{p_i\}$ with $i \in [1,4]$ $\big($$i \in [1,5]$$\big)$ for subprocesses involving two (three) particles in the final state. For instance, the assignment of partonic momenta for a real-emission QED subprocess is given by
\beq \label{eq:2to3}
\nu_\ell (p_1) + q(p_2) \to \ell^- (p_3) + q^{\hspace{0.25mm} \prime}(p_4) + \gamma(p_5) \,,
\eeq
where the incoming parton momentum is defined as
\beq \label{eq:p2}
p_{2} = x \hspace{0.25mm} p \,,
\eeq
with $x$ denoting the momentum fraction of the nucleon four-momentum $p$ carried by the incoming quark. It is also convenient to introduce invariants constructed from the partonic momenta
\beq \label{eq:sij}
\hat{s}_{ij} = 2 \hspace{0.25mm} p_i \cdot p_j \,.
\eeq
For massless incoming particles, $\hat{s}_{12}$ is the squared partonic COM energy.

\subsection{Dressed leptons}
\label{sec:dressed}

It is important to note that, in the presence of QED emissions, the momentum of the outgoing charged lepton $p_3$ can be altered by photon radiation. As a consequence, observables depending on $p_3$, such as $Q^2$, $x_B$, and $y$, are likewise modified by such emissions. These quantities are therefore not collinear safe with respect to photon radiation, and differential cross-section predictions in terms of these variables may acquire large logarithmic corrections involving the lepton mass, as already alluded to in~Section~\ref{sec:leptonfragmentation}.

To address this issue, one may introduce a collinear-safe recombination procedure in which leptons are first ``dressed'' with QED radiation, ensuring that the reconstructed lepton momentum $p_{\ell}^{\rm dressed}$ is insensitive to collinear emissions. Observables constructed from dressed momenta are collinear safe. This property can be exploited, for instance, to define a collinear-safe reference scale for PDF evaluations, which remains well-defined even when computing differential cross sections with respect to intrinsically unsafe observables such as the inelasticity~$y$.

An example of such a procedure was introduced in~Ref.~\cite{Diener:2003ss}, where the relative angle between the photon and the outgoing charged fermion, evaluated in the nucleon rest frame, is used as a measure of their separation. In particular, if the smallest angle in the event corresponds to the lepton-photon pair and satisfies $\theta_{\ell\gamma}<\theta_{\rm cut}$, the dressed lepton is defined by recombining the lepton and photon momenta
\beq \label{eq:dressed1}
p_{\ell}^{\rm dressed} = p_{\ell} + p_{\gamma} \,.
\eeq
More generally, other distance measures may be employed, such as those used in standard jet algorithms, for example JADE, general-$k_T$, etc. --- see~Ref.~\cite{Salam:2009jx} for a detailed review and references to the original literature. Using any such recombination prescription, we define
\beq \label{eq:dressed2}
Q^{2}_{\rm dressed} = -\left( p_1 - p_{\ell}^{\rm dressed} \right)^2 \,,
\eeq
which reduces to $Q^2$ as defined in~Eq.~(\ref{eq:Q2q}) in the absence of QED radiation.

In our full NLO~QCD$+$EW calculation of~CC~DIS, we have considered different definitions of ``dressed'' leptons and found that the precise recombination procedure leads to very similar results. We therefore adopt the simple prescription~(\ref{eq:dressed1}), using a lepton-photon angular cut of $\theta_{\rm cut} = 5^\circ$. Notice that the definition~(\ref{eq:dressed1}) ensures, up to photons radiated outside the recombination cone or off~other charged particles, that the four-momentum of the ``dressed'' lepton corresponds to that of the charged lepton produced in the hard interaction mediated by the $t$-channel exchange of the $W$ boson in the process~(\ref{eq:2to2}). This~feature allows for a direct comparison with the lepton fragmentation approach discussed in~Section~\ref{sec:leptonfragmentation}. A representative study of the impact of various dressing procedures is included in~Appendix~\ref{app:FSR}.

\subsection{Hadronic input}
\label{sec:hadinput}

As briefly mentioned in~Section~\ref{sec:DIScalc}, the full NLO QCD$+$EW DIS calculation requires the evaluation of partonic channels involving initial-state photons. The consistent inclusion of such contributions therefore necessitates the use of PDF sets that include a photon~PDF. Furthermore, when the calculation is performed in the massless lepton approximation, mass singularities can arise, requiring the introduction of lepton PDFs, for instance in the presence of initial-state $\gamma \to \ell^- \ell^+$ splittings. For these reasons, we choose the {\tt NNPDF31\_nlo\_as\_0118} PDF set~\cite{NNPDF:2017mvq} with $\alpha_s(m_Z) = 0.118$ at the $Z$-boson mass scale as our benchmark, as variants of this fit that include determinations of both the photon~PDF~\cite{Bertone:2017bme}, based on the {\tt LUXqed} approach~\cite{Manohar:2016nzj,Manohar:2017eqh}, as well as lepton PDFs~\cite{Buonocore:2020nai}, are available. Our calculations also evaluate the strong coupling constant $\alpha_s$ using the numerical grids provided with the PDF set.

The PDF sets mentioned above provide a description of the partonic content of free protons. In practice, however, one is interested in describing neutrino interactions with nucleons bound inside nuclei, which in principle requires knowledge of neutron PDFs as well as nuclear corrections. To our knowledge, no nuclear PDF fits are currently available that also include parametrisations of photon (and lepton) PDFs. Since our focus in this work is on EW corrections, we therefore choose to work with PDFs of free nucleons. To facilitate our computation, we construct neutron PDFs starting from the benchmark proton PDF set, assuming isospin symmetry between the up- and down-quark distributions and between the corresponding anti-quark distributions. This symmetry, i.e.~$f^n_u(x,Q_0^2) = f^p_d(x,Q_0^2)$, $f^n_d(x,Q_0^2) = f^p_u(x,Q_0^2)$ and $f^n_{\bar u}(x,Q_0^2) = f^p_{\bar d}(x,Q_0^2)$, $f^n_{\bar d}(x,Q_0^2) = f^p_{\bar u}(x,Q_0^2)$, is imposed at the initial scale $Q_0 = 1.65 \, {\rm GeV}$, after which the~PDFs are evolved including QCD and QED effects using {\tt HOPPET~v2}~\cite{Karlberg:2025hxk}, making use of its functionality to generate an {\tt LHAPDF} grid~\cite{Buckley:2014ana}. The photon (and lepton) PDFs for the neutron are thus generated purely perturbatively, as they are set to zero at the initial scale $Q_0$. This neglects the non-perturbative photon content of the neutron at low scales~\cite{Xie:2023qbn}, and hence underestimates photon-induced contributions for neutron and isoscalar targets.

The factorisation and renormalisation scales are chosen dynamically, i.e.~on an event-by-event basis, as $\mu^2 = \mu_F^2 = \mu_R^2 = Q^2_{\rm dressed}$, with $Q^2_{\rm dressed}$ given in~Eq.~(\ref{eq:dressed2}). There are two exceptions. For subprocesses involving a quark of mass $m_q$, the reference scale is taken to be $\mu^2 = Q^2_{\rm dressed} + m_q^2$, while for photon-initiated channels we instead use $\mu^2 = \hat{s}_{12}$, with~$\hat{s}_{12}$ defined in~Eq.~(\ref{eq:sij}). Scale uncertainties are estimated via a seven-point variation, obtained by independently varying~$\mu_F$ and~$\mu_R$ by a factor of two around the reference scale under the constraint $1/2 \leq \mu_F/\mu_R \leq 2$. For scales below the lower boundary of the PDF grid, $Q_0 = 1.65\,{\rm GeV}$, which are reached in particular in the downward variations, we rely on the default extrapolation provided by {\tt LHAPDF}.

The cross-section predictions for coherent and diffractive scattering processes require the knowledge of nuclear and nucleon form factors. For coherent scattering, we employ the analytic Woods--Saxon form factor, together with the dipole parametrisation of the electric and magnetic Sachs form factors for nucleons. Further details can be found in~Appendix~A of~Ref.~\cite{Ballett:2018uuc}.

\subsection{Input parameters}
\label{sec:input}

For the numerical values of the SM input parameters, we use those reported in the latest review~\cite{ParticleDataGroup:2024cfk} of the Particle Data Group (PDG). The PDG values for the on-shell $W$-, $Z$-, and Higgs-boson masses and widths are
\beq \label{eq:bosonproperties}
\begin{gathered}
m_W = 80.3692 \, {\rm GeV} \,, \qquad \Gamma_W = 2.085 \, {\rm GeV} \,, \\[2mm]
m_Z = 91.1880 \, {\rm GeV} \,, \qquad \Gamma_Z = 2.4955 \, {\rm GeV} \,, \\[2mm]
m_H = 125.2 \, {\rm GeV} \,, \qquad \Gamma_H = 3.7 \, {\rm MeV} \,.
\end{gathered}
\eeq
We further use the following values for the charged-lepton masses
\beq \label{eq:leptonmasses}
m_e = 0.510999 \, {\rm MeV} \,, \qquad m_\mu = 105.658 \, {\rm MeV} \,, \qquad m_\tau = 1.77693 \, {\rm GeV} \,.
\eeq
For the quark masses, we instead adopt the values used in the {\tt NNPDF31\_nlo\_as\_0118} PDFs and its QED variants discussed above and employ a VFNS with five active quark flavours. The on-shell top-quark mass is taken as
\beq \label{eq:topmass}
m_t = 172.5 \, {\rm GeV} \,.
\eeq

For the Cabibbo-Kobayashi-Maskawa~(CKM) matrix, we assume a diagonal structure for the third generation and use the following values for the relevant elements of the CKM matrix in our numerical simulations:
\beq \label{eq:CKM}
|V_{ud}| = |V_{cs}| = 0.9744 \,, \qquad |V_{us}| = |V_{cd}| = 0.2245 \,, \qquad |V_{tb}| = 1 \,. 
\eeq

Our fixed-order EW DIS calculations are performed consistently within the complex-mass scheme~\cite{Denner:2005fg,Denner:2019vbn,Denner:2006ic}, adopting the $G_F$ scheme for the EW input parameters~\cite{Denner:2000bj}. The~input set consists of the Fermi constant
\beq \label{eq:GF}
G_{F} = 1.1663785 \cdot 10^{-5} \, {\rm GeV}^{-2} \,,
\eeq
together with the $W$- and $Z$-boson masses and widths specified in~\eqref{eq:bosonproperties}. The electromagnetic coupling and the weak mixing angle are then derived from the pole-mass inputs according to
\beq \label{eq:alphasine}
\alpha_{G_F} = \frac{\sqrt{2} \hspace{0.25mm} G_F \left | s_w^2 \hspace{0.25mm} \mu_W^2 \right |}{\pi}\,, \qquad s_w^2 = 1 - \frac{\mu_W^2}{\mu_Z^2}\,, \qquad \mu_V^2 = m_V^2 - i \hspace{0.125mm} m_V \Gamma_V\,.
\eeq
Numerically, this yields $\alpha_{G_F}=1/132.156$ and $\left |s_w^2 \right | = 0.223195$. In the following, the subscript ``$G_F$'' indicates that the electromagnetic coupling is evaluated in the $G_F$ scheme.

\subsection{EW scheme dependence}
\label{sec:ewscheme}

The use of the $G_F$ scheme is standard in calculations of CC processes, where the cross section is typically expressed directly in terms of $G_F$ --- see, for example,~Refs.~\cite{Sirlin:1980nh,Sirlin:1981ie,Denner:1991kt}. This choice is well motivated, since the muon decay rate determines the strength of the CC interaction, i.e.~the combination $\alpha_{G_F}/s_w^2$, which, using the first relation in~Eqs.~(\ref{eq:alphasine}) and ignoring small width effects, depends only on $m_W$ and $G_F$. In contrast, NC interactions depend additionally on $s_w^2$, which is a derived parameter rather than a direct input.

A key advantage of the $G_F$ scheme is that the universal corrections associated with the running of the electromagnetic coupling from zero up to the EW scale are absorbed into the definition of $G_F$ via muon decay, together with the leading universal EW contributions entering the relation between $m_W$, $G_F$, and $\alpha_{G_F}$. As a result, the one-loop effects in muon decay encoded in $\Delta r$, as well as the dominant two-loop contributions enhanced by the top-quark Yukawa coupling, are effectively included in the Born-level normalisation. The genuine NLO~EW corrections therefore appear only as comparatively small residual contributions, leading to improved perturbative stability of CC predictions.

To assess the uncertainty associated with our choice of EW input scheme, we also present results for the NLO~EW calculation of the NC and CC DIS processes in the $\alpha(0)$ and $\alpha(m_Z)$ schemes, using as input
\beq \label{eq:otherEW}
\alpha(0) = \frac{1}{137.036} \,, \qquad \alpha(m_Z) = \frac{1}{128.0} \,,
\eeq
where the latter corresponds to the default value in {\tt OpenLoops}. The running of $\alpha$ from zero to $m_Z$ is then encoded in
\beq \label{eq:delalp}
\Delta \alpha(m_Z) = 1 - \frac{\alpha(0)}{\alpha(m_Z)} = 0.06594 \,.
\eeq

To allow for a fast EW scheme change, we implement the one-loop correction to muon decay, parametrised by $\Delta r$, in terms of two-point functions and renormalisation constants within the complex-mass scheme. Numerically, for our choice of EW input parameters detailed in~Section~\ref{sec:input}, we find
\beq
\label{eq:dr}
\Delta r = 0.03584 \,.
\eeq

Including the NLO~EW corrections in the $\alpha_{G_F}$ scheme, together with the LO DIS cross section $d\sigma_{\rm LO}$, $\Delta \alpha(m_Z)$ in~Eq.~(\ref{eq:delalp}), and $\Delta r$ in~Eq.~(\ref{eq:dr}), allows one to obtain results in alternative EW schemes through a simple shift:
\begin{equation}
\big ( \delta_{\rm NLO, EW} \left ( d  \sigma \right ) \big )_{\rm scheme}
= \big ( \delta_{\rm NLO, EW} \left ( d \sigma \right ) \big )_{G_F} + \Delta_{\rm scheme} \, d \sigma_{\rm LO} \,,
\label{eq:EWshift}
\end{equation}
where, in line with~Ref.~\cite{Buccioni:2019sur}, we find
\begin{equation}
\Delta_0 = 2 \hspace{0.25mm} \Delta r \,, \qquad \Delta_{m_Z} = 2 \hspace{0.25mm} \big [ \Delta r - \Delta \alpha (m_Z) \big ] \,,
\label{eq:Deltaschemes}
\end{equation}
with the subscripts denoting the corresponding EW schemes, and with all terms in~Eq.~(\ref{eq:EWshift}) evaluated with the couplings of the respective scheme. These shifts are defined such that the differential cross section $d\sigma$ is formally independent of the chosen scheme up to~$\mathcal{O}(\alpha^3)$ as far as EW corrections are concerned, i.e.~scheme differences arise only at $\mathcal{O}(\alpha^4)$. 

\section{Fixed-order NC and CC DIS predictions}
\label{sec:fixedorder}

Throughout this section, we present integrated DIS cross sections as a function of the incident neutrino energy $E_{\nu}$, and for the CC process we also study the mean value of the inelasticity parameter $y$ defined in~Eqs.~(\ref{eq:kin1}) and~(\ref{eq:kin2}). These observables are defined as
\beq \label{eq:observables}
\sigma = \int \! d\sigma(E_{\nu}) \, \Theta(Q^2 - Q_0^2)\,, \qquad 
\langle y \rangle = \frac{1}{\sigma} \int \! y \, d\sigma(E_{\nu}) \, \Theta(Q^2 - Q_0^2)\,,
\eeq
where the integration is performed over phase space with a minimal cut $Q^2 > Q_0^2$ imposed by the function $\Theta(Q^2 - Q_0^2)$. The~scale $Q_0$ corresponds to the low-energy boundary of the PDF grid, which for the benchmark PDF set {\tt NNPDF31\_nlo\_as\_0118} is $Q_0 = 1.65 \, {\rm GeV}$. 

The observables introduced in~Eqs.~(\ref{eq:observables}) depend not only on the underlying scattering process, but also on the target, the flavour of the incident neutrino or antineutrino, and, where applicable, additional experimental selection criteria. We indicate these dependencies in the legends of the plots according to
\beq \label{eq:ptfc}
\begin{gathered}
{\rm process} \in \{ {\rm NC}, {\rm CC}, {\rm coherent} \} \,, \qquad {\rm target} \in \{ p, n, {\rm isoscalar} \} \,, \\[2mm]
{\rm beam} \in \{ \nu_e, \nu_\mu, \nu_\tau, \bar{\nu}_e, \bar{\nu}_\mu, \bar{\nu}_\tau \} \,, \qquad {\rm cuts} \in \{ E_{\ell} > E_{\ell}^{\rm cut} \} \,.
\end{gathered}
\eeq
Unless stated otherwise, we consider an isoscalar target, such as oxygen, for which the predictions are obtained by averaging the corresponding proton and neutron results. For~coherent scattering, we instead take oxygen as the nuclear target. For CC DIS, we may additionally impose a minimum energy requirement,
$E_\ell > E_\ell^{\rm cut}$, on the final-state charged lepton, in particular on the muon. Such a selection is relevant for event classes in large-volume detectors characterised by a single high-energy muon signature, such as through-going track events. When no charged-lepton energy cut is imposed, the corresponding subscript is omitted.

As discussed in the introduction, neutrino-nucleon DIS has been studied including, for instance, NNLO~QCD corrections, nuclear effects, and small Bjorken-$x_B$ resummation at very high energies. While NLO~EW corrections to neutrino DIS have been computed before~\cite{Diener:2003ss,Diener:2005me}, the fully differential implementation at high energies presented here, including lepton-mass effects and photon-induced channels, is new, and we combine these corrections with the corresponding NLO~QCD predictions. This is sufficient to quantify EW effects and matches the QCD accuracy of the \NLOPS~results discussed later in~Section~\ref{sec:comparison}. For~reference, we also include the central NNLO~QCD predictions without~EW~corrections. Nuclear effects and small Bjorken-$x_B$ resummation are instead not considered in this~work.

\subsection{Inclusive DIS cross sections}
\label{sec:sigmainclusive}

The simplest observables to consider are the inclusive cross sections as defined by the first relation in~Eq.~(\ref{eq:observables}). We begin by studying the fixed-order DIS predictions for this observable over a wide energy range, $E_{\nu} \in [ 40, 10^9 ] \, {\rm GeV}$. This range covers the regime where the perturbative DIS calculation provides the dominant contribution to the cross section, with the upper limit chosen slightly above the highest-energy neutrino observed to~date, $2.2 \cdot 10^8 \, {\rm GeV}$, reported by the KM3NeT collaboration~\cite{KM3NeT:2025npi}.

\begin{figure}[t!]
\begin{center}
\includegraphics[width=0.49\textwidth]{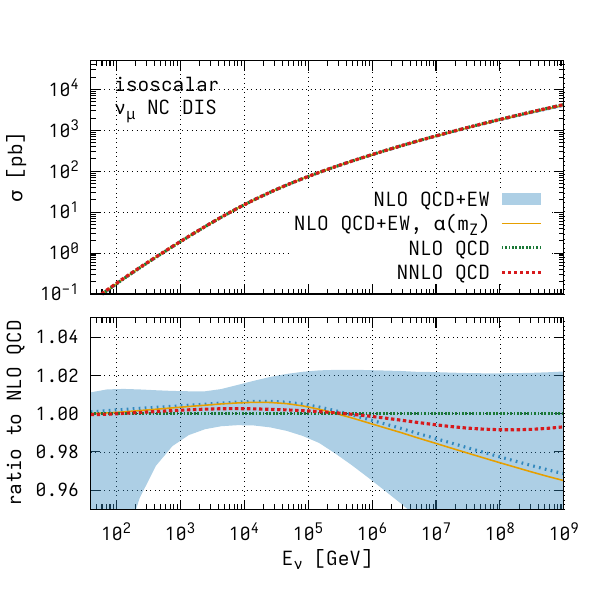}
\includegraphics[width=0.49\textwidth]{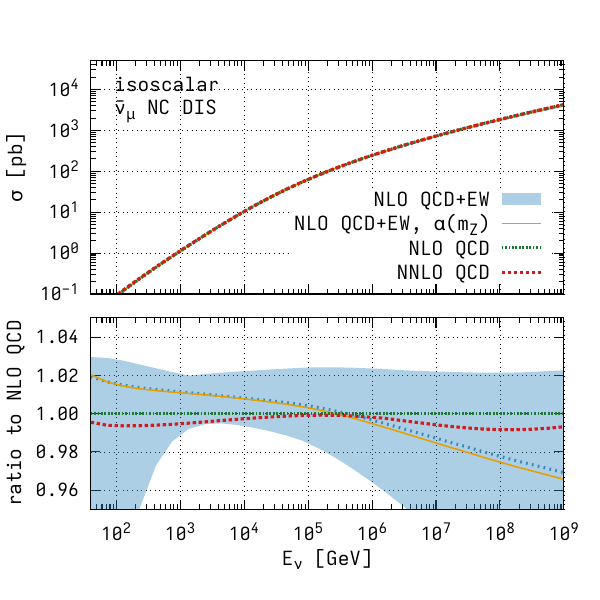} 
\vspace{0mm}
\caption{Inclusive NC DIS cross sections for $\nu_\mu$~(left) and $\bar{\nu}_\mu$~(right) scattering as defined in~Eq.~(\ref{eq:observables}), assuming an isoscalar target. The default NLO~QCD$+$EW predictions are shown in the $G_F$~scheme with scale uncertainties~(blue dotted lines and bands), together with the central NLO~QCD$+$EW result in the $\alpha(m_Z)$ scheme~(solid orange). For comparison, the NLO~(green dotted) and NNLO~(red dashed) QCD predictions are also included. The upper panels display the cross sections, while the lower panels show the ratios normalised to the NLO QCD result. See~text for further details. \label{fig:sigmaNC}}
\end{center}
\end{figure}

Our numerical results for the NC DIS process are shown in~Figure~\ref{fig:sigmaNC}, considering both~$\nu_\mu$~(left panel) and~$\bar{\nu}_\mu$~(right panel) scattering. The corresponding predictions for the other neutrino flavours are numerically almost identical and are therefore not shown. We assume a free isoscalar target, obtained by averaging proton and neutron contributions while neglecting nuclear effects, and compare the full NLO~QCD$+$EW prediction with the corresponding result without EW corrections. Unless stated otherwise, the displayed (N)NLO QCD predictions are obtained using the central {\tt NNPDF31\_(n)nlo\_as\_0118}~PDF~set, while the NLO~QCD$+$EW results are computed with NLO PDFs including the photon distribution in the {\tt LUXqed} formalism. By default, the NLO~QCD$+$EW predictions are obtained in the~$G_F$ scheme and the uncertainty from scale variation is shown, cf.~Sections~\ref{sec:hadinput} and~\ref{sec:ewscheme}. For comparison, we also include the central NLO~QCD$+$EW prediction in the $\alpha(m_Z)$ scheme. These reference predictions highlight the impact of the NLO~EW corrections, their associated uncertainties, and their size relative to higher-order QCD~effects.

From both panels it is evident that the central value of our default NLO~QCD$+$EW prediction deviates by no more than $4\%$ from the NLO~QCD result for the inclusive NC~DIS cross section over the considered energy range, with a negative and nearly universal correction at the highest energies. By varying the EW scheme from $G_F$ to $\alpha(m_Z)$, we estimate an intrinsic uncertainty of less than $0.5\%$ in the NLO~EW correction. The results obtained in the $\alpha(0)$ scheme are very similar to those in the $\alpha(m_Z)$ scheme and are therefore not shown. Interestingly, the size of the NLO~EW effects is larger than the difference between the NNLO and NLO QCD predictions, computed with their respective PDF sets, which is typically at the level of $0.5\%$. This indicates that already for inclusive neutrino-DIS predictions, achieving percent-level precision in general requires the inclusion of NLO~EW corrections in addition to higher-order QCD~effects.

At the lowest incident neutrino energies shown in~Figure~\ref{fig:sigmaNC}, differences between the $\nu_\mu$- and $\bar{\nu}_\mu$-induced NC DIS cross-section predictions become visible. While the squared amplitudes for $\nu_\mu + q \to \nu_\mu + q$ and $\bar{\nu}_\mu + \bar{q} \to \bar{\nu}_\mu + \bar{q}$ are identical, the corresponding hadronic cross sections differ due to the distinct PDF weightings, in particular the valence-sea asymmetry at high $x$. In the QCD calculation these effects are governed by the structure function~$F_3(x,Q^2)$, which encodes the parity-violating $q\bar q$ asymmetry in the nucleon --- see~Refs.~\cite{Dicus:2001kb,Gandhi:1998ri,Gluck:1998js,Cooper-Sarkar:2007zsa,Connolly:2011vc,Cooper-Sarkar:2011jtt,Bertone:2018dse,Candido:2023utz,Xie:2023suk} for further details. These differences are significant at low energies but become negligible in the high-energy regime. Another effect at low energies arises from the choice of nucleon target. For $\nu_\mu$-induced NC~DIS, the NLO~EW correction is negative~(positive) on a proton~(neutron) target, while for an isoscalar target, as considered in the figure, there is an approximate cancellation between the two contributions. In contrast, for the corresponding $\bar{\nu}_\mu$-induced process the NLO~EW correction yields a positive effect of about~$2\%$ at low energies, while at the highest energies the net impact of the NLO~EW corrections is a target-independent reduction of the cross section by a few percent. 

\begin{figure}[t!]
\begin{center}
\includegraphics[width=0.49\textwidth]{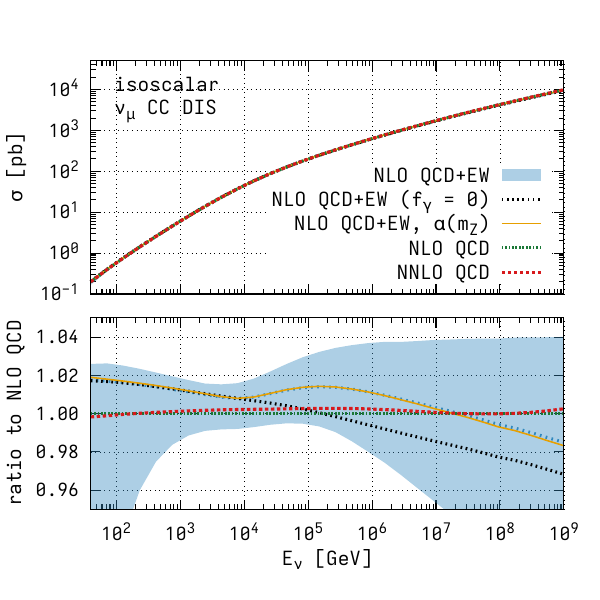}
\includegraphics[width=0.49\textwidth]{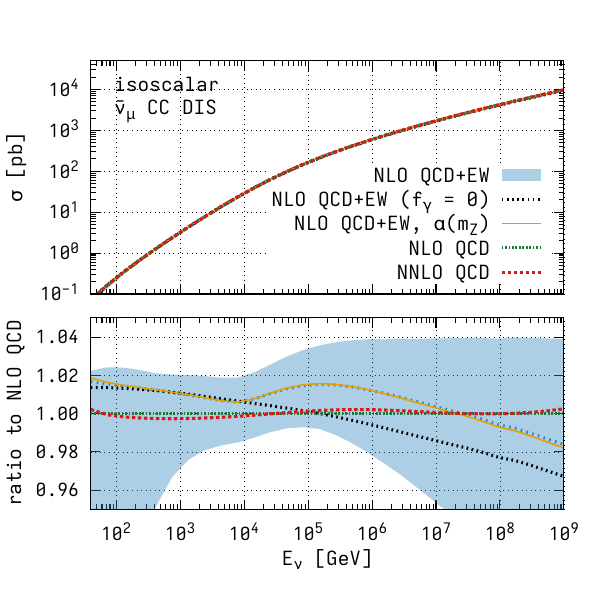}
\vspace{0mm}
\caption{As~Figure~\ref{fig:sigmaNC}, but for CC DIS, additionally showing NLO~QCD$+$EW predictions in the $G_F$ scheme without the photon~PDF (black dotted lines), denoted by the label ``NLO QCD$+$EW ($f_\gamma = 0$)''.  \label{fig:sigmaCC}}
\end{center}
\end{figure}

The corresponding results for the CC DIS process are shown in~Figure~\ref{fig:sigmaCC}, considering both~$\nu_\mu$~(left panel) and~$\bar{\nu}_\mu$~(right panel) flavours. The setup follows that of Figure~\ref{fig:sigmaNC}, but additionally includes NLO~QCD$+$EW predictions in the $G_F$ scheme without the photon~PDF. For both neutrino flavours, the NLO~EW corrections amount to a positive shift of about~$2\%$ at low energies and a small negative correction of roughly~$2\%$ at the highest energies. The residual EW~scheme dependence is at the level of about $0.2\%$. Numerically, this effect is subleading compared to the NLO~EW corrections themselves at the inclusive level, and for more differential observables --- where the NLO~EW corrections are typically larger, as will be shown below --- the scheme dependence becomes negligible compared to the dominant physical effects. We therefore present results only in the $\alpha_{G_F}$ scheme from this point onwards. As in the NC DIS case, one observes that the NLO~EW corrections are larger than the difference between the NNLO and NLO~QCD predictions, which in the case of CC DIS are nearly indistinguishable. As discussed in~Appendix~\ref{app:Top}, top-quark production becomes increasingly important at high neutrino energies, contributing an enhancement of nearly $10\%$ in the highest-energy region considered. Additional details on the impact of QCD and EW corrections to top-quark production can be found in~Appendix~\ref{app:Top}.

It is also apparent from both panels in~Figure~\ref{fig:sigmaCC} that for $E_{\nu} \in [10^4,10^6] \, {\rm GeV}$ a small~``hump'' appears in the NLO~EW corrections. From a comparison of the NLO~EW predictions with and without the photon~PDF, it is clear that this feature originates from the photon-induced subprocess. In fact, this channel receives an $s$-channel $W$-boson contribution together with an initial-state (pseudo-)collinear enhancement regulated by the charged-lepton mass. This feature has been referred to as a hidden Glashow resonance and was discussed in~Ref.~\cite{Alikhanov:2015kla}, with resummed effects studied in~Ref.~\cite{Gauld:2019pgt}. The energy dependence of this contribution to the cross section is approximately described by Eqs.~(\ref{eq:sigmavgDIS}) and~(\ref{eq:sigmavNpartonic}). As~this correction depends logarithmically on the charged-lepton mass, its impact on the cross section varies with the flavour of the neutrino beam, being largest in the electron channel. Due to the large hierarchy between the electron mass $m_e$ and the partonic COM energy of order~$m_W$, special care is required in the implementation of the process $\nu_e + \gamma \to e^- + q \bar q^{\hspace{0.25mm} \prime}$. Details of this issue and a validation of our procedure are discussed in~Appendix~\ref{app:glashow}.

\begin{figure}[t!]
\begin{center}
\includegraphics[width=0.49\textwidth]{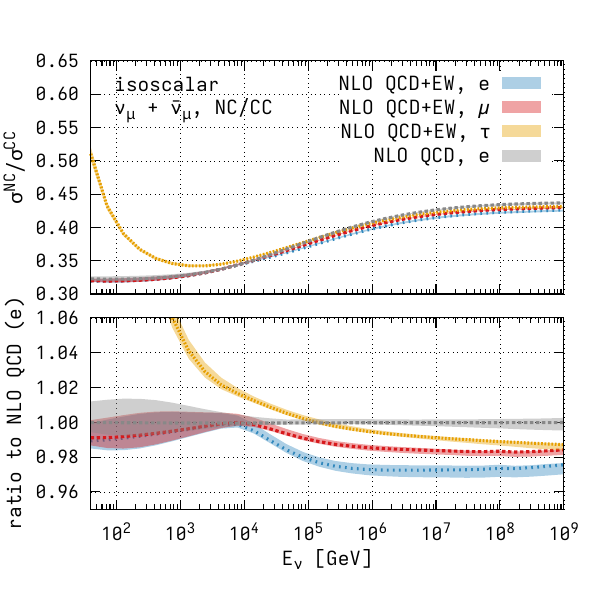}
\includegraphics[width=0.49\textwidth]{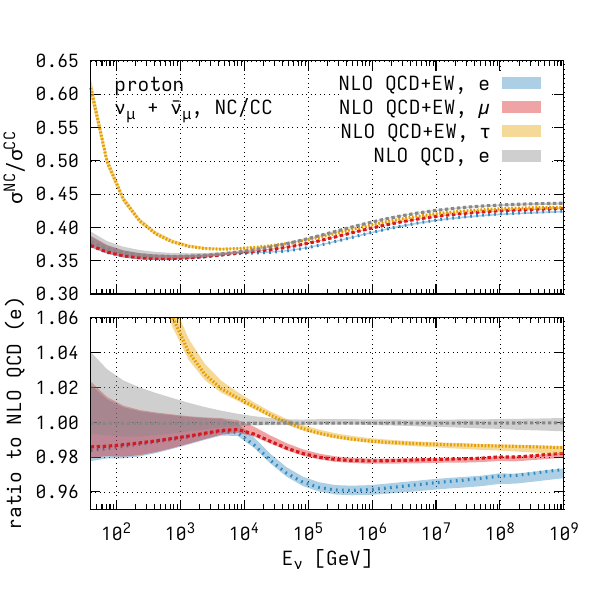}   
\vspace{0mm}
\caption{Ratio of inclusive NC and CC DIS cross sections for different neutrino flavours as defined in~Eq.~(\ref{eq:RNC}). An isoscalar (proton) target is assumed in the left~(right) panel. The NLO~QCD$+$EW predictions are shown for electron~(blue), muon~(red), and tau~(orange) flavours, together with the NLO~QCD-only prediction for the electron case~(grey). The~coloured bands correspond to scale uncertainties. Additional details are provided in the main text.}
\label{fig:ratioNCCC}
\end{center}
\end{figure}

To better illustrate the impact of the flavour-dependent corrections associated with the hidden Glashow process, Figure~\ref{fig:ratioNCCC} shows the ratio of NC to CC DIS cross sections for fixed neutrino flavours, summed over neutrino and antineutrino contributions, as a function of the incoming neutrino energy. In particular, we consider
\beq \label{eq:RNC}
R_{\nu_{\ell}}^{\rm target} = \frac{ \displaystyle \sum_{f = \nu_\ell, \bar{\nu}_\ell} \sigma^{{\rm NC}, {\rm target}}_{f}}{ \displaystyle \sum_{f = \nu_\ell, \bar{\nu}_\ell} \sigma^{{\rm CC}, {\rm target}}_{f}} \,,
\eeq
where the target in the left and right panels is an isoscalar and a proton, respectively.

From the panels in~Figure~\ref{fig:ratioNCCC}, one observes that at low neutrino energies the presence of a massive charged lepton in the CC DIS process leads to a suppression of the corresponding cross section and thus to an enhancement of the ratio defined in~Eq.~\eqref{eq:RNC}. This is a purely kinematic effect and is relevant for both the NLO QCD and NLO~EW corrections. The effect is most pronounced in the tau channel and becomes sizeable for $E_{\nu_\tau} \lesssim 10^3 \, {\rm GeV}$. In contrast, at high neutrino energies one observes a decrease in the ratio for the NLO~QCD$+$EW predictions, which is partly due to the different size of the NLO~EW corrections to NC and CC DIS, present for all flavours, and partly due to the resonantly enhanced hidden Glashow contribution, responsible for the flavour-dependent dip for $E_{\nu} \in [10^4,10^6]\,{\rm GeV}$. This effect is larger for a proton target than for a neutron (and hence also for an isoscalar target), reflecting the larger photon~PDF in the proton, a difference that is enhanced by our purely perturbative treatment of the neutron photon~PDF (see~Section~\ref{sec:hadinput}). Comparing the impact across flavours, the expected hierarchy associated with the logarithmic dependence on $m_e$, $m_\mu$, and $m_\tau$ is clearly visible.

To summarise, the NLO~EW corrections induce percent-level effects on the cross-section predictions, with a sign and magnitude that depend on the energy regime, the target, and in the CC DIS case also the beam flavour. At the highest energies considered, they are negative for both NC and CC DIS processes but remain below $4\%$ in magnitude. This behaviour is expected since the dominant contribution arises from the region $Q^2 \simeq m_V^2$, where $m_V$ is the mass of the exchanged EW boson: the propagator suppression $1/(Q^2+m_V^2)^2$ disfavours large $Q^2$, preventing the growth of EW Sudakov logarithms, while the PDFs increase only logarithmically with $Q^2$. In CC DIS, additional flavour-dependent effects originate from initial-state collinear logarithms involving the charged-lepton mass in photon-induced channels, whereas final-state QED radiation is negligible for inclusive observables, subject only to a minimal $Q^2 > Q_0^2$~cut. We recall that residual uncertainties from QCD modelling and nuclear effects range from about $2\%$ at low energies to roughly $5\%$ at the highest energies considered~\cite{Bertone:2018dse,AbdulKhalek:2022fyi}. The NLO~EW corrections to the inclusive neutrino-DIS cross sections are thus generically smaller, but not negligible, as compared to the present theoretical uncertainties.

\subsection{Other CC DIS predictions}
\label{sec:exclusive}

In this subsection we study the impact of the NLO~EW corrections on CC DIS observables that require an identified muon, as well as those imposing an additional minimum energy cut on the muon. Such selections are motivated by analyses of muon tracks in large-volume neutrino detectors such as IceCube and KM3NeT: minimum-energy requirements are typical of through-going tracks, while inelasticity measurements require starting tracks with the interaction vertex inside the detector. Since these requirements are not collinear safe with respect to QED radiation, logarithmically enhanced EW effects are expected in the corresponding predictions. These effects have recently been studied in~Ref.~\cite{Plestid:2024bva} at LL~accuracy for FSR of photons. In the following, we present results for incoming $\nu_\mu$ only, noting that the relative effects are very similar for $\bar{\nu}_\mu$.

\begin{figure}[t!]
\includegraphics[width=0.49\textwidth]{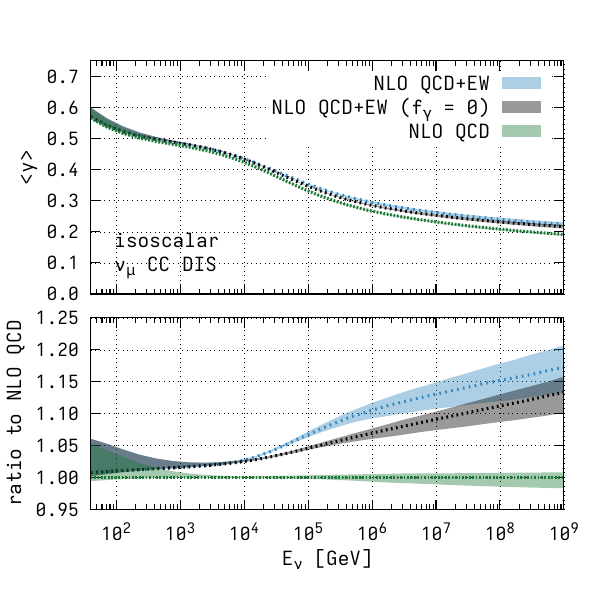}
\includegraphics[width=0.49\textwidth]{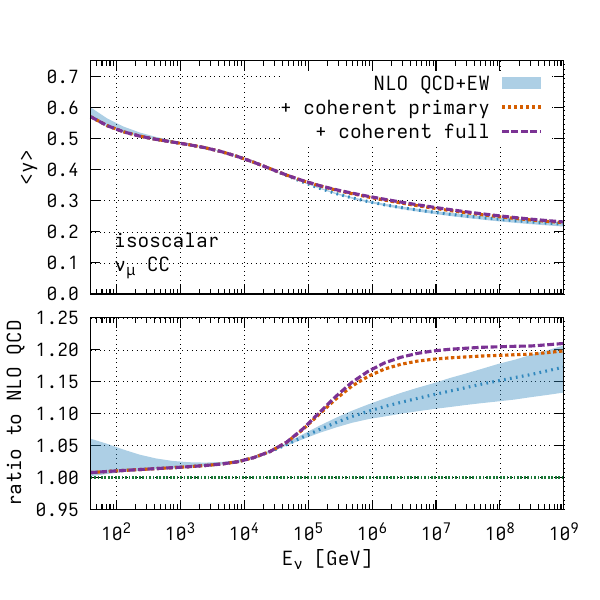}
\vspace{0mm}
\caption{Inclusive mean inelasticity $\langle y \rangle$ as a function of the incoming $\nu_\mu$ energy on an isoscalar target. The NLO~QCD predictions~(green) are compared to the default NLO~QCD$+$EW predictions in the $G_F$~scheme with~(blue) and without~(black) photon~PDF in the left panel, while the right panel includes predictions with coherent scattering effects on top of the NLO~QCD$+$EW corrections, without~(dotted vermillion) and with~(dashed purple) decay of the $W$~boson produced in a Glashow-like process. In both panels, the upper plot shows the mean inelasticity and the lower plot displays the ratio normalised to the NLO~QCD result. Consult the main text for further explanations. \label{fig:meany}}
\end{figure}

The IceCube collaboration has performed measurements~\cite{IceCube:2018pgc,IceCubeCollaborationSS:2025zgz} of the inelasticity in CC~DIS events with a final-state muon over a wide range of high energies. Of particular interest is the mean inelasticity, defined in the second relation of~Eq.~(\ref{eq:observables}), with the inelasticity in the laboratory frame given by the third relation in~Eq.~(\ref{eq:kin2}). Fixed-order predictions for this observable are shown in~Figure~\ref{fig:meany} as a function of the incoming $\nu_\mu$ energy on an isoscalar target. The left panel compares central predictions and scale uncertainties for the NLO~QCD$+$EW results in the $G_F$~scheme, both with and without the photon~PDF, against the pure NLO~QCD predictions. We note that the NLO~QCD$+$EW result without the photon~PDF includes final-state QED radiation --- in particular from the muon leg --- but excludes photon-induced contributions such as the hidden Glashow resonance discussed above. The right panel displays predictions that incorporate coherent scattering effects in addition to the NLO~QCD$+$EW corrections, shown both without and with the decay of the $W$~boson produced in a Glashow-like process. We~refer to these two cases as ``coherent primary'' and ``coherent full'', respectively.

The impact of the NLO~EW corrections on $\langle y \rangle$ is already visible in the left panel of~Figure~\ref{fig:meany} at neutrino energies as low as $E_{\nu_\mu} = 100 \, {\rm GeV}$, with positive deviations from the NLO~QCD prediction reaching up to $10\%$ at energies around $E_{\nu_\mu} = 10^6 \, {\rm GeV}$. The~role of photon-induced contributions can be further quantified by comparing the two depicted NLO~QCD$+$EW predictions, showing that these effects become relevant already for neutrino energies $E_{\nu_\mu} \gtrsim 10^4 \, {\rm GeV}$. For instance, at around $E_{\nu_\mu} = 10^8 \, {\rm GeV}$ we find relative corrections of about $10\%$ when photon-induced channels are excluded, increasing to $15\%$ when they are included. This behaviour can be understood as follows: collinear photon emission from the muon reduces its energy and correspondingly increases the inelasticity, while in photon-induced processes the outgoing muon is typically produced with relatively low momentum (and hence large inelasticity), since the resonantly produced $W$ boson carries most of the~energy. As shown in~Figure~\ref{fig:meany}, the hidden Glashow contributions also lead to a faster rise of~$\langle y \rangle$ for $E_{\nu_\mu} \in [10^4, 10^6] \, {\rm GeV}$. This feature arises from the characteristic logarithmic enhancement in~Eq.~(\ref{eq:sigmavNpartonic}), together with the approximate scaling of the photon~PDF in~Eq.~(\ref{eq:sigmavgDIS}) as $\alpha/(2 \pi) \hspace{0.25mm} \ln \left (\mu_F^2/m_N^2 \right )$ with $\mu_F^2 = \hat{s}_{12}$. While not shown here, we have also investigated the impact of including or neglecting the top-quark contribution in the prediction of $\langle y \rangle$. Removing the top-quark component leads only to a small relative change of about $1\%$ in the mean inelasticity. This small effect results from a partial cancellation of the top-quark contribution between the numerator and denominator entering the definition of $\langle y \rangle$.

We furthermore observe that the general behaviour of the NLO~EW corrections shown on the left-hand side in~Figure~\ref{fig:meany} is consistent with the fixed-order LL~approximation presented in~Ref.~\cite{Plestid:2024bva}, although differences in magnitude are observed. For example, Figure~4 of~Ref.~\cite{Plestid:2024bva} exhibits a relative positive correction of about $15\%$ at $E_{\nu_\mu}=10^8\,{\rm GeV}$, to be compared with our central prediction of about $10\%$ when the photon-induced channel is removed, allowing for a consistent comparison with their calculation. We have verified that our LL implementation of~Eq.~(\ref{eq:sigma_LL}) reproduces the behaviour found in~Ref.~\cite{Plestid:2024bva} when adopting their choice of hadronic scale, $\mu_h^2=s$, where $s$ denotes the squared COM energy of the neutrino--nucleon collision. However, this agreement also illustrates an important limitation of this scale choice. In contrast to the partonic scale $\hat{s}_{12}$, the hadronic scale $s$ does not reflect the kinematics of the hard neutrino--parton subprocess. In collinear factorisation, one has that $\hat{s}_{12}$ scales as $x\hspace{0.125mm} s$ such that the two scales can differ by several orders of magnitude at the small momentum fractions $x$ relevant for ultra-high-energy neutrino scattering. Consequently, replacing $\hat{s}_{12}$ by $s$ can lead to a sizeable modification of the logarithmic enhancement entering the LL approximation. Numerically, at $E_{\nu_\mu}=10^8\,{\rm GeV}$, the choice $\mu_h^2=s$ overestimates the QED radiation by approximately $50\%$ relative to the NLO~EW correction, which is independent of $\mu_h$. Varying $\mu_h$ by a factor of two changes the result by only about $\pm6\%$, and hence does not capture the potentially much larger uncertainty associated with the choice of the relevant hard scale. We therefore do not regard $\mu_h^2=s$ as a reliable implementation of the fixed-order LL expectation. Instead, the choice $\mu_h^2=\hat{s}_{12}$, with $\hat{s}_{12}$ the squared COM energy of the neutrino--parton collision, retains the partonic kinematics of the hard scattering and yields results consistent with the fixed-order NLO~EW calculation. We note that, even within a structure-function description of the QCD cross section, as used in~Ref.~\cite{Plestid:2024bva}, a partonic hard scale can be approximated at LO by $\hat{s} \simeq x_B \hspace{0.25mm} s$.

Given the sizeable impact of the NLO~EW corrections on the mean inelasticity, it is natural to also examine the effect of coherent scattering, shown in the right panel of Figure~\ref{fig:meany}. The coherent scattering contributions are computed in the narrow-width approximation for the $W$~boson, which allows a clean separation into two categories: the ``coherent primary'' contribution, which accounts for muons produced directly in $\nu_\mu + A \to \mu^- + A^\prime + W^+$, and ``coherent full'', which additionally includes muons from the subsequent $W$-boson decay --- the latter being accessible via electron-, muon-, or tau-neutrino-induced coherent scattering. The difference between the two isolates the contribution of muons from $W$-boson decays to the mean inelasticity. As is apparent from the figure, both coherent contributions exhibit a logarithmic enhancement associated with Glashow-like resonant processes. In this case, the muon-mass logarithm arises solely from~Eq.~(\ref{eq:sigmavNpartonic}), while the logarithmic correction $\alpha/(2 \pi) \hspace{0.25mm} \ln \left (\hat{s}_{12}/m_N^2 \right )$ appearing for the photon~PDF is replaced by a nuclear form factor suppression. This explains why the coherent scattering contributions saturate at neutrino energies $E_{\nu_\mu} \gtrsim 10^{7} \, {\rm GeV}$. Furthermore, a comparison of the ``coherent primary'' and ``coherent full'' results reveals that the additional inclusion of muons from $W$-boson decays has only a minor impact on $\langle y \rangle$ at the level of $1\%$, when the distribution is inclusive~in~$E_\mu$. The coherent contributions themselves, however, increase $\langle y \rangle$ by up to about $6\%$ for $E_{\nu_\mu} \gtrsim 10^6\,{\rm GeV}$.

\begin{figure}[t!]
\begin{center}
\includegraphics[width=0.49\textwidth]{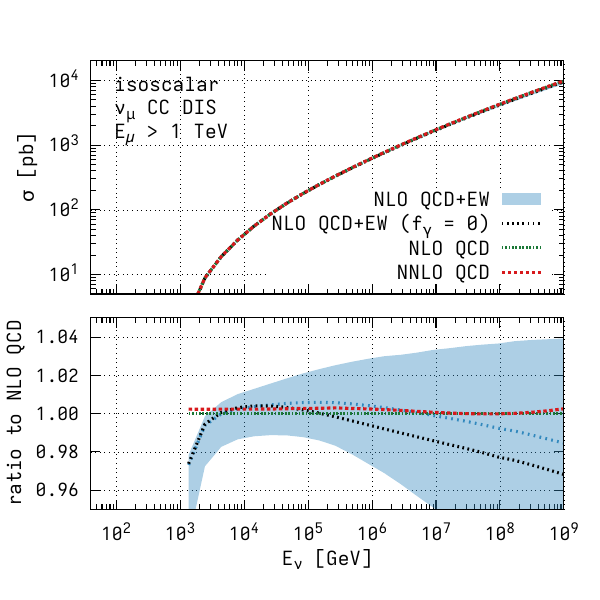}
\includegraphics[width=0.49\textwidth]{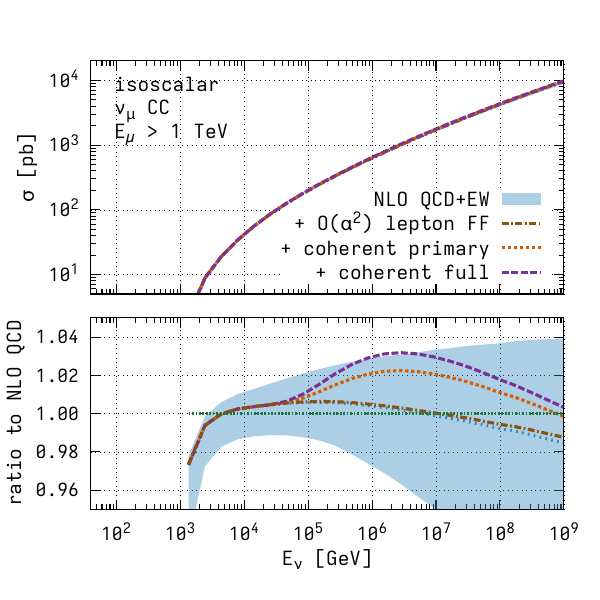}       
\vspace{0mm}
\caption{CC~DIS cross sections as a function of the incoming $\nu_\mu$ energy on an isoscalar target, imposing a cut $E_\mu > E_\mu^{\rm cut}$ with $E_\mu^{\rm cut} = 1 \, {\rm TeV}$. The NLO~QCD predictions~(green) are compared to the default NLO~QCD$+$EW predictions with~(blue) and without~(black) photon~PDF, and to the NNLO~QCD predictions~(red), in the left panel, while the right panel depicts the ``coherent primary''~(dotted vermillion), ``coherent full''~(dashed purple), as well as the $\mathcal{O}(\alpha^2)$ lepton~FF~contributions~(brown dotted-dashed) on top of the NLO~QCD$+$EW corrections. The upper panels display the cross sections and the lower panels show ratios normalised to the NLO~QCD result. See the main text for additional~details. \label{fig:sigmaEcut}}
\end{center}
\end{figure}

The NLO~EW and coherent scattering contributions to $\nu_\mu + A \to \mu^- + X$ enhance the production of low-energy muons, leading to a positive shift in the mean inelasticity. In large-volume detectors such as IceCube, muon-track selections are not based on a sharp cut on the true muon energy but instead rely on reconstructed quantities such as the muon energy proxy or the associated energy loss along the track, which effectively impose a minimum threshold at the TeV scale depending on the analysis --- for additional details, we refer to, for instance, Refs.~\cite{IceCube:2014rwe,IceCube:2015qii,IceCube:2016umi,IceCube:2017trr,IceCube:2021oqo}. From the theory perspective, such selections suppress events with soft muons and can therefore also shift the predicted mean inelasticity. For~theory-data comparisons, these requirements can be mimicked by imposing a cut on the produced muon energy, $E_\mu > E_\mu^{\rm cut}$. More differential selections may instead be formulated in terms of the visible energy or the muon energy loss, which more closely track the experimental reconstruction. While such a mapping is important for a precise assessment of NLO~EW effects in an experimental context, since QED radiation and photon-induced channels predominantly affect the low-energy muon region most strongly impacted by these cuts, in this work we adopt the simpler energy-cut prescription, as a more refined treatment would require a dedicated detector simulation of large-volume neutrino detectors and lies beyond the scope of this~study.

We first study the impact of imposing a cut $E_\mu > E_\mu^{\rm cut}$ with $E_\mu^{\rm cut} = 1 \, {\rm TeV}$ on the CC~DIS cross-section predictions. The effect of this requirement is shown in~Figure~\ref{fig:sigmaEcut} as a function of the $\nu_\mu$ energy for an isoscalar target. The left panel compares the default NLO~QCD$+$EW predictions, with and without the photon~PDF, to the pure NLO~QCD results, while the right panel displays the  ``coherent primary'' and ``coherent full'' effects, as well as the $\mathcal{O}(\alpha^2)$ contributions to the lepton~FF~defined in~Eq.~(\ref{eq:sigma_frag}) on top of the NLO~QCD$+$EW corrections.

\begin{figure}[t!]
\begin{center}
\includegraphics[width=0.49\textwidth]{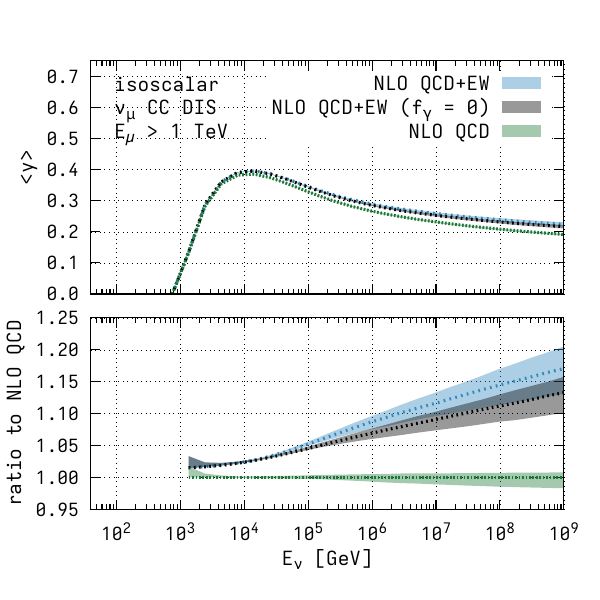}
\includegraphics[width=0.49\textwidth]{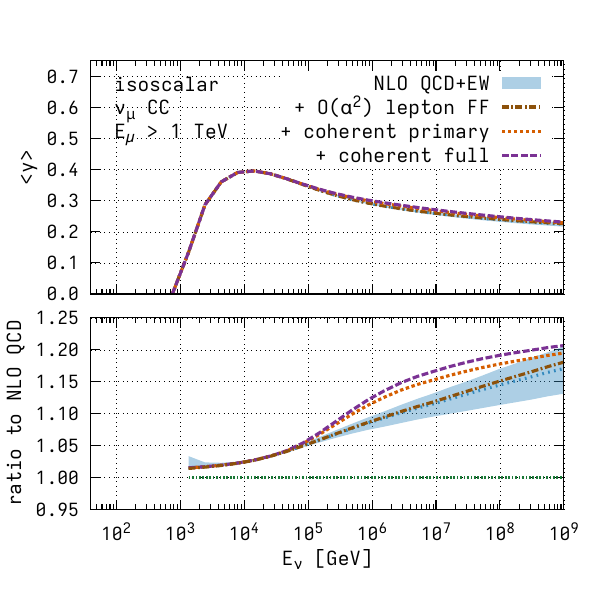}   
\caption{As in~Figure~\ref{fig:sigmaEcut}, but for the mean inelasticity $\langle y \rangle$. The left panel does not include the NNLO~QCD prediction, and shows scale uncertainties also for the NLO~QCD$+$EW predictions without photon~PDF and for the NLO~QCD predictions. Further details are provided in the main text. \label{fig:inelasticityEcut}}
\end{center}
\end{figure}

The left panel of Figure~\ref{fig:sigmaEcut} shows that, relative to the inclusive results in~Figure~\ref{fig:sigmaCC}, the NLO~EW corrections remain~essentially unchanged in the high-energy regime. This is expected: the energy cut preferentially removes low-energy muons and thus leaves the high-energy region largely unaffected. Deviations arise instead for $E_\nu \gtrsim E_\mu^{\rm cut}$, where the neutrino energy approaches the kinematic threshold imposed by the muon energy cut. In this regime, the available phase space becomes increasingly constrained, enhancing the sensitivity of the cross section to the energy partition between the final-state muon and the hadronic system and consequently distorting its energy dependence.

The right panel of Figure~\ref{fig:sigmaEcut} demonstrates that imposing $E_\mu > E_\mu^{\rm cut}$ affects the ``coherent primary'' and ``coherent full'' predictions differently, with the latter systematically exceeding the former by about $1\%$ for $E_{\nu_\mu} \gtrsim 10^5 \, {\rm GeV}$. This is readily understood, since the ``coherent full'' predictions include muons from $W$-boson decays in the Glashow-like process which tend to carry energies well above $E_\mu^{\rm cut} = 1 \, {\rm TeV}$ and therefore pass the cut, adding to the cross section relative to the ``coherent primary'' result. The same panel also demonstrates that the~$\mathcal{O}(\alpha^2)$ contributions to the lepton~FF~have only a minor impact on the CC~DIS cross section compared to the NLO~QCD$+$EW corrections. The LL~and next-to-leading-logarithmic~(NLL) $\mathcal{O}(\alpha^2)$ muon-mass logarithms encoded in $D_\ell(z, \mu_F^2)$, cf.~Eq.~(\ref{eq:sigma_frag}), are thus negligible in practice, at the level of $0.1\%$, compared to the LL contributions at~$\mathcal{O}(\alpha)$ of~Eq.~(\ref{eq:sigma_LL}). This is expected, since for $E_{\nu_\mu} \gg E_\mu^{\rm cut}$ the cut only removes events with $y \to 1$, so that this cross section is nearly inclusive in the muon kinematics.

Figure~\ref{fig:inelasticityEcut} displays the mean inelasticity $\langle y \rangle$ in the CC~DIS channel for $\nu_\mu$ scattering on an isoscalar target, imposing the cut $E_\mu > E_\mu^{\rm cut}$ with $E_\mu^{\rm cut} = 1 \, {\rm TeV}$. The left panel shows the default NLO~QCD$+$EW predictions with and without the photon~PDF together with the pure NLO~QCD results, while the right panel additionally includes the ``coherent primary'', ``coherent full'', and the $\mathcal{O}(\alpha^2)$ lepton fragmentation contributions on top of the NLO~QCD$+$EW corrections.

As shown in the left panel of~Figure~\ref{fig:inelasticityEcut}, compared to the inclusive predictions displayed in~Figure~\ref{fig:meany}, the cut
$E_\mu^{\rm cut} = 1 \, {\rm TeV}$ has only a minor impact on the NLO~EW corrections, both with and without the photon~PDF.  This behaviour is understood from the fact that for $E_{\nu_\mu} \gg E_\mu^{\rm cut}$ the requirement $E_\mu > E_\mu^{\rm cut}$ only removes events with $y > 1 - E_\mu^{\rm cut}/E_{\nu_\mu}$, i.e.~a small region close to $y = 1$. A TeV-scale cut therefore does not significantly distort the relevant phase space, as it mainly removes events with very low-energy muons that contribute only weakly to the inelasticity average after integration over the full kinematic range. Photon-induced contributions, in particular those associated with the hidden Glashow mechanism, are somewhat more sensitive to this selection, which explains the slightly larger --- though still small --- effect observed in the predictions including the photon~PDF.

The right panel of Figure~\ref{fig:inelasticityEcut} further illustrates that, for the mean inelasticity $\langle y \rangle$, the difference between the ``coherent primary'' and ``coherent full'' predictions is significantly smaller than in the case of the cross section, reaching about $1\%$ at the highest energies. Moreover, the~LL and~NLL muon-mass logarithms at $\mathcal{O}(\alpha^2)$  induce modifications of a comparable magnitude. This indicates that computing the coherent effects for a stable $W$ boson, while retaining only the LL~contributions at~$\mathcal{O}(\alpha)$ of~Eq.~(\ref{eq:sigma_LL}) with respect to muon-mass logarithms, already provides a good approximation to corrections of this kind.

\section{Comparison to NLO\texorpdfstring{$\bm{+}$}{+}PS predictions}
\label{sec:comparison}

In the previous section the fixed-order calculations including both QCD and EW corrections have been used to evaluate mostly inclusive observables, or those with only minimal requirements on an identified charged lepton. These calculations can also readily be applied to more differential quantities in the outgoing particle kinematics, such as charged-lepton inelasticity distributions. From an experimental perspective, however, a fully exclusive event simulation is generally preferred, as it provides a hadron-level description of all final-state particles and thus forms the basis for a realistic detector simulation. Such a description can be obtained using the {\tt POWHEG-BOX} DIS generator~\cite{FerrarioRavasio:2024kem}, matched to {\tt Pythia~8.3}, thereby providing an \NLOPS~accurate simulation of fully exclusive final states. While the hard scattering is described with NLO QCD accuracy, the matching to the PS ensures the resummation of LL~contributions and retains a fully exclusive description of the hadron-level final-state kinematics. The purpose of this section is to benchmark the new fixed-order NLO QCD$+$EW calculation against the \NLOPS~prediction and to provide guidance for the optimal use of the latter in experimental analyses, such as those performed by IceCube and KM3NeT. Throughout, we consider a fixed neutrino beam energy and focus on the inelasticity distribution $y$, defined in~Eq.~\eqref{eq:kin1}, which is a key observable in experimental studies and exhibits a pronounced sensitivity to EW corrections.

\subsection{NLO QCD benchmarking and momentum reshuffling}
\label{sec:POWHEGsettings}

As a first step towards benchmarking the description of EW effects in the fixed-order and \NLOPS~approaches, we first compare the two frameworks at NLO QCD accuracy,\footnote{A recent benchmark comparison of DIS generators in the context of the FPF was presented~in~\cite{Hernandez-Sainz:2026rxc}.} prior to the inclusion of any EW corrections. The {\tt POWHEG-BOX} simulations have been configured to reproduce the setup of the fixed-order calculation, employing the EW input parameters defined in~Section~\ref{sec:input} and the {\tt NNPDF31\_nlo\_as\_0118} PDF set. Following Appendix~A of~Ref.~\cite{FerrarioRavasio:2024kem}, all simulations are performed in the neutrino-hadron COM frame. In the {\tt POWHEG-BOX} DIS implementation, this frame choice is enabled by setting
\begin{equation*}
{\tt fixed\_lepton\_beam~0}
\end{equation*}
in the {\tt powheg.input} file. For ultra-high-energy neutrino beams, this setup is found to be numerically more stable than the corresponding fixed-target configuration. In practice, the relevant Lorentz transformation must then be applied when constructing observables that depend on the choice of reference frame.

A potential source of discrepancy between the fixed-order CC DIS calculation described in~Section~\ref{sec:calculation} and the corresponding {\tt POWHEG-BOX} implementation of Ref.~\cite{FerrarioRavasio:2024kem} stems from the treatment of particle masses. In the fixed-order calculation, all mass effects are retained exactly, whereas the {\tt POWHEG-BOX} implementation generates events with massless final-state particles. Physical masses are restored only after the events are passed to the PS. By~default, {\tt Pythia~8.3} performs an internal ``momentum reshuffling'' to recover the physical masses of the outgoing particles, while the {\tt POWHEG-BOX} framework offers an alternative reshuffling prescription that acts directly on the hard-process final state. We further note that {\tt Pythia~8.3} restores the target nucleon mass. As a result, the reshuffling procedure constitutes a potential source of difference between the MC~simulation and the fixed-order calculation, where the latter treats all masses exactly from the outset.

When the default {\tt POWHEG-BOX} reshuffling procedure is applied, the daughter particles associated with a branching are first transformed into the rest frame of the mother system through a longitudinal boost followed by a transverse boost. Subsequently, the magnitudes of their three-momenta are rescaled by a common factor, leaving their directions unchanged. The rescaling parameter $\lambda$ is obtained numerically using the Newton--Raphson algorithm by requiring the conservation of energy
\beq \label{eq:oldshuffle}
\sum_{i\in{\rm daughters}}
\sqrt{\lambda \, |\overline{\vec p}_i|^2+m_i^2} = E_{\rm mother} \,,
\eeq
where $\overline{\vec p}_i$ denotes the three-momentum of daughter particle $i$ before reshuffling, evaluated in the rest frame of the mother system, and $E_{\rm mother}$ is the corresponding total energy. While~this procedure guarantees four-momentum conservation, it modifies the kinematics of the final-state particles and can therefore distort differential distributions of observables that depend sensitively on them, such as the DIS variables introduced in~Eq.~\eqref{eq:kin1}.

\begin{figure}[t!]
\begin{center}
\includegraphics[width=0.49\textwidth]{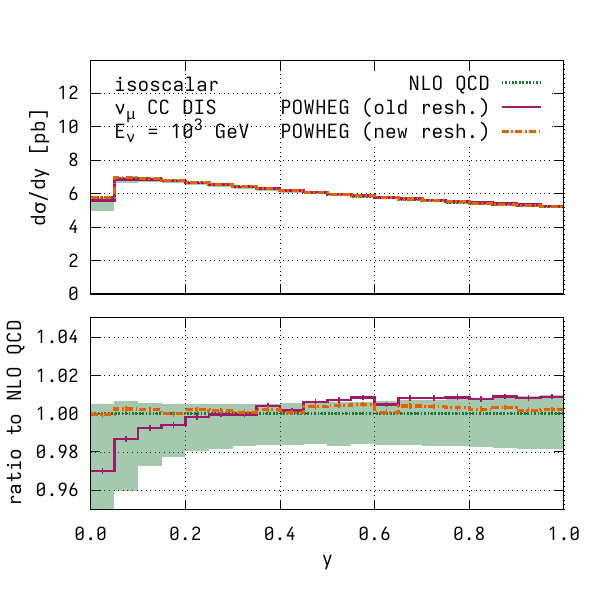}
\includegraphics[width=0.49\textwidth]{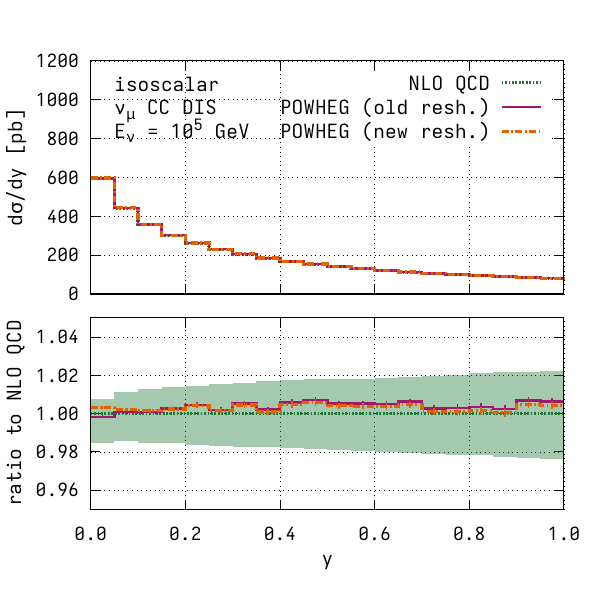} 
\vspace{0mm}
\caption{Inelasticity distributions for CC DIS in $\nu_\mu$ scattering at $E_{\nu_\mu}=10^3\,{\rm GeV}$~(left) and $E_{\nu_\mu}=10^5\,{\rm GeV}$~(right). The fully massive NLO QCD predictions~(green dotted) are compared to the {\tt POWHEG-BOX} results obtained using both the original~(magenta solid) and the new reshuffling procedure~(orange dashed-dotted). The upper panels show the differential cross sections as functions of the inelasticity, while the lower panels display ratios normalised to the fully massive NLO QCD prediction. The green bands indicate scale variations. Further details can be found in the main text. \label{fig:reshuff}}
\end{center}
\end{figure}

To overcome this issue, we have introduced a new process-specific momentum reshuffling procedure that preserves the DIS observables exactly. This is achieved with a Frixione--Kunszt--Signer~(FKS)~parametrisation of the final-state lepton momentum in both the massless, $\left(\bar{\xi}_\ell,\bar{y}_\ell,\bar{\phi}_\ell\right)$, and massive, $\left(\xi_\ell,y_\ell,\phi_\ell\right)$, cases, where $\xi_\ell$ denotes the lepton energy in units of $\sqrt{s}/2$, and $y_\ell$ and $\phi_\ell$ the cosine of its polar angle and its azimuthal angle in the neutrino-nucleon COM frame, neglecting the nucleon mass. Expressing $Q^2$, $x_B$, and $y$ in terms of the corresponding~FKS~variables and requiring the two descriptions to yield identical values of the DIS observables leads to the following relations:
\beq \label{eq:DIS-to-FKS-LO-massive}
s = \bar{s} \,, \qquad \xi_\ell = \bar{\xi}_\ell+\frac{m_\ell^2}{s} \,, \qquad y_\ell = \frac{s \hspace{0.25mm} \bar{y}_\ell \hspace{0.25mm} \bar{\xi}_\ell-m_\ell^2} {\sqrt{m_\ell^4 - 2 \hspace{0.125mm} m_\ell^2 \hspace{0.25mm} s \left(2-\bar{\xi}_\ell\right) + s^2 \hspace{0.25mm} \bar{\xi}_\ell^2}} \,.
\eeq
Since the DIS observables are independent of the lepton azimuthal angle, no condition is imposed on $\phi_\ell$. In the implementation used here, we simply retain the azimuthal angle generated by the massless {\tt POWHEG} DIS generator. Because the DIS observables depend exclusively on the final-state lepton kinematics, the relations in~Eq.~\eqref{eq:DIS-to-FKS-LO-massive} apply equally to~LO and~NLO events. They therefore define a mapping between the massless and massive~FKS~variables that preserves both the DIS observables and the total COM~energy.

A~comparison between the fixed-order NLO QCD predictions and those obtained with {\tt POWHEG-BOX} using the new momentum reshuffling procedure is presented in~Figure~\ref{fig:reshuff}. Specifically, we compare the fixed-order NLO QCD predictions for the inelasticity distribution in $\nu_\mu$-induced CC DIS with the corresponding {\tt POWHEG-BOX} results obtained using both the standard and the new reshuffling prescriptions. The left panel shows the results for $E_{\nu_\mu}=10^3\,{\rm GeV}$ and illustrates that the original reshuffling procedure leads to deviations from the fixed-order NLO QCD prediction at the percent level, most notably in the low-$y$ region. In contrast, the new process-specific reshuffling prescription provides substantially improved agreement with the fully massive calculation. The same comparison is shown in the right panel for $E_{\nu_\mu}=10^5\,{\rm GeV}$. At this higher energy, both the original and the improved reshuffling prescriptions reproduce the fixed-order NLO QCD prediction with high accuracy, and the differences between the two approaches become negligible. This~behaviour is expected, as lepton-mass power corrections become increasingly suppressed at higher neutrino energies, where the relevant kinematics approach the ultra-relativistic limit. In this regime, the specific prescription used to restore particle masses through reshuffling has only a minimal impact on the reconstructed event kinematics, resulting in an effective convergence of the different approaches.

\subsection{Ordering variable}
\label{sec:ordering}

As already explained in~Ref.~\cite{Banfi:2023mhz} and discussed in more detail in~Appendices B and C of that work, the {\tt POWHEG-BOX} DIS generator employs a definition of the ordering variable $k_t$ that differs from the standard {\tt POWHEG} implementation. Consequently, matching to {\tt Pythia~8.3} requires the use of the {\tt PowhegHooks} and {\tt PowhegHooksVincia} classes, which are provided in the {\tt py8} directory of the {\tt POWHEG-BOX} DIS repository~\cite{POWHEG_DIS}. To activate the veto procedure implemented in these classes, we use the settings
\begin{align*}
&{\tt SpaceShower:pTmaxMatch = 2}\,,\\
&{\tt TimeShower:pTmaxMatch = 2}\,,\\
&{\tt Vincia:pTmaxMatch = 2}\,,\\
&{\tt POWHEG:veto = 1}\,,\\
&{\tt POWHEG:vetoCount = 3}\,,\\
&{\tt POWHEG:pTemt = 0}\,,\\
&{\tt POWHEG:QEDveto = 2}\,.
\end{align*}
With these settings, the PS is started at the kinematic limit, and all shower emissions harder than the hardness scale of the original {\tt POWHEG} event are vetoed.  An adjustment to the hooks is made to exclude hard photons from the veto cascade (i.e. they are not vetoed).
To ensure a consistent definition of this hardness scale, we further set
\begin{align*}
&{\tt POWHEG:pThard = 0}\,,\\
&{\tt POWHEG:pTdef = 1}\,,\\
&{\tt POWHEG:nFinal = 2}\,,
\end{align*}
where {\tt POWHEG:nFinal} specifies the number of final-state particles in the Born process, $\nu_\ell + q \to \ell^- + q^{\hspace{0.25mm}\prime}$. To ensure that mass reshuffling effects are minimised when using the {\tt VINCIA} algorithm, we also set the number of flavours treated as massless to
\begin{align*}
&{\tt Vincia:nFlavZeroMass = 2}\,.
\end{align*}
A comprehensive list of matching options can be found in~Ref.~\cite{Pythia8317manual}. As a technical remark, our {\tt VINCIA} simulations use a modified version of the source file {\tt VinciaQED.cc} of {\tt Pythia~8.317} (the same behaviour is present in version~8.318). In the default implementation of the function {\tt QEDemitElemental::generateTrial}, a trial QED emission is only cached, i.e.~the flag {\tt hasTrial} is set, if its scale {\tt q2TrialNow} lies above the lower cutoff of the evolution window. Otherwise a new trial is generated each time the function is called. We instead set {\tt hasTrial} independently of the value of {\tt q2TrialNow}, such that a QED emitter that has failed to produce a trial above the cutoff is not re-sampled.

\subsection{Shower options}
\label{sec:showeroptions}

When interfacing with {\tt Pythia~8.3}, different PS algorithms can be selected. These algorithms employ different treatments of QCD and QED radiation, which may lead to differences in QED-sensitive observables such as the inelasticity $y$. Comparing the available~PS~algorithms therefore provides a useful assessment of their impact on the kinematic distributions relevant for neutrino-induced DIS.

\paragraph{Simple shower:} The default {\tt Pythia~8.3} shower is the transverse-momentum-ordered ``simple shower'' algorithm~\cite{Sjostrand:2004ef}, which combines a global recoil scheme for ISR with a dipole recoil scheme for FSR. Although this prescription provides a satisfactory description for a wide range of collider observables, it has been found to be inadequate for DIS-like processes~\cite{Jager:2014vna,Rauch:2016upa,Jager:2020hkz,Covarelli:2021gyz,Hoche:2021mkv}. This limitation originates primarily from the global recoil strategy employed for ISR, which modifies the kinematics of the DIS system and can consequently distort observables that are sensitive to the underlying Born-level configuration.

\paragraph{Dipole shower:} To overcome these limitations, Ref.~\cite{Cabouat:2017rzi} introduced an extension of the ``simple shower'' employing a local dipole recoil scheme. In this approach, QCD emissions are treated with a combined dipole evolution, such that recoil effects are absorbed locally and the DIS kinematics are better preserved throughout the shower evolution. The~corresponding option can be enabled by setting
\begin{equation*}
{\tt SpaceShower:dipoleRecoil = on}\,.
\end{equation*}
A limitation of the dipole-recoil implementation is that photon emission from the incoming lepton leg is not yet supported. For lepton-induced DIS, initial-state QED radiation should therefore be switched off. This issue does not arise for neutrino-induced DIS, since neutrinos do not radiate photons.

\paragraph{Antenna shower:} In addition to the standard shower algorithms, {\tt Pythia~8.3} also provides the {\tt VINCIA} shower~\cite{Brooks:2020upa}, which is based on the antenna formalism~\cite{Gehrmann-DeRidder:2011gkt}. Unlike conventional dipole showers, the antenna approach treats radiation as arising coherently from each colour-connected parton pair as a whole, providing an improved description of soft and collinear emissions and a more faithful implementation of QCD coherence effects.

\subsection{Benchmarking of PS algorithms}
\label{sec:PSbench}

To benchmark the various PS algorithms introduced above, we consider the inelasticity distribution of bare muons produced in the CC process $\nu_\mu + N \to \mu^- + X$, where $N$ denotes an isoscalar target, at fixed neutrino energies of $E_{\nu_\mu}=10^{3}\,{\rm GeV}$ and $E_{\nu_\mu}=10^{5}\,{\rm GeV}$. Our primary interest is in the logarithmic corrections associated with photonic FSR. We~therefore compare the PS predictions with fixed-order QCD results supplemented by the corresponding LL~QED correction, denoted by ``LO$+$LL'' and ``NLO$+$LL'', respectively, and defined in~Eq.~\eqref{eq:sigma_LL}. A~comparison with the full NLO~EW calculation, which includes several additional effects, is presented below. Although the PS algorithms resum the LL~corrections to all orders, their dominant qualitative effects are already captured by the first LL~correction. The LO$+$LL and NLO$+$LL predictions therefore provide useful benchmarks for assessing the performance of the different shower algorithms. For~the LO$+$LL prediction, we choose $\mu_h^2=\hat{s}_{12}$, corresponding to the total partonic COM energy. Unless stated otherwise, all results below are obtained using a multiplicative matching of the QCD and QED corrections, i.e.~the LL correction of~Eq.~(\ref{eq:sigma_LL}) is applied to the full partonic QCD cross section, as in~Eq.~(\ref{eq:sigma_frag}).

\begin{figure}[t!]
\begin{center}
\includegraphics[width=0.49\textwidth]{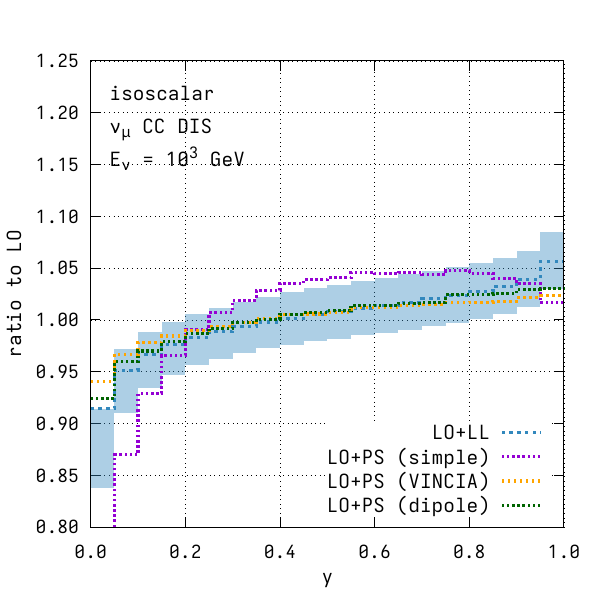}
\includegraphics[width=0.49\textwidth]{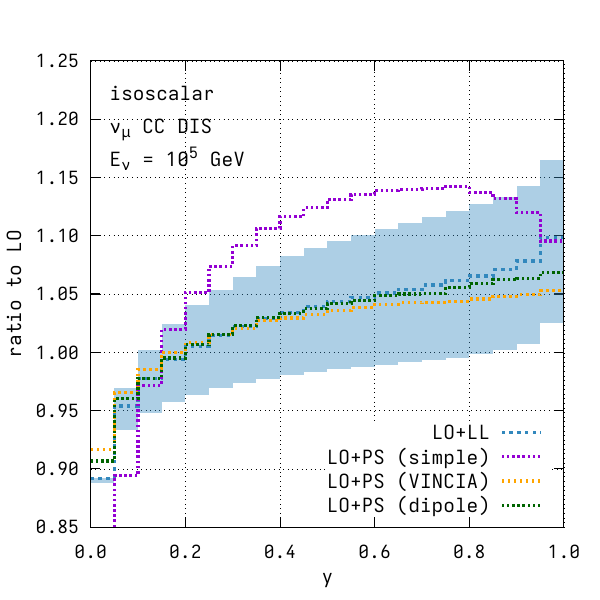}
\vspace{0mm}
\caption{Inelasticity distributions for $\nu_\mu$-induced CC DIS at $E_{\nu_\mu}=10^{3}\,{\rm GeV}$ (left) and $E_{\nu_\mu}=10^{5}\,{\rm GeV}$ (right). The LO QCD fixed-order prediction supplemented by the LL~corrections from photonic FSR (blue) is compared with the corresponding LO QCD$+$PS predictions obtained using different shower algorithms. All distributions are normalised to the LO QCD fixed-order prediction. The blue band indicates the associated scale uncertainty of the LO$+$LL result. Further details are given in the main text.
\label{fig:showerLO}}
\end{center}
\end{figure}

We first consider a hard process described at LO QCD. The corresponding results are shown in~Figure~\ref{fig:showerLO} for $E_{\nu_\mu}=10^{3}\,{\rm GeV}$ (left panel) and $E_{\nu_\mu}=10^{5}\,{\rm GeV}$ (right panel). At~both beam energies, the dipole and {\tt VINCIA} algorithms closely reproduce the LO$+$LL prediction, both in shape and normalisation. By contrast, the ``simple shower'' exhibits visible deviations already at $E_{\nu_\mu}=10^{3}\,{\rm GeV}$, which become increasingly pronounced at higher energies. In particular, it distorts the shape of the $y$~distribution, most notably by depleting the low-$y$ region. We also note that the relative impact of the QED corrections is approximately twice as large at $E_{\nu_\mu}=10^{5}\,{\rm GeV}$. This behaviour is expected. Although the reference scale $\hat{s}_{12}$ entering the logarithm is typically larger in this regime, this increases the logarithm by only about $25\%$. The main reason is that the QCD prediction for the~$y$ distribution, which is nearly flat at $E_{\nu_\mu}=10^{3}\,{\rm GeV}$, becomes strongly concentrated at small $y$ at $E_{\nu_\mu}=10^{5}\,{\rm GeV}$ (see~Figure~\ref{fig:reshuff}), which enhances the relative impact of the QED corrections.

\begin{figure}[t!]
\begin{center}
\includegraphics[width=0.49\textwidth]{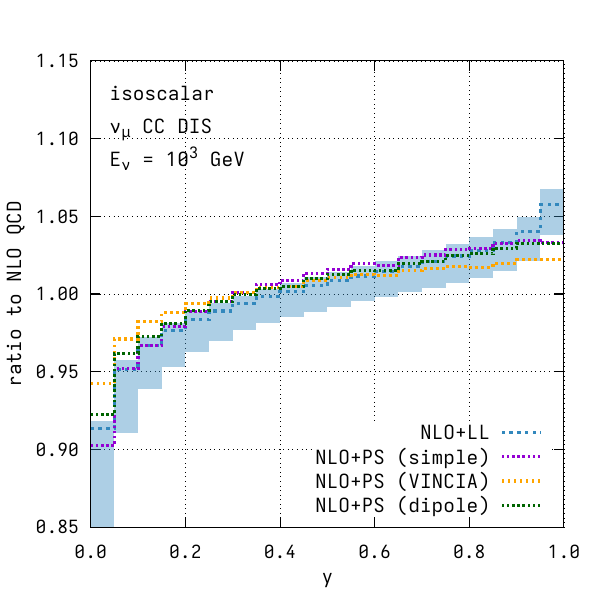}
\includegraphics[width=0.49\textwidth]{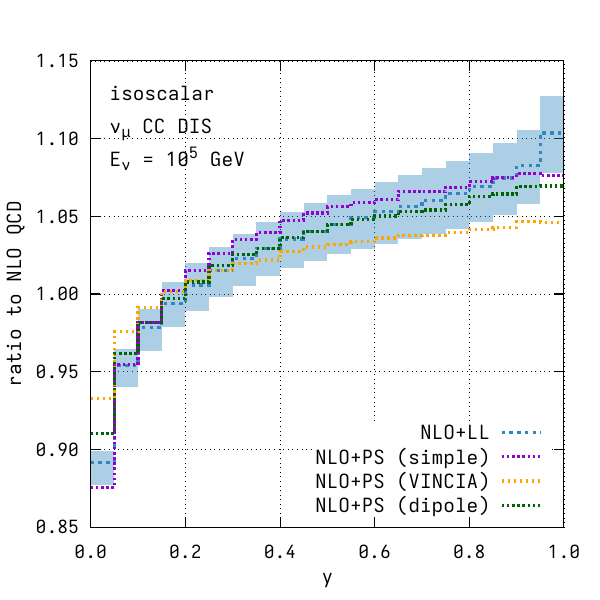}
\vspace{0mm}
\caption{Same as~Figure~\ref{fig:showerLO}, but with NLO QCD used as the fixed-order baseline for all predictions. See main text for additional details. 
\label{fig:showerNLO}}
\end{center}
\end{figure}

We next repeat the comparison using the NLO QCD prediction as the hard-process baseline. Figure~\ref{fig:showerNLO} shows the corresponding NLO$+$LL prediction together with the results obtained after matching to the different PS algorithms. At NLO, the matching procedure introduces an additional scale that separates the fixed-order calculation from the PS. For~the~LO~setup, this scale is simply identified with the kinematic limit $\sqrt{\hat{s}_{12}}$, which also serves as the shower starting scale. The dipole prediction again shows good agreement with the NLO$+$LL baseline at both energies, with deviations at the percent level for $0.05 \lesssim y \lesssim 0.9$, increasing to about $2\%\hspace{0.25mm}$--$\hspace{0.25mm}3\%$ in the bin closest to $y=1$. Once matched to an NLO hard process, the ``simple shower'' reaches a similar level of agreement, in contrast to the LO comparison in~Figure~\ref{fig:showerLO}. The {\tt VINCIA} prediction instead shows a tilt relative to the NLO$+$LL baseline that grows with energy, reaching about $\pm5\%$ at the edges of the~$y$~range at $E_{\nu_\mu}=10^{5}\,{\rm GeV}$.

Taken together, these comparisons demonstrate that the dipole shower provides a reliable description of the logarithmically enhanced QED radiation in neutrino-induced DIS processes. It reproduces the LL~benchmark over the energy range $10^{3}\,{\rm GeV}\hspace{0.25mm}$--$\hspace{0.25mm}10^{5}\,{\rm GeV}$ tested~here and remains stable when combined with an NLO QCD hard process. We~therefore adopt the dipole shower as our default choice for the remainder of this work and recommend its use in phenomenological studies and experimental analyses of neutrino-induced~DIS~processes.

\subsection{Non-universal and subleading EW effects}
\label{sec:eweffects}

Before presenting a comprehensive comparison of the fixed-order and NLO+PS predictions obtained with the dipole shower, it is useful to first quantify the impact of the non-universal corrections to the inelasticity distribution that are included in the full NLO~EW calculation but are not captured by the NLO+PS predictions.

\begin{figure}[t!]
\begin{center}
\includegraphics[width=0.49\textwidth]{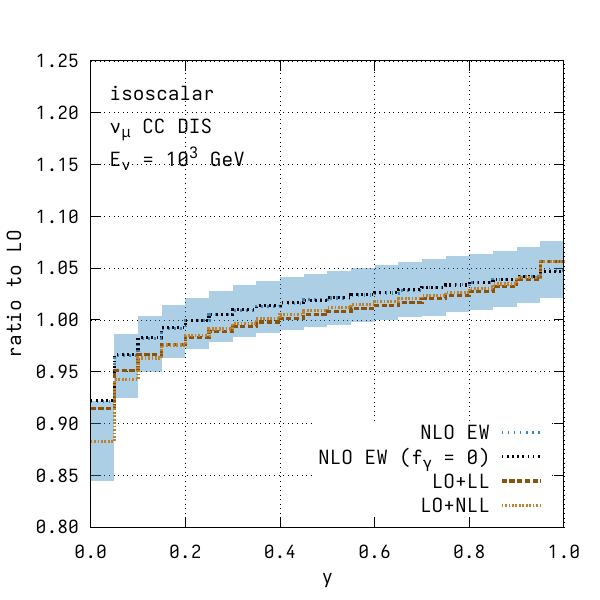}
\includegraphics[width=0.49\textwidth]{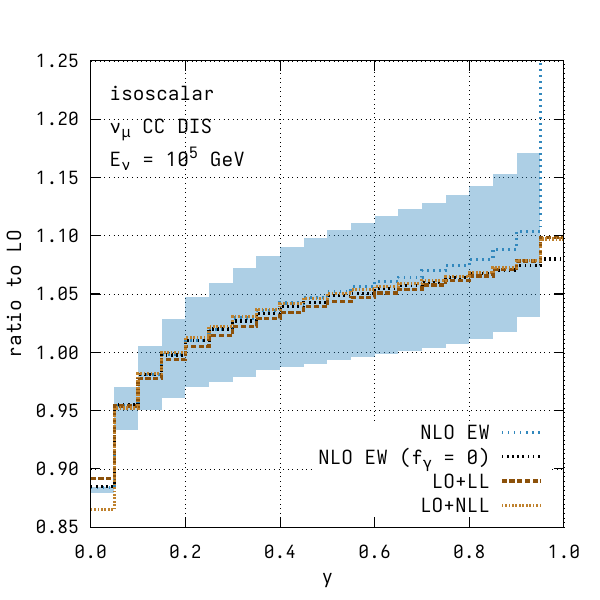}
\vspace{0mm}
\caption{Same as~Figure~\ref{fig:showerLO}, but with the full NLO~EW fixed-order prediction, including contributions from a non-zero photon~PDF, $f_\gamma (x,\mu_F^2)\neq 0$, as the baseline. This is compared with the corresponding NLO~EW prediction obtained by setting $f_\gamma (x,\mu_F^2)=0$, as well as with the LO prediction supplemented by the LL~and NLL photonic corrections, denoted by ``NLO~EW ($f_\gamma=0$)'', ``LO$+$LL'', and ``LO$+$NLL'', respectively. All results are normalised to the LO prediction. Further details are provided in the main text.
\label{fig:inelasticityEW}}
\end{center}
\end{figure}

The relative importance of these effects can be assessed from~Figure~\ref{fig:inelasticityEW}, which compares the full NLO~EW prediction, both including and excluding photon-induced contributions, with the corresponding LO$+$LL and LO$+$NLL predictions based on the~FF~approach. A~comparison of the two NLO~EW predictions allows us to isolate the impact of photon-induced contributions. These corrections are negligible at $E_{\nu_\mu}=10^{3}\,{\rm GeV}$, but become increasingly relevant at $E_{\nu_\mu}=10^{5}\,{\rm GeV}$, where they are strongly enhanced towards $y\to1$, corresponding to events with relatively low muon energies. In the last bin, the photon-induced contribution enhances the prediction by about $50\%$ relative to the LO result and is therefore not shown in the figure. This behaviour is consistent with the results obtained for the more inclusive observable $\langle y \rangle$, shown in~Figure~\ref{fig:meany}. Imposing a realistic energy requirement on the final-state muon, such as $E_{\mu} \gtrsim 1\,{\rm TeV}$, as considered in~Figure~\ref{fig:inelasticityEcut}, substantially reduces the photon-induced contribution in the final bin. We return to this point below. Turning to the~FF~predictions, comparing the LO$+$LL and LO$+$NLL results with the NLO~EW prediction for $f_\gamma(x,\mu_F^2)=0$ allows us to assess the size of the non-universal corrections, i.e. contributions not captured by the~FF~approach. These effects are typically found to be no larger than about $2\%$ in magnitude. The impact of subleading, yet still universal, corrections within the FF framework can be assessed by comparing the LO$+$LL and LO$+$NLL predictions, where NLL refers to the inclusion of the non-logarithmic $\mathcal{O}(\alpha)$ initial condition of the lepton~FF. Their difference originates from the initial condition of the~FF~\cite{Mele:1990cw} and is most pronounced in the limit $y\to0$. Numerically, the inclusion of these~NLL universal corrections can reduce the inelasticity distribution by up to $3\%$. Finally,~including the soft-exponentiated corrections, relative to the LO$+$LL prediction, leads to permille-level effects across the inelasticity distribution (not shown), with effects reaching the percent level in the $y\to 0$ region.

\subsection{Final comparison}
\label{sec:finalcomparison}

Having established the generator setup and PS configuration that provide the most accurate description of the dominant corrections to CC DIS within the {\tt POWHEG-BOX} implementation~\cite{FerrarioRavasio:2024kem}, we now compare these predictions with the fixed-order calculation and assess the impact of coherent scattering contributions.

To this end, we first define our reference fixed-order prediction for the process $\nu_{\ell} + A \to \ell^- + X$. It combines the exact NLO QCD$+$EW corrections, photon-induced contributions, finite top-quark mass effects, universal fragmentation corrections to the lepton energy distribution, and coherent scattering contributions. Specifically,
\beq \label{eq:best}
d \hat \sigma_{\rm best} = d \hat \sigma_{\rm LO} + d \hat \sigma_{\text{NLO~QCD$+$EW}} + d \hat \sigma_{\rm frag,QED}^{\rm add} + d \hat \sigma_{\rm coherent} \,,
\eeq
where the first two terms describe the DIS process including the exact NLO QCD$+$EW corrections to the partonic differential cross section. 
The third term $d \hat{\sigma}_{\rm frag,QED}^{\rm add}$ accounts for the additional universal lepton fragmentation effects that are not already included in the NLO~EW calculation.
The last term, $d \hat{\sigma}_{\rm coherent}$, denotes the full coherent scattering contribution --- see Section~\ref{sec:coherent} for further details.

The precise implementation of the fragmentation correction is not unique, in particular as to whether it is combined additively or multiplicatively with the fixed-order prediction. Here, we adopt the multiplicative prescription. Avoiding double counting with the NLO~EW correction, the additional fragmentation contribution can then be written, using the notation introduced in~Eq.~(\ref{eq:sigma_frag}), as
\beq \label{eq:sigma_frag_add}
\begin{split}
d \hat \sigma_{\rm frag,QED}^{\rm add} & = d \hat \sigma_{\rm LO} \int_0^1 \! dz \, \left[ D_\ell\left(z,\mu_h^2\right) - D^{(0)}_\ell(z) - \frac{\alpha}{4 \pi} \hspace{0.25mm} D^{(1)}_\ell\left(z,\mu_h^2\right) \right] \Theta_{\rm selection}\left(z p_\ell\right) \\[2mm]
& \phantom{xx} + \frac{\alpha_s}{4 \pi} \hspace{0.5mm} d \hat \sigma_{\rm NLO,QCD} \int_0^1 \! dz \, \left[ D_\ell\left(z,\mu_h^2\right) - D^{(0)}_\ell(z) \right] \Theta_{\rm selection}\left(z p_\ell\right) \,,
\end{split}
\eeq
where $D^{(n)}_{\ell}(z,\mu_h^2)$ denotes the coefficient of $(\alpha/(4 \pi))^n$ in the perturbative expansion of the lepton fragmentation function, with $D^{(0)}_{\ell}(z) = \delta(1-z)$. For observables subject to a cut on the charged-lepton energy (or, equivalently, an upper restriction on the inelasticity $y$), we use the fixed-order hard-collinear fragmentation function $D_{\ell}$~\cite{Arbuzov:2002cn}. In the absence of such cuts, we instead employ the soft-exponentiated fragmentation function $D_{\ell}^{\rm soft}$.

\begin{figure}[t!]
\begin{center}
\includegraphics[width=0.49\textwidth]{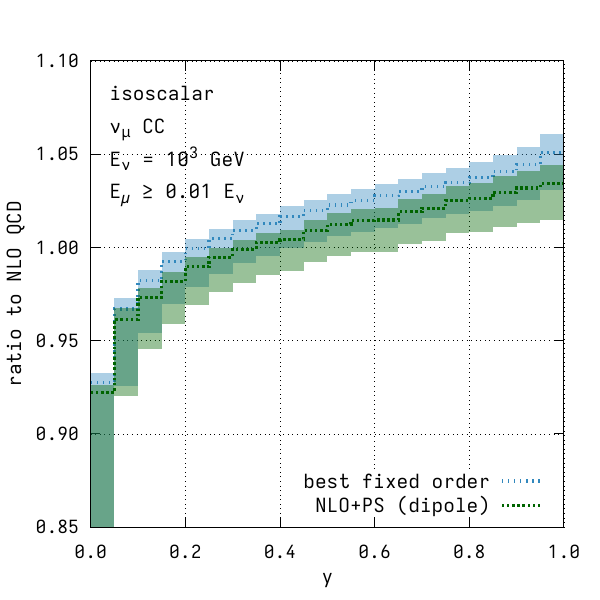}
\includegraphics[width=0.49\textwidth]{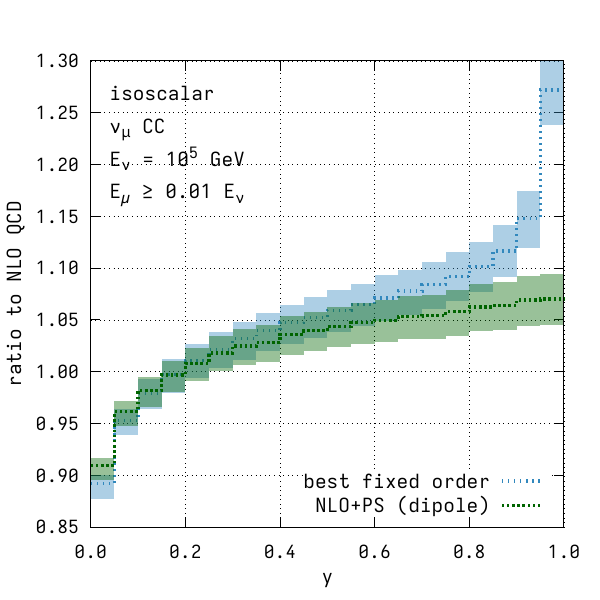}
\vspace{0mm}
\caption{Same as Figure~\ref{fig:showerLO}, but with the additional requirement $E_{\mu} \geq 0.01\hspace{0.25mm}E_{\nu_\mu}$, and now comparing the best fixed-order prediction shown in blue, as defined in~Eq.~(\ref{eq:best}), with the NLO QCD prediction matched to the dipole shower in green. All predictions are normalised to the corresponding NLO QCD fixed-order result, while the bands indicate the scale uncertainties of the respective predictions. Further details are given in the main text. \label{fig:pwhg+py8-QCD$+$EW-qcd+ff}
}
\end{center}
\end{figure}

The final comparison of our best fixed-order prediction with the NLO QCD prediction matched to the {\tt Pythia~8.3} dipole shower is shown in~Figure~\ref{fig:pwhg+py8-QCD$+$EW-qcd+ff}. The figure displays the inelasticity distributions for $\nu_\mu$-induced scattering on an isoscalar target for the neutrino energies $E_{\nu_\mu}=10^{3}\,{\rm GeV}$~(left panel) and $10^{5}\,{\rm GeV}$~(right panel). All predictions are normalised to the corresponding NLO QCD fixed-order result, and the bands indicate their respective scale uncertainties. We additionally impose a minimal energy requirement, $E_{\mu} \geq 0.01\hspace{0.25mm}E_{\nu_\mu}$, in both results shown in~Figure~\ref{fig:pwhg+py8-QCD$+$EW-qcd+ff}. This requirement affects the final inelasticity bin by restricting the kinematic range to $y<0.99$. Similarly to the fixed cut $E_\mu > 1\,{\rm TeV}$ discussed in~Section~\ref{sec:exclusive}, this cut is motivated by experimental considerations, and coincides with it at $E_{\nu_\mu} = 10^{5}\,{\rm GeV}$.

At $E_{\nu_\mu}=10^{3}\,{\rm GeV}$, the EW corrections induce a non-trivial modification of the $y$ distribution, ranging from about $-10\%$ at low $y$ to approximately $+5\%$ at high $y$. The~NLO+PS dipole prediction agrees closely with the best prediction defined in~Eq.~\eqref{eq:best}, with central values differing by at most $2\%$ and remaining compatible within the respective scale uncertainties. This agreement is expected since the kinematic thresholds for on-shell $W$-boson and top-quark production are not reached at this energy, rendering photon-induced processes and coherent scattering negligible. At $E_{\nu_\mu}=10^{5}\,{\rm GeV}$, visible differences in the shape of the $y$ distribution emerge, particularly towards $y\to1$. These deviations can exceed $10\%$ and are primarily due to resonantly enhanced $W$-boson production subprocesses in both DIS and coherent scattering, which enhance the best prediction. Consequently, for $y\gtrsim0.8$, the two predictions differ beyond their respective scale-uncertainty bands. 

\begin{figure}[t!]
\begin{center}
\includegraphics[width=0.6\textwidth]{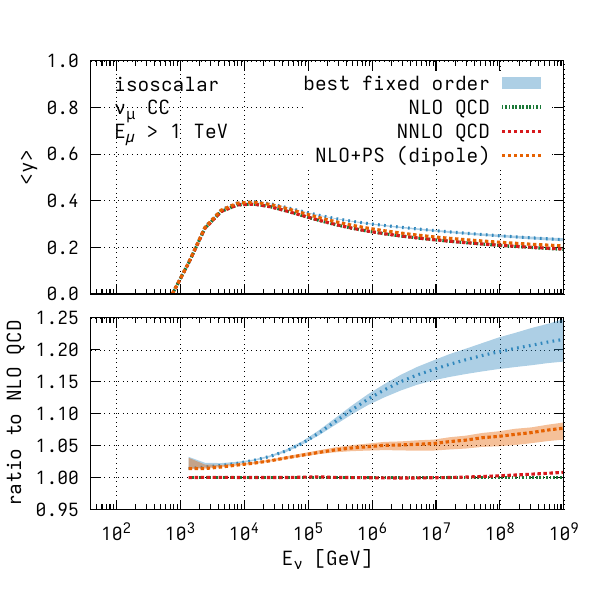}
\vspace{0mm}
\caption{Final comparison of the predictions for the mean inelasticity $\langle y \rangle$ of $\nu_\mu$-induced CC~DIS on an isoscalar target, imposing $E_\mu > 1 \, {\rm TeV}$, as a function of the incoming $\nu_\mu$ energy. The best fixed-order prediction of~Eq.~(\ref{eq:best})~(blue with scale uncertainties) is compared with the NLO~QCD$+$PS prediction obtained with the {\tt POWHEG-BOX} DIS event generator matched to the {\tt Pythia~8.3} dipole~(orange dashed) shower, and with the NLO~(green dotted) and NNLO~(red dashed) QCD predictions. The upper panel shows $\langle y \rangle$ and the lower panel the ratio to the NLO~QCD result, including the scale uncertainties of the NLO$+$PS prediction. The NLO$+$PS prediction does not include the photon-PDF, coherent, and top-quark contributions, each of which ranges from $1\%$ to $4\%$ depending on the incident neutrino energy~$E_{\nu_\mu}$. At the highest energies these contributions account for only about half of the difference to the best prediction. See the main text for~details.}
\label{fig:meanyBest}
\end{center}
\end{figure}

The energy dependence of these differences is shown in~Figure~\ref{fig:meanyBest}, which compares the~best prediction for the mean inelasticity, obtained by imposing $E_\mu > 1\,{\rm TeV}$, with the NLO$+$PS prediction based on the dipole shower as a function of $E_{\nu_\mu}$. For reference, the NLO and NNLO QCD predictions are also shown. The hidden Glashow contributions, from either DIS or coherent scattering, give rise to a large part of the deviations from the NLO$+$PS prediction. In addition, the shift of $\langle y \rangle$ induced by QED radiation off~the muon is smaller in the NLO$+$PS prediction, amounting to about $+8\%$ at $10^9\,{\rm GeV}$, compared to about $+13\%$ in the fixed-order calculation without photon-induced contributions (see~Figure~\ref{fig:inelasticityEcut}). Top-quark production has a smaller impact in the $10^6\,{\rm GeV}$ energy range, reflecting the different kinematics of this process.

Overall, our study demonstrates that the combination of the {\tt POWHEG-BOX} DIS NLO~QCD calculation with the {\tt Pythia~8.3} dipole shower captures the dominant features of the full fixed-order NLO QCD$+$EW prediction up to neutrino energies of approximately $10^4\,{\rm GeV}$. At higher energies, around $10^5\,{\rm GeV}$, the treatment of hard photonic corrections associated with the outgoing muon remains in reasonable agreement with the NLO$+$LL benchmark, with~differences typically at the few-percent level, as illustrated in~Figure~\ref{fig:showerNLO}, while at even higher energies the NLO$+$PS predictions underestimate this effect (see~Figure~\ref{fig:meanyBest}). However, $W$-boson production in DIS and coherent scattering becomes phenomenologically relevant at these energies. Its omission affects both the normalisation and the shape of the inelasticity distribution, as illustrated in~Figure~\ref{fig:pwhg+py8-QCD$+$EW-qcd+ff}.

We note that the approach adopted here represents the current state of the art and relies on publicly available and validated methods. At the same time, this is an active area of research, and ongoing developments may motivate alternative approaches in the future. In particular, the continued development of the {\tt Pythia} and {\tt VINCIA} PS algorithms towards an improved description of DIS processes~\cite{Helenius:2024wjg,Helenius:2026odp}, as well as recent efforts to develop PS frameworks with NLL accuracy~\cite{vanBeekveld:2022zhl,vanBeekveld:2022ukn,Preuss:2024vyu,Hoche:2024dee}, are noteworthy in this context. Such~developments may further improve the theoretical description of DIS observables and could ultimately lead to alternative and more accurate predictions.

\section{Public fixed-order grids}

\label{sec:realistic}

In the previous section, we presented theoretical predictions for the energy distributions of muons produced in interactions of high-energy neutrinos with matter at fixed neutrino beam energies. These predictions were obtained within the fixed-order framework defined in~Eq.~\eqref{eq:best} and benchmarked against results from a flexible event generator for exclusive neutrino-DIS simulations. Relative to the event-generator approach, the fixed-order framework has the advantage of treating EW and quark-mass effects exactly at NLO accuracy, including, for instance, the effects relevant to top-quark production, and of incorporating additional contributions from coherent scattering processes. The neutrino-DIS event generator, on the other hand, provides a realistic description of hadron-level final states, making it possible to interface the simulation with experimental analyses that account for detector-response~effects.

The aim of this section is to make both sets of predictions readily accessible for future experimental and theoretical studies of neutrino-matter scattering. We pursue two complementary approaches. First, we provide the fixed-order predictions in the form of a numerical grid for the inelasticity distribution of the outgoing muon at a set of incoming neutrino energies. Second, we compute the same grid with the event generator and incorporate an energy-dependent neutrino flux into the exclusive event simulation and study its impact using the neutrino-DIS event generator. We discuss these two approaches in this section and in Section~\ref{sec:astroflux}, respectively.

\subsection{Double-differential fixed-order cross-section predictions}

The fixed-order predictions defined in~Eq.~\eqref{eq:best}, with the coherent contribution provided separately, are obtained with a private MC~framework in which the individual contributions to the cross section are calculated separately and combined at the histogram level. For instance, the NLO EW corrections require separate evaluations of the real and virtual contributions and their corresponding subtraction terms according to~Eq.~\eqref{eq:sigma_NLO}, for both massless final-state quarks and top-quark production. Analogous calculations are performed for the QCD and QED FSR corrections. These~contributions, together with coherent scattering, must then be combined, and the procedure repeated for the different scale choices. This involves a non-negligible computational and bookkeeping effort. We therefore provide the final results as pre-computed numerical tables rather than the underlying private code, allowing them to be readily used in future phenomenological and experimental studies. 

The table is constructed by computing the binned differential $\nu_\mu + A \to \mu^- + X$ cross section with respect to the inelasticity
\beq \label{eq:d2sig}
\frac{d\sigma_{\nu_{\mu}A}(E_{\nu_{\mu}})}{dy}\,,
\eeq
at discrete values of $E_{\nu_{\mu}}$. In practice, we calculate the binned inelasticity distribution over the range $y\in[0,1]$, divided linearly into 40 bins. This calculation is repeated for~41~discrete values of $E_{\nu_{\mu}}$ in the range $E_{\nu_{\mu}}\in[10^3,10^7]\,{\rm GeV}$ with logarithmic spacing in energy. With~this setup, the uniform inelasticity bins can be directly converted into energy bins of the outgoing muon. The complete results, including an estimate of the associated uncertainties and a breakdown of the individual contributions, such as the coherent contributions from the different neutrino flavours, are available upon request.

\begin{figure}[t!]
\begin{center}
\includegraphics[width=0.49\textwidth]{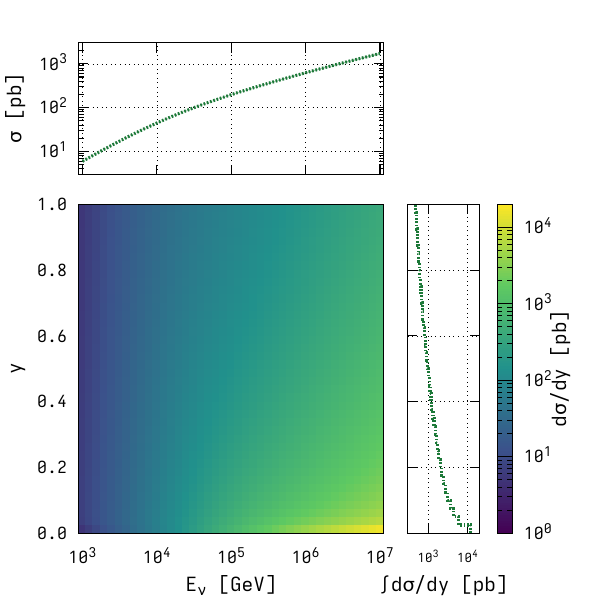}
\includegraphics[width=0.49\textwidth]{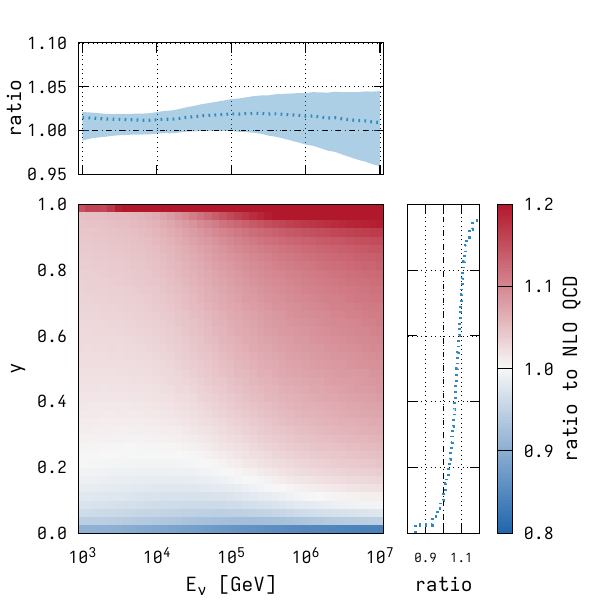}
\vspace{0mm}
\caption{Inelasticity distribution $d\sigma/dy$ as a function of the incoming muon-neutrino energy and the muon inelasticity for an isoscalar target. The left panel shows the central NLO~QCD prediction, while the right panel shows the NLO~QCD$+$EW prediction including the additional FSR corrections of the lepton~FF normalised to the corresponding NLO~QCD result. Projections are shown along the corresponding axes: onto $\log_{10} E_{\nu_\mu}$~(top), i.e.~the cross section integrated over $y$, together with the theoretical uncertainties from scale variation, and onto $y$ (right), i.e.~$d\sigma/dy$ integrated over $\log_{10} E_{\nu_\mu}$ in the range $10^3 \, {\rm GeV} \leq E_{\nu_\mu} \leq 10^7 \, {\rm GeV}$, for which only the central prediction is shown. Each bin is coloured according to the value of the central prediction in that bin. See main text for further explanations.}
\label{fig:dsig2_RG}
\end{center}
\end{figure}

To visualise part of this result, the left panel of~Figure~\ref{fig:dsig2_RG} shows the NLO QCD component of the resulting cross section for the central scale choice in the $E_{\nu_\mu}$--$\hspace{0.5mm}y$ plane. The~absolute value of the cross section at each point of this two-dimensional grid is indicated by the colour of the corresponding cell, while the projections onto the $\log_{10} E_{\nu_\mu}$ and $y$ axes, corresponding to the respective integrated cross sections, are shown alongside. Note that the integration over $\log_{10} E_{\nu_\mu}$ corresponds to a neutrino flux $dN/dE_{\nu_\mu} \propto 1/E_{\nu_\mu}$. The~right~panel displays the NLO QCD$+$EW cross section normalised to the corresponding NLO QCD result, thereby illustrating the relative impact of the full NLO EW corrections. The results exhibit the same features observed previously: QED FSR corrections are positive (negative) at large (small) values of the inelasticity, with their relative impact increasing with~$E_{\nu_\mu}$, while the resonant $W$-boson production subprocess gives rise to a positive correction for $E_{\nu_\mu}\gtrsim 10^4\,{\rm GeV}$, which becomes particularly pronounced at large inelasticities. Overall,~this calculation provides a state-of-the-art prediction for this class of double-differential distributions and is readily applicable to experimental analyses and phenomenological simulations.

\section{Double-differential event-generator cross-section predictions}
\label{sec:astroflux}

We now repeat the grid computation of~Section~\ref{sec:realistic} with the {\tt POWHEG-BOX} DIS generator, whose limitations relative to our fixed-order framework were discussed in Section~\ref{sec:finalcomparison}. This~serves both to benchmark the two approaches and to quantify the impact of these missing contributions across the full $E_{\nu_\mu}$--$\hspace{0.5mm} y$ plane.

\begin{figure}[t!]
\begin{center}
\includegraphics[width=0.49\textwidth]{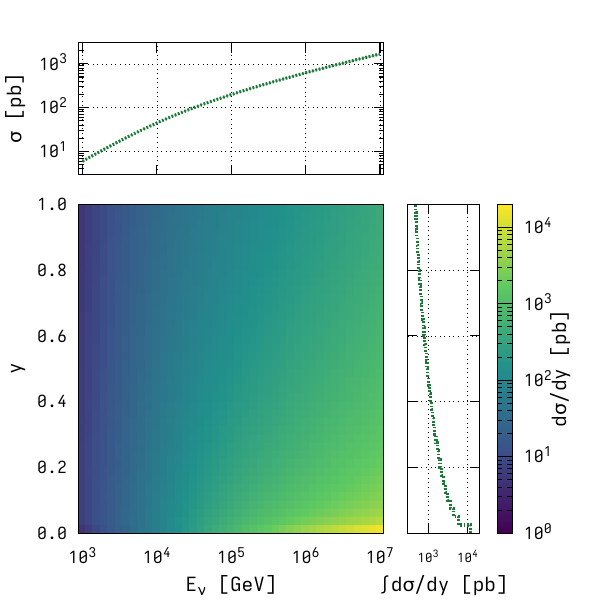}
\includegraphics[width=0.49\textwidth]{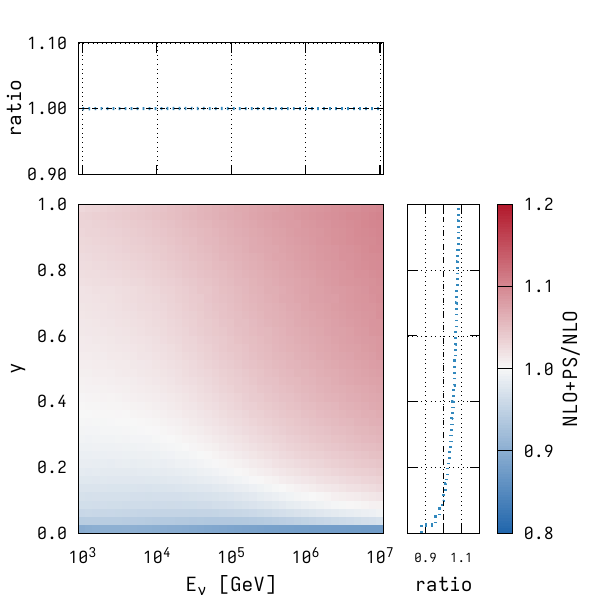}
\vspace{0mm}
\caption{
{As in~Figure~\ref{fig:dsig2_RG}, but obtained with the {\tt POWHEG-BOX} DIS generator. The left panel shows the NLO~QCD prediction, while the right panel shows the prediction matched to the {\tt Pythia~8.3} dipole shower, normalised to the NLO~QCD result.}}
\label{fig:dsig2_PWG}
\end{center}
\end{figure}

The {\tt POWHEG-BOX} NLO~QCD grid is shown in the left panel of~Figure~\ref{fig:dsig2_PWG} and is qualitatively similar to the corresponding fixed-order result in the left panel of~Figure~\ref{fig:dsig2_RG}. The~right panel displays the effect of matching to the {\tt Pythia~8.3} dipole shower, normalised to the NLO~QCD {\tt POWHEG-BOX} prediction. The behaviour follows that observed in~Figure~\ref{fig:pwhg+py8-QCD$+$EW-qcd+ff} at the fixed neutrino energies $E_{\nu_\mu} = 10^3 \, {\rm GeV}$ and $10^5 \, {\rm GeV}$: QED radiation off the muon leg, generated by the shower, depletes the region of small inelasticity, by about $10\%\hspace{0.25mm}$--$\hspace{0.25mm}13\%$ in the lowest $y$~bin, and populates larger values of~$y$, with an effect that grows with the neutrino~energy.

\begin{figure}[t!]
\begin{center}
\includegraphics[width=0.49\textwidth]{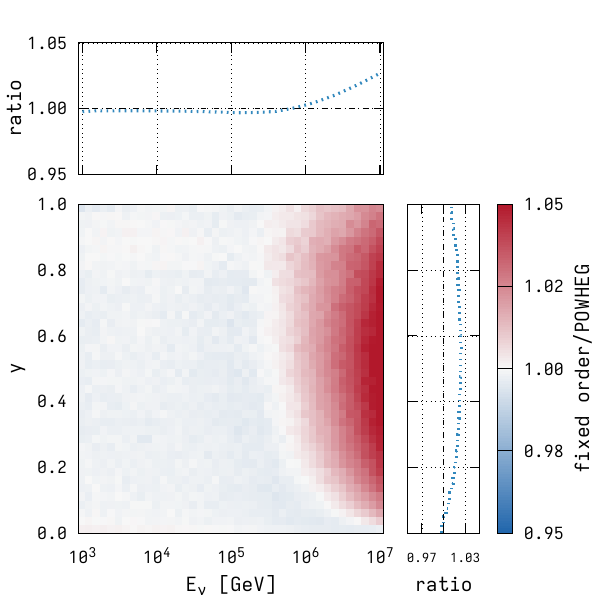}
\includegraphics[width=0.49\textwidth]{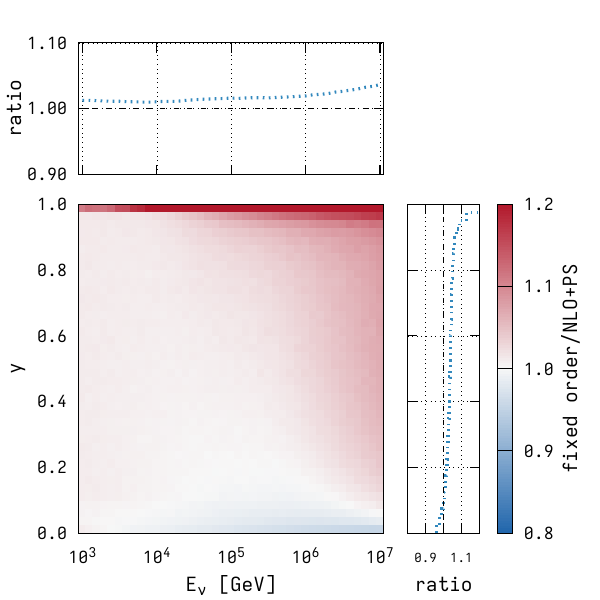}
\vspace{0mm}
\caption{Comparison of the fixed-order and {\tt POWHEG-BOX} grids, shown as the ratio of the former to the latter. Left: NLO~QCD predictions from the two approaches, with the missing top-quark contribution in {\tt POWHEG-BOX} becoming visible at high energies. Right: Fixed-order NLO~QCD$+$EW prediction compared with the {\tt POWHEG}$+${\tt Pythia~8.3} result obtained with the dipole shower, including the additional FSR corrections from the lepton~FF.}
\label{fig:dsig2_PWGvsFO}
\end{center}
\end{figure}

The two approaches are compared directly in~Figure~\ref{fig:dsig2_PWGvsFO}, where the plots show the ratio of the fixed-order prediction to the corresponding {\tt POWHEG-BOX} result. At NLO~QCD~(left panel), the fixed-order and {\tt POWHEG-BOX} grids agree at the percent level up to approximately $E_{\nu_\mu} = 10^6 \, {\rm GeV}$. At~higher energies, the ratio increases, with the fixed-order prediction exceeding the {\tt POWHEG-BOX} result by up to about $5\%$ at intermediate values of~$y$, owing to the absence of top-quark production in {\tt POWHEG-BOX}. For details see~Appendix~\ref{app:Top}. The right panel compares the {\tt POWHEG-BOX}$+${\tt Pythia~8.3} prediction with the fixed-order NLO~QCD$+$EW result, which additionally includes the fragmentation contribution~$d \hat \sigma_{\rm frag,QED}^{\rm add}$ defined in~Eq.~(\ref{eq:sigma_frag_add}). Over most of the distribution, the ratio remains within a few percent of unity, indicating that the dipole shower captures the dominant QED FSR effects. At small~$y$, the ratio falls below unity, due to the NLL and non-universal corrections discussed in~Section~\ref{sec:eweffects}. At large~$y$ and high energies, by contrast, the fixed-order prediction receives additional photon-induced contributions, most notably from the resonant hidden Glashow contribution. These contributions are absent in the {\tt POWHEG-BOX} simulation and lead to ratios exceeding $1.2$ in the bin closest to $y=1$ for $E_{\nu_\mu} \gtrsim 10^4 \, {\rm GeV}$. Coherent scattering is included in neither grid, so that the comparison is restricted to the~DIS~component. In principle, the coherent contribution could be incorporated into both predictions in the~same~way.

\subsection{Implementation and implications of an astrophysical neutrino flux}

Realistic astrophysical neutrino analyses involve energy-dependent fluxes spanning several orders of~magnitude. In the \texttt{POWHEG-BOX}, such a flux can be incorporated through a fictitious neutrino PDF, which effectively samples the incoming neutrino energy according to the chosen flux parametrisation. This approach enables an event generator, particularly when interfaced with a detector simulation, to provide a realistic description of neutrino event rates in an experimental setting. As a proof of concept, we implement an astrophysical neutrino flux and present the resulting distributions differential in both the incident neutrino energy and the inelasticity, following the double-differential cross-section presentation discussed above. In principle, this setup also allows one to assess the impact of QED FSR corrections in a realistic flux~scenario. The current limitations are discussed below.

A detailed phenomenological study of neutrino-induced DIS at the forward-physics facilities of the LHC was presented in~Ref.~\cite{vanBeekveld:2024ziz}, using the \texttt{POWHEG-BOX} DIS generator supplemented with neutrino PDFs. Here, we adopt the same formalism for our analysis. Within~this framework, the number~$N_{\rm int}^{\nu_\mu}$ of muon neutrinos undergoing CC DIS interactions in the detector is given by
\beq
N_{\rm int}^{\nu_\mu}=n_T\hspace{0.5mm}L_T\,\int_{Q^2_{\rm min}}^{Q^2_{\rm max}}\int_{x_B^{\rm min}}^{x_B^{\rm max}}\int_{E^{\rm min}_{\nu_\mu}}^{E^{\rm max}_{\nu_\mu}} dQ^2\hspace{0.5mm}dx_B\hspace{0.5mm}dE_{\nu_\mu}\; \frac{dN}{dE_{\nu_\mu}}\,\left(\frac{d\sigma_{\nu_\mu A}}{dx_B \hspace{0.5mm} dQ^2}\right) \,.
\label{eq:nint}
\eeq
Here, the factor $n_T$ denotes the nucleon density of the detector material and $L_T$ its effective length. For a flux given per unit area, time, and solid angle, Eq.~(\ref{eq:nint}) has to be multiplied by the corresponding exposure. The integrations extend over the relevant ranges of the neutrino energy~$E_{\nu_\mu}$, the Bjorken variable~$x_B$, and the momentum transfer~$Q^2$. The quantity $dN/dE_{\nu_\mu}$ represents the incident neutrino flux, while $d\sigma_{\nu_\mu A}/(dx_B\hspace{0.25mm}dQ^2)$ is the corresponding differential CC DIS cross section per nucleon of the target nucleus~$A$.

To simulate a variable neutrino energy spectrum in the \texttt{POWHEG-BOX} DIS generator, we set \texttt{fixed\_lepton\_beam~0} in the \texttt{powheg.input} file, as described at the beginning of~Section~\ref{sec:POWHEGsettings}. A realistic astrophysical flux can then be incorporated by modifying the \texttt{pdf\_lepton\_beam} subroutine in \texttt{lepton\_flux.f}, following the procedure outlined in detail in~Ref.~\cite{FerrarioRavasio:2024kem}. We assume an isotropic single power-law (SPL) astrophysical neutrino flux, consistent with the IceCube~publication~\cite{IceCube:2024fxo}, with flavour composition $\nu_e:\nu_\mu:\nu_\tau=0:1:0$, i.e. with the $\nu_e$ and~$\nu_\tau$ contributions switched off, and a neutrino-to-antineutrino ratio, $\nu_\ell:\bar{\nu}_\ell=1:1$, at the Earth's surface. In practice, we adopt
\beq \label{eq:Phiastro}
\Phi_\mathrm{astro}^{\nu_\mu + \bar{\nu}_\mu}=\phi_\mathrm{astro}^{\text{per-flavour}}\cdot\left(\frac{E_{\nu_\mu}}{100\,\mathrm{TeV}}\right)^{-\gamma} \,,
\eeq
where
\beq \label{eq:Phiastropara}
\phi_\mathrm{astro}^{\text{per-flavour}}=1.68 \cdot 10^{-18}\,\mathrm{GeV}^{-1}\,\mathrm{cm}^{-2}\,\mathrm{sr}^{-1}\,\mathrm{s}^{-1}\,,\qquad\gamma=2.58\,.
\eeq
The values of $\phi_\mathrm{astro}^{\text{per-flavour}}$ and $\gamma$ correspond to the central values extracted in~Ref.~\cite{IceCube:2024fxo} for a single neutrino flavour.

We parametrise the flux as a neutrino PDF of flavour~$\ell$, interpreted as a probability density for observing a neutrino (or antineutrino) with fractional energy~$x_{\nu_\ell}$ with respect to a maximum energy~$E^\mathrm{max}_{\nu_\ell}$:
\beq \label{eq:nupdf}
x_{\nu_\ell}=\frac{E_{\nu_\ell}}{E^\mathrm{max}_{\nu_\ell}}\,,\qquad f(x_{\nu_\ell})=E^\mathrm{max}_{\nu_\ell} \, \frac{dN}{dE_{\nu_\ell}}\,,
\eeq
where the normalisation for the neutrino PDF $f(x_{\nu_\ell})$ is chosen in such a way that
\beq
N^\mathrm{flux}_{\nu_\ell} = \int_{x^\mathrm{min}_{\nu_\ell}}^{1} \! dx_{\nu_\ell}\,f(x_{\nu_\ell})=\int_{E^\mathrm{min}_{\nu_\ell}}^{E^\mathrm{max}_{\nu_\ell}}dE_{\nu_\ell}\,\frac{dN}{dE_{\nu_\ell}}\,.
\eeq
Since we are interested in high-energy astrophysical neutrinos, we generate events in the interval $[E^\mathrm{min}_{\nu_\ell} ,E^\mathrm{max}_{\nu_\ell} ]=[10^3,10^7]\,\mathrm{GeV}$. The event generator does not include top-quark production (see Section~\ref{sec:finalcomparison}), whose effects are small for the SPL flux considered here.

\begin{figure}[t!]
\begin{center}
\includegraphics[width=0.49\textwidth]{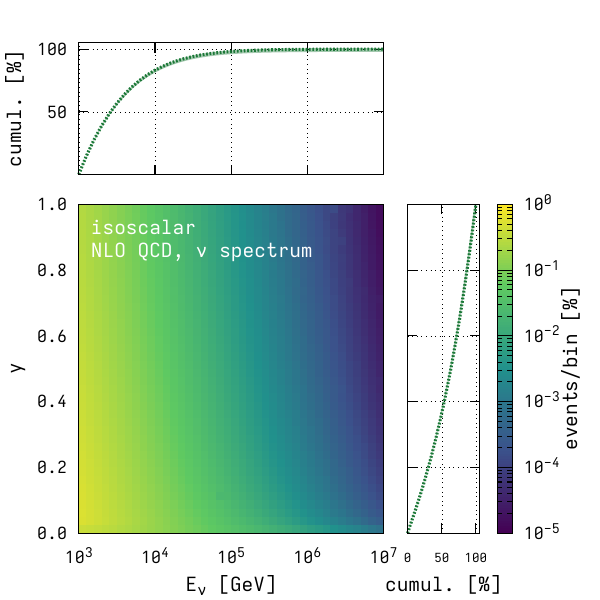}
\includegraphics[width=0.49\textwidth]{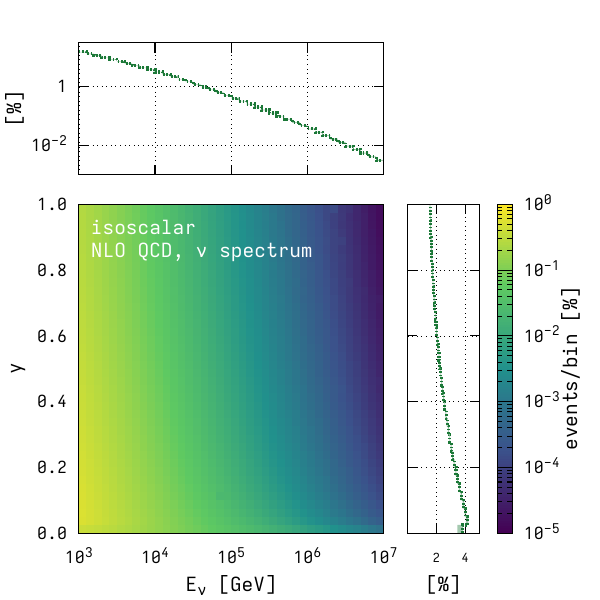}
\vspace{0mm}
\caption{Distribution of interacting events as a function of the incoming neutrino energy and the inelasticity for the astrophysical $\nu_\mu + \bar{\nu}_\mu$ flux of~Eq.~(\ref{eq:Phiastro}) and an isoscalar target, given in percent of the total number of events. Shown is the central value of the NLO~QCD prediction obtained with the {\tt POWHEG-BOX} DIS generator at the level of Les Houches events, i.e.~without PS. The side panels show the projections onto $E_\nu$ (top) and $y$ (right) as cumulative distributions~(left panel) and per bin~(right panel).}
\label{fig:dsig2_LG}
\end{center}
\end{figure}

Using the generator at the level of Les Houches events, i.e.~without a subsequent PS, we obtain the results shown in~Figure~\ref{fig:dsig2_LG}, which display the percentage of interacting events in each bin. Since the distribution is normalised to the total number of events, it does not depend on the detector geometry or exposure. As in~Figure~\ref{fig:dsig2_RG} (left panel), we show the central value of the NLO QCD prediction. The two panels show the same double-differential distribution and differ only in the one-dimensional projections: the cumulative distributions in the left panel directly give the fraction of events below a given neutrino energy or inelasticity, while the projections per bin in the right panel display the shape of the distributions. Unlike the results in~Figure~\ref{fig:dsig2_RG}, which correspond to discrete values of~$E_{\nu_\mu}$, the predictions shown here incorporate a continuous range of incident neutrino energies. Relative to~Figure~\ref{fig:dsig2_RG}, the flux weighting shifts the bulk of the interacting events towards the lowest neutrino energies: the steeply falling flux of~Eq.~\eqref{eq:Phiastro} more than compensates for the increase in the cross section with energy, so that the number of events per bin decreases by almost four orders of magnitude between $10^3\,{\rm GeV}$ and $10^7\,{\rm GeV}$. As a result, about $83\%$~($98\%$) of the events have $E_\nu < 10^4\,{\rm GeV}$ ($10^5\,{\rm GeV}$). The projection onto~$y$ rises towards small inelasticities, apart from a slight decrease in the lowest bin, with about $63\%$ of the events at $y < 0.5$. This behaviour reflects the suppression of~$d\sigma/dy$ as $y \to 0$ at the low neutrino energies that dominate the flux-weighted sample. See~Figure~\ref{fig:reshuff} for~details.

\paragraph{Current limitations:} When attempting to shower the events underlying~Figure~\ref{fig:dsig2_LG}, we encountered a technical limitation of {\tt Pythia~8.3}, which does not support neutrino beams whose energy varies on an event-by-event basis. This limitation can be circumvented by reinitialising {\tt Pythia~8.3} for each event. While this allows the simulation of a variable-energy neutrino beam, it substantially increases the computational cost: after the first initialisation, generating a single event in this way takes about $1000$ times longer than for a fixed-energy beam. Given this overhead, we do not perform a full PS simulation of the flux-weighted distributions shown in~Figure~\ref{fig:dsig2_LG}. Native support for variable neutrino-beam energies in {\tt Pythia~8.3} would remove this limitation.

\paragraph{Future improvements:} Our flux implementation assumes an isotropic SPL astrophysical neutrino flux, which suffices to validate the key components of the event-simulation tool chain, namely the generation of the hard process and its matching to the PS. A more realistic description of the flux could be obtained by replacing the SPL parametrisation with broken power-law or log-parabola models, which have been shown to describe the IceCube data more accurately~\cite{IceCube:2014rwe,IceCube:2020acn,IceCube:2025tgp}. Furthermore, our simulation does not account for detector~effects. Incorporating detector-specific acceptance factors that capture the geometry and response of a given experimental setup would enable more realistic event-rate predictions and facilitate more direct comparisons with experimental observations. We~leave these extensions to future studies by the individual experiments.

As shown in Section~\ref{sec:finalcomparison}, the NLO$+$PS predictions miss several contributions whose impact grows with the incident neutrino energy. Incorporating all these effects into an exclusive event simulation will therefore be essential for improving the accuracy of future predictions. Some of these contributions can already be simulated with existing tools. For~example, on-shell $W$-boson production in coherent scattering, including an interface to {\tt Pythia~8.3}, was studied in~Ref.~\cite{Garcia:2020jwr} --- see also~Ref.~\cite{GENIE:2021npt}. Top-quark production can be simulated with the DIS generator presented in~Ref.~\cite{Buonocore:2024pdv}. By contrast, an implementation of the exact NLO QCD$+$EW corrections matched to a PS, particularly an interleaved QCD$+$QED PS, is not currently available for neutrino-induced CC~DIS, whereas for charged-lepton--proton DIS exclusive $\mathcal{O}(\alpha)$ QED corrections have long been available~\cite{Kwiatkowski:1990es,Charchula:1994kf}. Developing such a framework would also be valuable for NC~DIS, especially in the context of lepton-hadron collider phenomenology. We leave these developments to future work.

\section{Conclusions}
\label{sec:conclusions}

To fully exploit the increasing experimental precision of upcoming neutrino telescopes, such as GRAND~\cite{GRAND:2018iaj} and IceCube-Gen2~\cite{IceCube-Gen2:2020qha,IceCube-Gen2:2021rkf}, as well as future accelerator-based neutrino detectors, including SHiP~\cite{SHiP:2015vad,SHiP:2025ows} and the FPF~\cite{Anchordoqui:2021ghd,Feng:2022inv}, precise theoretical predictions for neutrino-hadron interactions, in particular DIS, will be essential. Achieving percent-level theoretical uncertainties for neutrino-induced~CC~DIS distributions in realistic exclusive experimental setups requires the inclusion of QED effects from collinear photon emission off high-energy leptons. These corrections are enhanced by terms of the form $\alpha/\pi \hspace{0.25mm} \ln \left(E_\ell/m_\ell \right)$ and can reach $10\%$ or more. It is therefore not sufficient to account for QCD corrections alone~\cite{Dicus:2001kb,Gandhi:1998ri,Gluck:1998js,Cooper-Sarkar:2007zsa,Connolly:2011vc,Cooper-Sarkar:2011jtt,Bertone:2018dse,Candido:2023utz,Xie:2023suk} in order to reach the desired precision. Although QED and EW corrections to~DIS~have been investigated over several decades~\cite{DeRujula:1979grv,Sigl:1997cq,Seckel:1997kk,Diener:2003ss,Diener:2005me,Alikhanov:2015kla,Ballett:2018uuc,Gauld:2019pgt,Zhou:2019vxt,Zhou:2019frk,Garcia:2020jwr,Xie:2023qbn,Plestid:2024bva,Weigel:2024gzh}, the interplay of collinear QED radiation, photon-induced contributions, and nuclear effects in realistic exclusive observables has so far prevented a fully quantitative assessment of their impact on neutrino-detector~signatures.

In this work, we have taken a step towards precision predictions for exclusive~DIS processes by combining higher-order QCD and EW effects in a fully differential framework and enabling direct comparisons with present and future (ultra-)high-energy neutrino measurements. We~performed a fully differential DIS calculation including NLO QCD and EW corrections, implemented in a dedicated fixed-order MC~program and supplemented by higher-order, logarithmically enhanced QED corrections through the lepton~FF~of~Ref.~\cite{Arbuzov:2002cn}, thereby controlling terms up to $\mathcal{O} \big( \alpha^2 \hspace{0.25mm} \ln \left (\mu_h/m_\ell \right) \big)$. QED contributions from coherent and diffractive nuclear scattering --- including on-shell $W$-boson production with subsequent decay --- are incorporated using the formalism of~Refs.~\cite{Ballett:2018uuc,Gauld:2019pgt,Garcia:2020jwr}. We~have further compared our predictions with those of the \NLOPS~QCD event generator of~Ref.~\cite{FerrarioRavasio:2024kem}, which provides a fully exclusive simulation of neutrino-induced DIS. In~this~{\tt POWHEG} implementation, logarithmically enhanced QED effects from multiple soft or collinear FSR photons are resummed by the~PS, whereas non-logarithmic QED and EW effects are not included.

\begin{figure}[t!]
\begin{center}
\includegraphics[width=0.6\textwidth]{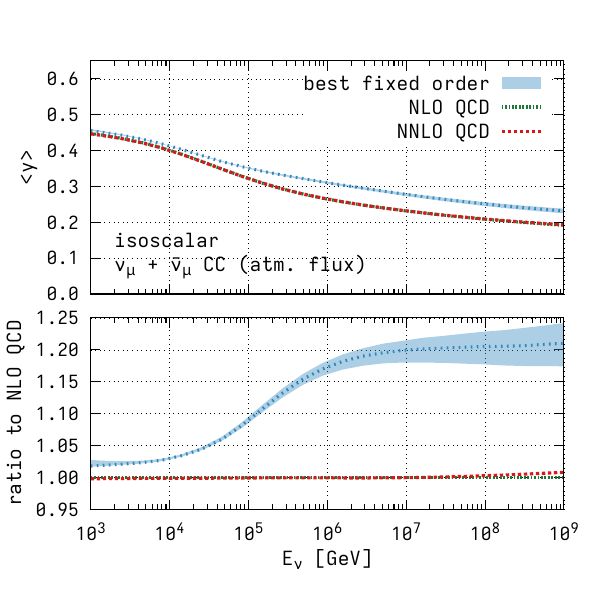}
\vspace{0mm}
\caption{Comparison of the flux-averaged mean inelasticity $\langle y \rangle$ on an isoscalar target. The best fixed-order prediction of~Eq.~(\ref{eq:best}), shown in blue with scale uncertainties, is compared with the fixed-order NLO (green dotted) and NNLO (red dashed) QCD predictions. The upper panel shows $\langle y \rangle$, while the lower panel shows the ratio to the NLO~QCD result. See the main text for further explanations.}
\label{fig:moneyplot}
\end{center}
\end{figure}

Our analysis systematically disentangles the effects of higher-order QED and~EW corrections on key DIS observables and identifies the kinematic regions in which enhanced FSR and fixed-order EW effects are most relevant. We find that NLO~EW corrections to inclusive neutrino-DIS cross sections are at the percent level, typically larger than NNLO~QCD effects but still below current QCD and nuclear uncertainties. In~inclusive CC~DIS, additional flavour-dependent effects arise from photon-induced channels associated with the hidden Glashow resonance, leading to a characteristic energy dependence and enhanced corrections, particularly in the electron channel where lepton-mass logarithms are more pronounced. For~$\nu_\mu$-induced CC~DIS, the NLO~EW corrections to the mean inelasticity~$\langle y \rangle$ reach the $15\%$ level at ultra-high energies, driven by the combined effects of collinear photon FSR and photon-induced contributions. The latter become relevant already at neutrino energies $E_{\nu_\mu} \gtrsim 10^4 \, {\rm GeV}$. 

We~are able to reproduce the results of~Ref.~\cite{Plestid:2024bva} when adopting their prescription for the scale choice in the evaluation of the LL~FSR corrections, $\mu_h^2 = s \simeq 2 \hspace{0.25mm} m_N E_{\nu_\mu}$. With~this choice, at $E_{\nu_\mu} = 10^8\,{\rm GeV}$, the LL correction is about $1.5$ times larger than the full NLO~EW correction, corresponding to an overestimate of the QED radiation by roughly $50\%$. The~origin of this difference can be traced to the use of the hadronic COM energy $s$ rather than the partonic hard-scattering scale $\hat{s}_{12}$. In~collinear factorisation, $\hat{s}_{12}$ scales as $x\hspace{0.125mm}s$ and can therefore be several orders of magnitude smaller than $s$ for the small momentum fractions $x$ relevant to ultra-high-energy neutrino scattering. Using~instead $\mu_h^2=\hat{s}_{12}$ accounts for the relevant partonic kinematics and yields an LL estimate that substantially improves the description of the fixed-order NLO~EW result. Including the initial condition for the lepton fragmentation function and higher-order effects leads to percent-level modifications in selected regions of phase space, but these contributions are generally subleading. 

Imposing a minimum muon energy threshold of $1 \, {\rm TeV}$ to emulate the experimental reconstruction of muon-track events leaves the NLO~EW corrections to the inclusive CC~DIS cross section essentially unchanged and induces only a negligible shift in $\langle y \rangle$, demonstrating the robustness of our predictions with respect to a simple muon-energy selection. A dedicated study of reconstruction effects is left to the experimental collaborations. Finally, coherent contributions modify the cross section at the percent level and $\langle y \rangle$ by up to about $6\%$ and exhibit a characteristic Glashow-like logarithmic enhancement, while lepton~FF~effects at~$\mathcal{O}(\alpha^2)$ remain at the permille level.

The main phenomenological implications of our results are summarised in the ``money plot'' of~Figure~\ref{fig:moneyplot}, which compares our best prediction of~Eq.~(\ref{eq:best}) with the NLO and NNLO QCD results for the flux-averaged mean inelasticity~$\langle y \rangle$ on an isoscalar target. The~predictions use the atmospheric neutrino flux, including both conventional and prompt components, computed with the help of \texttt{MCEq~1.3.0}~\cite{Fedynitch:2015zma}, using the hadronic interaction model \texttt{SIBYLL~2.3c}~\cite{Riehn:2017mfm,Fedynitch:2018cbl} and the H3a primary cosmic-ray spectrum~\cite{Gaisser:2011klf}. The resulting flux is averaged over the down-going hemisphere ($0 \le \cos\theta \le 1$), for which absorption in the Earth is absent. Since $\langle y \rangle$ is shown as a function of $E_{\nu}$, the flux enters only through the relative weights of the different neutrino and antineutrino flavours. For reference, our QCD predictions should be compared with the red line in~Figure~8 of~Ref.~\cite{IceCube:2018pgc}, based on the NLO QCD DIS calculation of~\cite{Cooper-Sarkar:2011jtt}. The EW corrections to $\langle y \rangle$ remain below about $+3\%$ for neutrino energies below $10^4\,{\rm GeV}$ but reach up to $+20\%$ at $10^9\,{\rm GeV}$, exceeding the theoretical uncertainties of the QCD predictions. Their~present phenomenological impact is nevertheless limited by the experimental uncertainties, which are about~$70\%$ above $10^5\,{\rm GeV}$ in the IceCube inelasticity measurement~\cite{IceCube:2018pgc} after symmetrisation. This measurement is~based on approximately five years of data taking, whereas the more recent study~\cite{IceCubeCollaborationSS:2025zgz},~based on more than nine years of data, measures the inelasticity distribution in the energy range $[80,560]\,{\rm GeV}$. An~updated measurement extending to higher energies would therefore be particularly valuable for testing the impact of EW corrections as experimental precision improves.

\acknowledgments{
The~Feynman diagrams in this article were generated using {\tt TikZ-Feynman}~\cite{Ellis:2016jkw}. 
We acknowledge Alexander~Karlberg and Gavin~Salam for their assistance with {\tt HOPPET~v2}~\cite{Karlberg:2025hxk}, and thank Giulia Zanderighi for numerous insightful comments throughout the project. 
We additionally thank Silvia Ferrario Ravasio for comments on the use of shower algorithms in the DIS process.
RG is grateful to Subir~Sarkar for introducing him to this topic and for many stimulating discussions over the years, and also acknowledges valuable discussions with Philipp~Eller, Colton~Hill, and Alexandra~Trettin. 
Finally, the authors thank Alois~Kabelschacht for keeping the spirits high throughout this work and for valuable input on the treatment of subtraction terms for QED radiation. 
The final version of this manuscript was reviewed by artificial intelligence~(AI) in the form of the large language model Claude~Opus~5.5~(Anthropic). It~assisted with improving the language and with identifying inconsistencies between the text, the figures, and the equations. The authors reviewed and edited all AI-assisted content and remain fully responsible for all results presented in this article, including any remaining mistakes, which are therefore of entirely natural origin.}

\begin{appendix}

\section{Hidden Glashow process}
\label{app:glashow}

In~Section~\ref{sec:sigmainclusive}, we computed the flavour-dependent NLO~EW effects in CC~DIS associated with the hidden Glashow resonance for electron, muon, and tau flavours. While a fully massive calculation is straightforward for the muon and tau channels, the electron case is more subtle: the process $\nu_{e} + \gamma \to e^- + q \bar{q}^{\hspace{0.25mm} \prime}$ develops large logarithms in the limit $m_e \to 0$, which can lead to numerical instabilities if not treated with care.

\begin{figure}[t!]
\begin{center}
\includegraphics[width=0.49\textwidth]{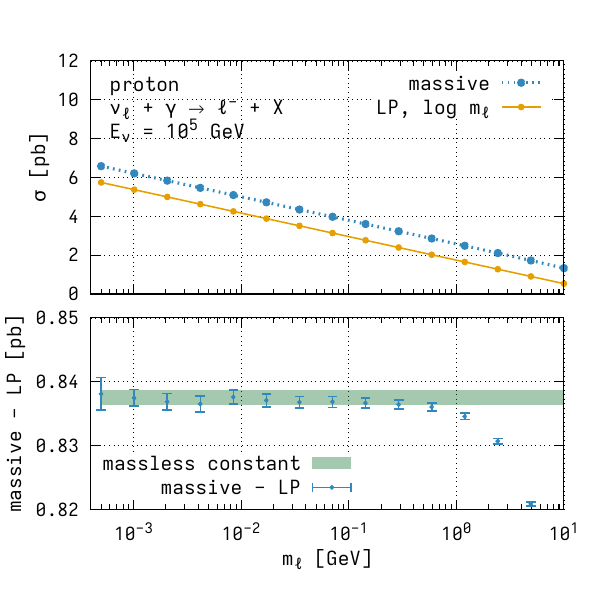}
\includegraphics[width=0.49\textwidth]{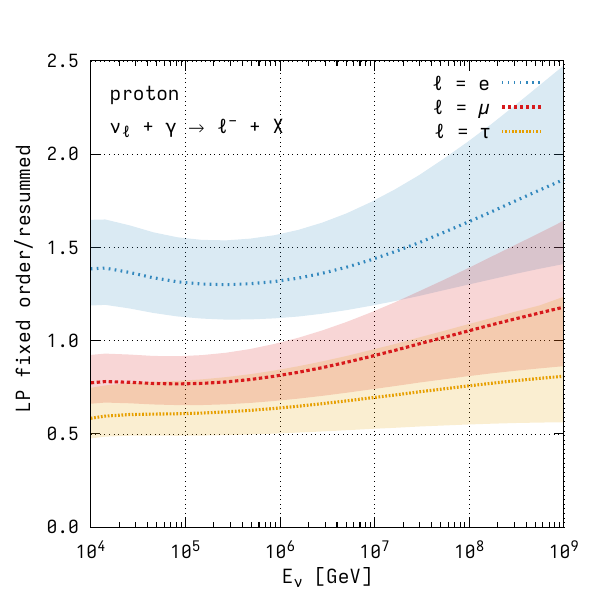}
\vspace{2mm}
\caption{Left: Test of the small lepton-mass limit for the photon-induced contribution in CC~DIS on a proton target at fixed neutrino energy $E_{\nu_{\ell}} = 10^5 \, {\rm GeV}$. The massive calculation is compared to the LP result. The lower panel shows their difference, which approaches a constant in the massless limit, reproduced by the mass-factorisation result. Right: Ratio of the fixed-order to the resummed LP predictions for electron, muon, and tau flavours. The~resummed result is obtained at LO using the lepton PDFs of~Ref.~\cite{Buonocore:2020nai}. The bands indicate the scale uncertainties. Additional information is provided in the main text. \label{fig:GlashowTest}}
\end{center}
\end{figure}

To test the small lepton-mass limit, we compare the exact massive calculation to the approximation
\beq \label{eq:sigmaLL}
d\sigma_{{\rm LP},{\rm QED}} = \int \! dx \, f^{(1)}_{\ell}(x,\mu_F^2)\, d\hat{\sigma}_{\rm LO}\,,
\eeq
where the fixed-order lepton PDF is defined by
\beq \label{eq:lepPDF}
f^{(1)}_{\ell}(x,\mu_F^2) = \frac{\alpha}{2\pi} \ln\!\left(\frac{\mu_F^2}{m_\ell^2}\right) \int_x^1 \! \frac{dz}{z} \, P^{(0)}_{\ell \gamma}(z)\, f_{\gamma}\!\left(\frac{x}{z},\mu_F^2\right)\,, \quad P^{(0)}_{\ell \gamma}(z) = z^2 + (1 - z)^2 \,.
\eeq
This approximation will be referred to as leading-power (LP). Here, $d\hat{\sigma}_{\rm LO}$ denotes the differential Born-level cross section for $\nu_{\ell} + \ell^{+} \to q + \bar{q}^{\hspace{0.25mm} \prime}$, $f_{\gamma}(x,\mu_F^2)$ is the photon~PDF, and $P^{(0)}_{\ell \gamma}(z)$ is the LO splitting function for $\gamma \to \ell^+ \ell^-$. This test is performed in the left panel of~Figure~\ref{fig:GlashowTest} for a fixed neutrino energy of $E_{\nu_{\ell}} = 10^5 \, {\rm GeV}$. In the upper panel, the fully massive calculation is compared to the LP approximation, while in the lower panel the difference between the massive and LP predictions is shown. In the massless limit, this difference approaches a constant, which can also be obtained directly after applying lepton mass factorisation. The corresponding prediction is shown as a solid green band in the lower panel, with its width reflecting MC~integration uncertainties, and is consistent within errors with the small-mass limit of the massive calculation. Based on these results, in~Section~\ref{sec:sigmainclusive} we use the massive calculation for muons and taus, while in the electron case we employ the LP approximation with the constant extracted from the left panel of Figure~\ref{fig:GlashowTest}.

The LP prediction defined in~Eq.~(\ref{eq:sigmaLL}) is evaluated using the fixed-order lepton PDF given in~Eq.~(\ref{eq:lepPDF}). Alternatively, one may employ a resummed description in which the collinear logarithms are absorbed into a lepton PDF obtained by QED evolution from the photon~PDF. Comparing these two approaches provides an estimate of the accuracy of the LP approximation. In the right panel of Figure~\ref{fig:GlashowTest} we perform such a comparison for the different lepton flavours. The~resummed prediction is obtained at LO using the evolved lepton PDFs of~Ref.~\cite{Buonocore:2020nai}. For muons, the fixed-order and resummed predictions agree within the scale uncertainties for $E_{\nu_\ell} \gtrsim 10^5\,{\rm GeV}$, with central values differing by up to about $20\%$. For tau leptons, the fixed-order result underestimates the resummed prediction by $20\%\hspace{0.25mm}$--$\hspace{0.25mm}40\%$, a difference that is covered by the scale uncertainties only at the highest energies. In the electron channel, where the collinear logarithms are largest, the fixed-order result exceeds the resummed prediction by $30\%\hspace{0.25mm}$--$\hspace{0.25mm}85\%$, well outside the scale uncertainties. Missing higher-order collinear logarithms therefore induce a sizeable uncertainty in the photon-induced contribution for electrons, which should be kept in mind when interpreting the corresponding results in~Figure~\ref{fig:ratioNCCC}. For muons, which are the focus of this work, the fixed-order treatment is adequate within its uncertainties.

\section{FSR QED corrections and lepton dressing}
\label{app:FSR}

The fixed-order predictions for processes of the form $\nu_\ell + q \to \ell^- + q^{\hspace{0.25mm} \prime} + \gamma$ in this work have been performed with finite lepton masses. For collinear-unsafe observables, such as the inelasticity distribution $y$ of muons, analysed in detail in~Sections~\ref{sec:exclusive} and~\ref{sec:comparison}, the cross section acquires a logarithmic sensitivity to the lepton mass, which has to be handled in the phase-space integration over the final-state particles. Since this fixed-order calculation provides an important baseline for benchmarking comparisons with the \NLOPS~predictions, it is also essential to test the numerical stability, efficiency, and correctness of the implementation.

\begin{figure}[t!]
\begin{center}
\includegraphics[width=0.49\textwidth]{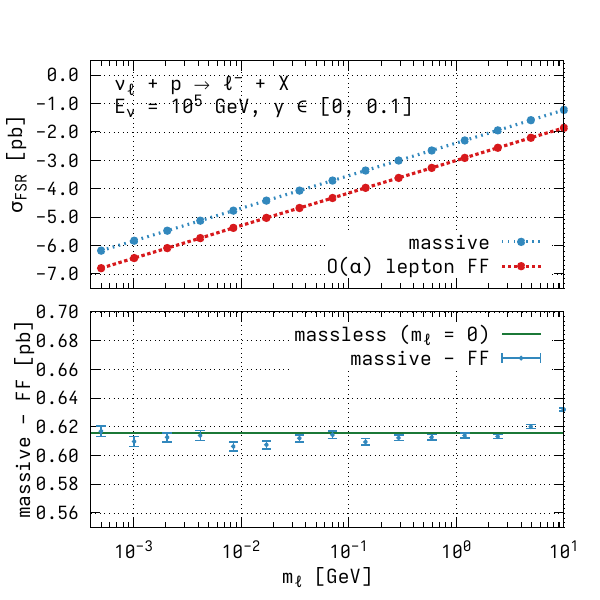}
\includegraphics[width=0.49\textwidth]{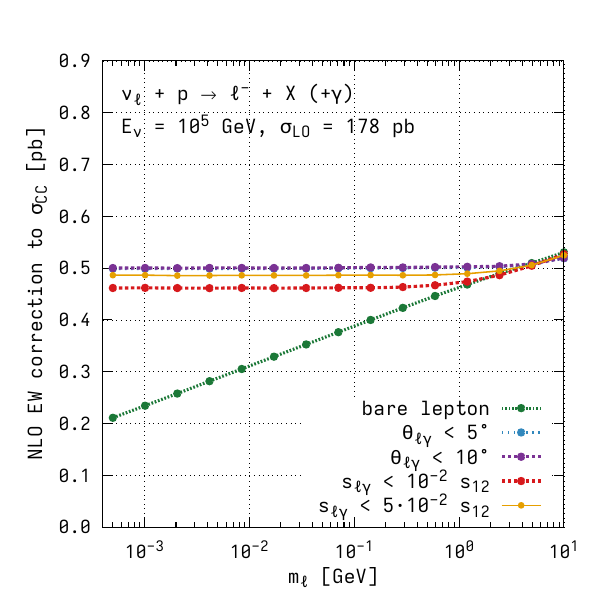}
\vspace{2mm}
\caption{Left: Test of the small lepton-mass limit for the QED FSR contribution in~CC~DIS at fixed neutrino energy $E_{\nu_{\ell}} = 10^5 \, {\rm GeV}$. The massive calculation is compared to the result of the lepton~FF at~$\mathcal{O}(\alpha)$. The lower panel shows their difference, which approaches a constant in the massless limit, compared with the independent calculation for a massless lepton~(green band, MC uncertainty). Right: The coefficient of the total NLO~EW correction is shown for various lepton-dressing schemes, including bare leptons. Consult the main text for additional details. \label{fig:FSRTest}}
\end{center}
\end{figure} 

To perform this comparison, we consider the FSR-sensitive subprocess $\nu_{\ell} + q \to \ell^- + q^{\hspace{0.25mm} \prime} + \gamma$ at a fixed neutrino energy of $E_{\nu_{\ell}} = 10^5 \, {\rm GeV}$ and vary the lepton mass across several orders of magnitude, from the electron mass up to the \GeV~range. The photon-induced process is not included in this comparison (see the discussion in the previous appendix), and we impose only a kinematic selection~$y \in [0.0,0.1]$ to ensure sensitivity to collinear~QED emissions off~the final-state lepton leg. The result of the comparison between the fixed-order prediction and that obtained using the lepton~FF~at $\mathcal{O}(\alpha)$ is shown on the left-hand side in~Figure~\ref{fig:FSRTest}. In the upper panel, we compare the fixed-order result with the lepton~FF prediction at ${\cal O}(\alpha)$, including its initial condition. The lower panel shows their difference alongside an independent massless calculation based on the ``many identified particles'' formalism of the dipole subtraction method~\cite{Catani:1996vz}. The difference between the massive and lepton~FF predictions approaches a constant as $m_{\ell}\to 0$, in agreement with the direct massless result. The numerical implementation remains stable down to the physical electron mass. Each data point is obtained with the same number of integrand evaluations, indicating that the efficiency of the code is not significantly degraded for small lepton masses. This stability can be achieved, for example, by constructing the phase-space integration such that the invariant mass of the combined~$\ell^- \gamma$ system is used as one of the integration variables.

As an additional remark, the lepton~FF~in~Figure~\ref{fig:FSRTest} includes its initial condition at ${\cal O}(\alpha)$. Retaining only the LL approximation of the~FF instead, the difference to the massive calculation changes by a constant: the initial condition of the muon~FF contributes $-0.71\,{\rm pb}$. Expressed relative to the LO cross section in the bin $y \in [0, 0.1]$, this amounts to a correction of nearly $-2\%$, indicating that the effect is numerically non-negligible. This arises from the fact that the initial condition contains a $\ln \left ( 1-z \right )$ behaviour, which enhances the region where the collinear splitting variable approaches $z \to 1$, while the Born cross section is also peaked towards small values of $y$. It is worth emphasising that this term, which is by construction part of the massive calculation, is not included in the \NLOPS~predictions, which contain only LL~corrections. The impact of this difference can be clearly seen in the comparisons in~Section~\ref{sec:comparison}.

To conclude this discussion, we compare different procedures for dressing the outgoing charged lepton in the presence of photon radiation. As described in~Section~\ref{sec:dressed}, we adopt the lepton-dressing prescription introduced in~Ref.~\cite{Diener:2003ss}. The right panel of~Figure~\ref{fig:FSRTest} illustrates the impact of several dressing schemes on the integrated cross section for the same incident neutrino energy. As distance measures for the recombination procedure, we consider both the angular separation defined in the laboratory frame, introduced in~Section~\ref{sec:dressed}, and the invariant dot product normalised to the partonic COM energy. All dressing prescriptions lead to very similar results and yield finite cross sections in the limit $m_\ell \to 0$. The residual differences between the various schemes amount to only about $0.02\%$ of the LO prediction. By contrast, employing the collinear-unsafe reference scale $\mu^2=Q^2$, computed from the bare lepton momentum (labelled ``bare lepton'' in~Figure~
\ref{fig:FSRTest}), results in an inclusive cross section that retains a logarithmic dependence on the charged-lepton mass.

\section{Top-quark production}
\label{app:Top}

\begin{figure}[t!]
\begin{center}
\includegraphics[width=0.49\textwidth]{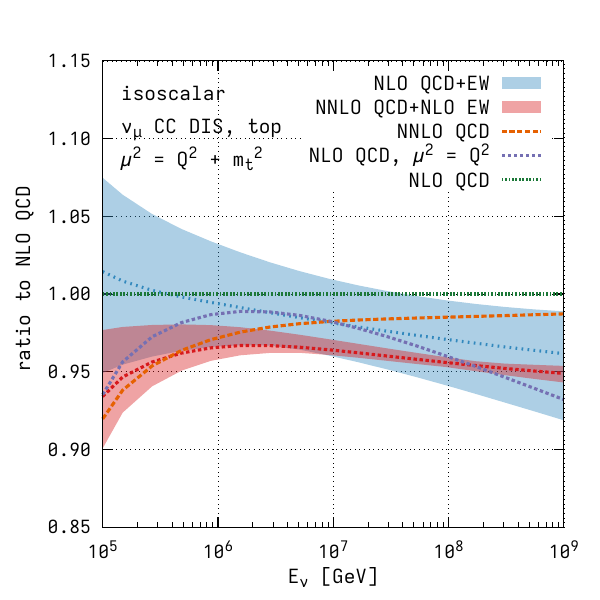}
\includegraphics[width=0.49\textwidth]{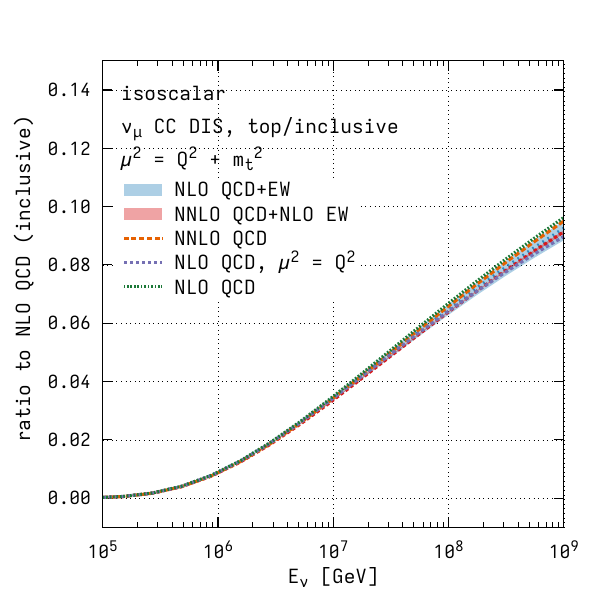}
\vspace{2mm}
\caption{Left:~Impact of higher-order corrections to the process $\nu_{\mu} + N \to \mu^- + t + X$ for an isoscalar target. The cross-section predictions are shown normalised to the NLO~QCD result~(green dotted). The NLO~QCD$+$EW~(blue) and NNLO~QCD$+$NLO~EW~(red) predictions are shown with their scale uncertainties, together with the central NNLO~QCD prediction~(orange dashed). The NLO~QCD prediction obtained using $\mu^2 = Q^2$ (violet dashed) is also shown for reference. Right:~The~same predictions shown normalised to the flavour-inclusive NLO QCD cross section. Further explanations can be found in the main~text. \label{fig:Top}}
\end{center}
\end{figure} 

As noted in~Section~\ref{sec:DIScalc}, top-quark production, $\nu_{\ell} + N \to \ell^- + t + X$, where $X$ denotes the outgoing hadronic system, becomes kinematically accessible above $E_{\nu_\ell} \simeq m_t^2/(2 \hspace{0.25mm} m_N) \simeq 16\,{\rm TeV}$ and numerically relevant for $E_{\nu_\ell} \gtrsim 100\,{\rm TeV}$. This contribution has been studied in several works using different approximations~\cite{Cooper-Sarkar:2011jtt,Barger:2016deu,Bertone:2018dse,Garcia:2020jwr}. Here we present a fixed-order calculation of top-quark production including both QCD and EW corrections, which retains the full dependence on the top-quark mass and, in the EW corrections, on the lepton mass. The relative impact of the QCD and EW corrections is quantified in~Figure~\ref{fig:Top} for the process $\nu_{\mu} + N \to \mu^- + t + X$ on an isoscalar target. As default, we use the scale choice $\mu^2 = Q^2_{\rm dressed} + m_t^2$ introduced in~Section~\ref{sec:hadinput}, and for comparison we also show the central NLO QCD result obtained with $\mu^2 = Q^2$. NNLO QCD predictions for massive-quark production in CC~DIS are available in~Refs.~\cite{Berger:2016inr,Gao:2017kkx,Caola:2026kvl}. We take the NNLO QCD corrections to top-quark production from the ancillary files of~Ref.~\cite{Caola:2026kvl}, using the same PDFs, top-quark mass, and scale choice as in our NLO predictions, and combine them additively with the NLO EW corrections. Scale uncertainties are shown for both the NLO QCD$+$EW and the NNLO QCD$+$NLO EW predictions.

For the considered energy range, we observe that the EW corrections to top-quark production range from $-4\%$ to $+1\%$, and that at high energies these corrections exceed the scale-variation band. Unlike the calculation with massless quarks, the photon-induced channel in top-quark production does not receive a resonant enhancement, since $m_t > m_W$. The NNLO QCD corrections are negative, reducing the cross section relative to NLO QCD by about $6\%$ at the lowest energies shown, $E_{\nu_\mu} \simeq 1.5 \cdot 10^5 \, {\rm GeV}$, where the NNLO QCD result lies below the NLO scale-variation band, and by about $1.5\%$ for $E_{\nu_\mu} \gtrsim 10^7 \, {\rm GeV}$. At~the same time, the scale uncertainty is substantially reduced with the inclusion of the NNLO corrections. As the top-quark component of the cross section reaches only about $10\%$ of the total cross section at $E_{\nu_\mu} = 10^9 \, {\rm GeV}$, the EW corrections to top-quark production translate into an overall correction of approximately $-0.4\%$ on the total rate, and the combined NNLO QCD$+$NLO EW corrections into approximately $-0.5\%$. The EW calculation of top-quark production presented here can also be adapted to the case of differential charm-quark production, which is relevant for the description of charm production in the multi-GeV to TeV energy range. Such an application will be considered in a separate work.

\end{appendix}


\providecommand{\href}[2]{#2}\begingroup\raggedright\endgroup

\end{document}